\documentclass[a4paper,12pt]{report}
\usepackage{amsfonts,amssymb,amsmath,graphicx,multirow}
\begin{document}

\def\CHECK{{\bf (CHECK)}}

\pagestyle{empty}
\begin{center}

\textbf{\LARGE Adventures in Radio Astronomy \linebreak \linebreak Instrumentation and Signal Processing} \linebreak \linebreak
by \linebreak \linebreak
{ \large Peter Leonard McMahon } \linebreak \linebreak
Submitted to the Department of Electrical Engineering \linebreak
in partial fulfillment of the requirements for the degree of \linebreak \linebreak
Master of Science in Electrical Engineering \linebreak \linebreak
at the \linebreak \linebreak
\textsc{University of Cape Town} \linebreak \linebreak
July 2008 \linebreak \linebreak

Supervisor: Professor Michael Inggs \linebreak \linebreak
Co-supervisors: \linebreak
Dr Dan Werthimer, {\it CASPER\footnote{Center for Astronomy Signal Processing and Electronics Research}, University of California, Berkeley} \linebreak
Dr Alan Langman, {\it Karoo Array Telescope} \linebreak \linebreak

\end{center}
 
\newpage
\pagestyle{plain}
\pagenumbering{roman}
\setcounter{page}{1}

\begin{abstract}
This thesis describes the design and implementation of several instruments for digitizing and processing analogue astronomical signals collected using radio telescopes. \\

Modern radio telescopes have significant digital signal processing demands that are typically best met using custom processing engines implemented in Field Programmable Gate Arrays. These demands essentially stem from the ever-larger analogue bandwidths that astronomers wish to observe, resulting in large data volumes that need to be processed in real time. \\

We focused on the development of spectrometers for enabling improved pulsar\footnote{Pulsars are rapidly rotating neutron stars that emit periodic broad band electromagnetic radiation.} science on the Allen Telescope Array, the Hartebeesthoek Radio Observatory telescope, the Nan\c{c}ay Radio Telescope, and the Parkes Radio Telescope. We also present work that we conducted on the development of real-time pulsar timing instrumentation. \\

All the work described in this thesis was carried out using generic astronomy processing tools and hardware developed by the Center for Astronomy Signal Processing and Electronics Research (CASPER) at the University of California, Berkeley. We successfully deployed to several telescopes instruments that were built solely with CASPER technology, which has helped to validate the approach to developing radio astronomy instruments that CASPER advocates. \\

\end{abstract}

\newpage

\textbf{Plagiarism Declaration} \\

I know the meaning of plagiarism and declare that all the work in this document, save for that which is properly acknowledged, is my own.

\newpage

\tableofcontents

\newpage

\listoffigures

\newpage

\listoftables

\newpage

{\it \Large To my teachers.}

\newpage

\section*{Acknowledgements}

An oft-used introduction to thesis acknowledgements is that there is an irony that somehow research papers list as authors anyone who contributes, whereas their far more lengthy brethren, theses, unfairly list only one author when in fact even greater debt is owed to many people. This is particularly apt in my case. This thesis would quite literally not have been possible without the assistance of many people, and would be considerably worse off were it not for the help of many more. I have relied heavily on the great generosity of my collaborators and the science and engineering communities at large. \\

The first pivotal figure in the creation of this thesis is Professor Inggs, who introduced me to Dr Alan Langman at the Karoo Array Telescope in September 2006, on the completion of my undergraduate project. Had that meeting not taken place, I likely wouldn't have undertaken this thesis. Prof. Inggs has been terrific in taking care of funding and in creating opportunities for me to collaborate with other groups, and has in this role been a great enabler for my research. \\

Dr Alan Langman at KAT has given me many hours of excellent advice, but most importantly, he provided me with the opportunity to spend 11 months with his collaborators in the CASPER\footnote{Center for Astronomy Signal Processing and Electronics Research, University of California, Berkeley} group at Berkeley. I hope that his faith in me has paid off in some small measure. The bulk of the work reported in this thesis was done while I was at Berkeley, so Alan's support of my visit made it possible for me to do the work that you can see presented to you here shortly. \\

Dan Werthimer at CASPER is the third individual to whom this thesis owes its existence, and hence to whom I am extremely grateful. Dan generously agreed to allow Jason Manley and I to join his group despite our lack of credentials and experience. He then proceeded to patiently teach me many of the fundamentals of building analogue/digital instruments that I should have known, but didn't, to get me to a point where I could carry out the projects that are described in these pages. Dan is an outstanding mentor and supervisor, and I think it's fairly safe to say that because of him I learnt more in my year in Berkeley that I've learnt in any year prior. Dan provided a superb combination of direction when I needed it and freedom for me to tinker otherwise, and this supervision style let me both learn and be semi-productive\footnote{The ``semi-'' caveat is required purely due to my inadequacies!}. \\

I also owe Dan a debt of gratitude for his willingness to make connections for me with his collaborators when I needed project ideas or help. Dan's extensive network of contacts made a massive difference in my ability to solve problems that I encountered. In this section I will try to thank all the people who have helped me over the past year and a half, but it is a great many, so please forgive me if I make any omissions. \\

At Berkeley, I was privileged to work with Prof. Don Backer and Dr Joeri van Leeuwen on instrumentation for pulsar science, and particularly a spectrometer for the Allen Telescope Array. I benefited greatly from the many discussions Don and I had; Don was an invaluable source of practical knowledge and techniques to meet challenges that routinely crop up in the development of pulsar instrumentation. Joeri was my day-to-day contact with the science user community, and he did an excellent job of keeping me focused on building ``something people can use''. He also put up with my stream of questions about pulsars as well as pulsar signal processing, so what little I know of pulsar science and the techniques for finding and timing pulsars I owe mostly to Joeri. \\

Dr Melvyn Wright and Prof. Geoffrey Bower, also from the Astronomy Department, both provided very useful advice, and kindly supported the efforts I was involved in to deploy instruments to the ATA. \\

Oren Milgrome, Dave MacMahon, Matt Dexter and Colby Kraybill from the Berkeley Radio Astronomy Lab\footnote{RAL is responsible for much of the digital electronics at the ATA, amongst many other aspects of the ATA's design and development.} (RAL) provided invaluable technical help that cumulatively probably saved me a couple of months. They also provided a great deal of entertainment during debugging sessions in the RAL basement. \\

Dr Rick Forster, the ATA's resident scientist at the Hat Creek Radio Observatory, was extremely helpful, and cheerfully reset servers for us in the middle of the night, moved disks around, and otherwise made maintaining our instruments at the ATA relatively painless. I'd also like to thank Rick and the rest of the staff at HCRO for all their help and hospitality during our site visits. \\

Dr Billy Barott, who is now an assistant professor at Embry-Riddle Aeronautical University, co-built the ATA beamformer, on which one of my instruments relies. Billy provided support to Joeri and I during the ATA pulsar machine planning phases, and while we were learning how to use the beamformer. Without his and Oren's help, we would undoubtedly have had great difficulty getting the Berkeley ATA Pulsar Processor (BAPP) system to work. \\

I'd like to thank Nancy Wang at Berkeley and Dr Rachael Padman at the University of Cambridge for providing help with some Stokes parameter statistics questions I had (thanks also to Mel for putting me in touch with Rachael). \\

CASPER's chief engineer, Henry Chen, provided many hours of excellent advice and free digital design tuition. I can't thank him enough. Without his help, I would have gone nowhere fast. \\

I entered the CASPER group as Aaron Parsons was leaving for Puerto Rico, but I managed to get from him an excellent introduction to the CASPER DSP libraries, of which he was the primary developer. Aaron did an excellent job with the libraries, and it's partially a testament to his workmanship that so many people are now using them. \\

I gratefully acknowledge Dr Chen Chang and Pierre-Yves Droz for their work on developing the BEE2 and the Simulink toolflow. They had both left Berkeley before my tenure, but their work lives on. \\

I had an excellent time in Berkeley courtesy of the students at the Berkeley Wireless Research Center and CASPER. I also benefitted from their help, and learnt from them all. Ben Blackman, Daniel Chapman, Terry Filiba, Griffin Foster, Greg Gibeling, Alex Krasnov, Vinayak Nagpal, Arash Parsa, Andrew Siemion, Mark Wagner, and my South African travel partner, Jason Manley, all contributed towards this thesis in meaningful ways. I also spent many fun non-work-related evenings with CASPERites -- thank you all! \\

The staff at the BWRC also provided excellent help and support -- I'd like to thank Brian Richards, Fred Burghardt, Dan Burke, Sue Mellers, Kevin Zimmerman and Brad Krebs. I also benefited greatly from many lunchtime discussions with Brian and Dan in particular.\\

It was a pleasure to work with Glenn Jones from Caltech. Glenn has been an excellent collaborator, and has acted as a very patient sounding-board for ideas on spectrometer system design. He also provided many library blocks that I have made use of in my projects. I had a useful and fun\footnote{Based on a visit there with Glenn, I can highly recommend the Western Sizzlin' Wood Grill Buffet in Harrisonburg, VA, to anyone making the trip from the East.} trip to NRAO in Green Bank, WV, with Glenn as well. \\

John Ford, Randy McCullough and Dr Glen Langston at NRAO Green Bank have been fun to work with, and I'd like to thank John especially for inviting me to his GUPPi pulsar workshop in 2007. I had interesting discussions with Dr Scott Ransom, Dr Paul Demorest, Prof. Maura McLaughlin and Prof. Dunc Lorimer at that meeting. I'd like to thank Maura in particular for help she subsequently gave me with {\it sigproc}, and Paul for his helpful advice on coherent dedispersion systems. \\

Glen Langston at NRAO, and Andrew Jameson and Willem van Straten at the Swinburne University of Technology kindly beta-tested the Parkes Spectrometer design (at Green Bank, and on the Parkes Radio Telescope, respectively). Glen has also been a valuable source of design ideas and techniques. \\

I'd like to thank Prof. Roy Booth, Dr Michael Gaylard, Dr Jonathan Quick and Sarah Buchner at the Hartebeesthoek Radio Astronomy Observatory for supporting my efforts to install and test a spectrometer on their main telescope. Jon and Sarah in particular deserve mention for all the effort they put into getting the spectrometer set up. \\

I had several very helpful e-mail exchanges with Dr Ismael Cognard and Dr Gilles Theureau from CNRS Orleans about the signals and systems available at the Nan\c{c}ay Radio Telescope, and about Ismael's work on GPU implementations of coherent dedispersion. \\

I was fortunate to have thesis examiners who improved the quality of this thesis through their careful reading of it, and their subsequent comments. I am grateful to Prof. Michael Inggs, Dr Alan Langman and Prof. Matthew Bailes for the suggestions in their examination reports. \\

I'd like to thank Profs. Birgitta Whaley and Yun Song for allowing me to sit in on their classes at Berkeley, lest I forget what attending lectures is like. I had a fun time learning about quantum computing and population genetics from them. \\

Dan's wife, Mary-Kate, and Joeri's wife, Annemieke, both kindly shared their homes with me, and provided Jason and I with a respite from the International House's dining hall delights. \\

Prior to going to Berkeley, I spent a couple of months at EPCC\footnote{Formerly ``EPCC'' stood for ``Edinburgh Parallel Computing Centre''.} at the University of Edinburgh. I'm grateful to Prof. Inggs and Dr Mark Parsons at EPCC for facilitating this visit, and to Dr Rob Baxter and James Perry for teaching me all about the ``Maxwell'' reconfigurable supercomputer. I had a great time in Edinburgh with my fellow traveler, Drew ``That's a Girl's Name\footnote{Mario Antonioletti refused to believe that ``Drew'' can be a masculine name, and Joeri van Leeuwen was confused by it too.}'' Woods. Dr Mario Antonioletti provided entertainment at the office and at Wednesday ``Pints'' at KB House. Our European flatmates, Arturo, Gara, Heinreich, Leonardo, Luca, Mario S., Milda and Steffen added greatly to the experience. \\

As part of my training on Maxwell, I spent a week in Bristol at Nallatech's UK Design Office with Drew. Allan Cantle went out of his way to arrange this for us, and we had many useful discussions with Robin Bruce, Dan Denning, Gildas Genest and Eric Lord while we were there. \\

I'd like to thank Dr Mike Keith for a useful discussion I had with him at the Manchester Reconfigurable Supercomputing Conference during my visit to the UK, on his work with Jodrell Bank to do pulsar data processing using grid technology. \\

During my Masters I've had outstanding administrative support from Regine Lord at RRSG, Lee-Ann Poggenpoel and Niesa Burgher at KAT, Catherine Inglis at Edinburgh, Tom Boot at BWRC and Stacey-Lee Harrison at the UCT Postgraduate Funding Office. \\

Alan Langman and Andrew Siemion kindly proof-read this thesis and caught several errors, hence helping me avoid considerable embarrassment. Needless to say, any remaining errors are my own fault. \\

I thank my friends and family for their support. \\

The work in this thesis has been generously supported by the National Research Foundation in the form of a KAT Masters bursary, an NRF M.Sc. Scarce Skills Prestigious SET bursary, and KAT travel funding. I also received UCT Postgraduate Funding Office support, and the CASPER group is supported by U.S. National Science Foundation Grant No. 0619596 and Infrastructure Grant No. 0403427. I would also like to acknowledge Xilinx for donating the FPGAs that I used, and Kees Vissers from Xilinx, for his continued support.

\newpage

\pagenumbering{arabic}
\setcounter{page}{1}

\chapter{Introduction}

This thesis presents the designs of several instruments developed for radio astronomy applications using generic reconfigurable computing hardware and toolflows. In this introduction, we provide a brief background and motivation, provide details of the objectives of the thesis, and outline the contents of the thesis.

\section{Background}

Radio astronomy is concerned with the study of the universe using radio frequency electromagnetic signals that are emitted as a result of physical processes and can be detected on or near\footnote{Radio telescopes can be mounted on spacecraft, and there are proposals for building radio telescopes on Earth's moon (to reduce radio frequency interference).} Earth using radio receivers.\\

Digital signal processing technology has enabled great advances in radio astronomy, but astronomers have an insatiable appetite for digital signal processing capacity, and the instruments described in this thesis have primarily helped to increase the {\it bandwidth} that can be observed. A simple implication of Nyquist's sampling theorem to DSP is that in order to increase the observable bandwidth, it is necessary to proportionately increase the sampling rate, and hence the data rate increases. This increased data rate typically necessitates the development of hardware that can firstly sample at the required rate and secondly process the data in real time\footnote{The requirement of real time processing is a result of the high sampling rates that are used; it is not feasible to store unprocessed data for any significant length of time. For example, if an astronomer wishes to observe for 10 hours, and the bandwidth is 1GHz, with 8-bit sampling the data rate will be 2 GBytes/sec, and hence approximately 70 TBytes of storage would be required.}.\\

The Center for Astronomy Signal Processing and Electronics Research (CASPER) at the University of California, Berkeley, has developed a common set of hardware, tools, libraries and processing software \cite{CASPER} that are intended to allow for the development of a wide range of astronomy signal processing instruments. This development has been an effort to promote reuse of hardware, gateware\footnote{FPGA firmware} and software whereever possible, whereas in the past most radio astronomy instruments have been developed from scratch, with very little reuse between projects at a single observatory, let alone between teams at different observatories. A partial explanation for this lack of reuse in the past is that most projects have used custom interfaces that are specific to a particular observatory at best, and often to an individual project. CASPER advocates the use of industry standard interconnect and interface technologies, particularly XAUI and 100Mbit, 1Gbit and 10Gbit Ethernet. The use of these standard interfaces not only makes it relatively simple to interface different instruments built using CASPER hardware, but also allows for easy interfacing with external devices such as control computers and data recorder computers.

\section{Objectives}

In this thesis we describe the design and development of several instruments, and the preliminary results from their deployments that verify their functionality. Broadly the development and investigations we carried out were as follows:

\subsection{Development of Spectrometers for Incoherent Dedispersion Applications}

We were tasked with the development of fast-readout spectrometers for the Parkes Radio Telescope, the Hartebeesthoek Radio Telescope and the Allen Telescope Array. The Parkes, HartRAO and BAPP\footnote{Berkeley ATA Pulsar Processor, a fast-readout spectrometer and associated infrastructure at the ATA.} spectrometers required 10GbE output. The ATA Fly's Eye spectrometers needed slower read-out (via 100MbE), but more spectrometers per processing board (four versus two). Incoherent dedispersion applications require power spectra, and the spectra can be accumulated.

\subsection{Development of Spectrometers for Coherent Dedispersion Applications}

We aimed to develop a system for performing coherent dedispersion-based pulsar studies at Nan\c{c}ay. Such a spectrometer outputs raw FFT complex data that cannot be accumulated, which for the bandwidths currently in use, implies a large\footnote{A data rate of approximately 10Gbits/sec for a single polarization with a bandwidth of approximately 600MHz is typical.} data rate. Thus a key target was the development of a system for distributing the spectrometer output to a cluster of compute nodes for further real time processing.



\section{Thesis Outline and Summary}

This thesis is organized in the following manner: \\

Chapter 2 provides an overview of radio astronomy instrumentation and pulsar instrumentation in particular. The necessary science background and terminology are also introduced. We describe CASPER's generic instrumentation hardware and tools, including the current-generation BEE2 and IBOB processing boards, and the next-generation ROACH board. \\

Chapter 3 presents our development of the ATA Fly's Eye quad spectrometer system. We present details of the gateware design, and of the control and data capture software design. We deployed a 44 spectrometer system using 11 CASPER IBOBs to the Allen Telescope Array, and here we provide test results that demonstrate the functionality of the system. We successfully detected giant pulses from the pulsar in the Crab Nebula in our single pulse search mode, which showed conclusively that the system is capable of detecting bright transient signals. \\

Chapter 4 presents our development of the Parkes, HartRAO and BAPP fast-readout (10GbE output) dual spectrometers. We present results from observations of PSR B0329+54 (a bright pulsar) using BAPP, of PSR J1028-5820 and PSR B1937+21 using the Parkes Spectrometer, and of PSR B0833-45 (Vela) using the HartRAO Spectrometer. \\

Chapter 5 presents our development of a prototype spectrometer system for coherent dedispersion observations at the Nan\c{c}ay Radio Telescope. We show that it is possible to statically load balance data using a commercial 10GbE-to-1GbE switch\footnote{We used the HP ProCurve series.}. \\

Chapter 6 presents our investigation into the design of a system for performing coherent dedispersion in real time using FPGAs. We present a high-level design to dedisperse a 400MHz bandwidth using a system with four Virtex 2 Pro FPGAs, and we predict that it will be possible to dedisperse a bandwidth of 1GHz (dual polarization) using four Virtex 5 FPGAs. Such a solution may have a considerable advantage in price-performance over the use of compute cluster implementations, which can currently process 8MHz\footnote{Matthew Bailes's group at the Swinburne University of Technology claims that a $\sim$2GHz dual quad-core system can coherently dedisperse a 64MHz dual polarization signal in real-time.} per CPU core. \\

Chapter 7 concludes this thesis, and provides remarks on the CASPER technologies, and what can be expected for pulsar instrumentation with CASPER's next generation ROACH board.

\section{Contributions}

As the Acknowledgements section indicates, much of the work described in this thesis resulted from collaborations that the author had with other students, and astromomers and engineers. In this section, I\footnote{In this thesis the plural personal pronoun, ``we'' is typically used, but in this section the singular seems more appropriate.} shall endeavour to explain what specific contributions I made, and who was largely responsible for the other related work that I reference in this thesis. In all cases, I worked closely with Dan Werthimer and Henry Chen -- Dan provided advice on telescope integration and general assistance with digital design matters, and Henry provided a great deal of help with the CASPER tools. All the digital design work in this thesis was done using the MSSGE toolflow, targeting the BEE2 and IBOB platforms, which were created at BWRC by Chen Chang, Pierre-Yves Droz, Hayden So, Henry Chen, Brian Richards and Dan Werthimer. I made extensive use of the CASPER DSP library, which was primarily developed by Aaron Parsons, with contributions from Glenn Jones. \\

\begin{enumerate}
  \item \textbf{ATA Fly's Eye}: The Fly's Eye spectrometer design was based on the ``Pocket Correlator'' design by Aaron Parsons. The idea for the experiment came from Prof. Jim Cordes, and Dan Werthimer was responsible for overseeing the execution of the engineering work. I was the developer primarily responsible for modifying the Pocket Correlator design to make it function as a quad-spectrometer instead. Andrew Siemion contributed to this effort. Andrew and I jointly developed the high-reliability data recorder system. We both assembled the full hardware system. We also jointly wrote the software to control the instrument, and to automate the data-taking process. Andrew, Joeri van Leeuwen, Griffin Foster, Mark Wagner and I jointly developed the data processing flow. Joeri served as the team's expert on {\it sigproc}\footnote{{\it sigproc} is a set of tools developed primarily by Dunc Lorimer for processing pulsar data.}. Griffin wrote much of the visualization code, and Mark worked on setting up scripts for running the processing flow on multiple computers in parallel\footnote{We initially intended to do processing on a workstation grid, but ultimately moved to a cluster computer at LBNL's NERSC facility.}. Andrew and I worked on techniques for rejecting RFI with help from Griffin and Mark. Andrew also wrote much of the data conversion code. Weekend observing runs were generally looked after by Andrew, Joeri, Griffin and I; I was directly responsible for approximately 100 hours of observing. Joeri was responsible for system tests with pulsar observations. Dan Werthimer, Geoffrey Bower and Melvyn Wright provided guidance, and we benefitted from technical help with hardware and tools from Matt Dexter, Dave MacMahon, Oren Milgrome and Colby Kraybill at Berkeley's Radio Astronomy Laboratory. \\
  
  \item \textbf{Berkeley ATA Pulsar Processor}: Dan Werthimer, Don Backer and Joeri van Leeuwen were responsible for the project idea. I was solely responsible for the development of the gateware for the FPGA. Oren Milgrome, Joeri van Leeuwen and I collaborated on the interface between the pulsar spectrometer and the ATA beamformer. Oren implemented the beamformer ``combiner'' that combines two polarization beams and sends them to my spectrometer. I set up the data recorder, and wrote the spectrometer control scripts. Joeri was responsible for the data processing, and for running our test observations (including beamformer control). \\
  
  \item \textbf{Parkes Spectrometer}: Dan Werthimer was responsible for the project idea, based on requirements from Matthew Bailes. I was solely responsible for the development of the gateware for the FPGA, and the control scripts. \\
  
  \item \textbf{HartRAO Spectrometer}: This is a ``Full Stokes'' version of the Parkes Spectrometer, requested by HartRAO. I was solely responsible for the development of the gateware for the FPGA, and the control scripts. \\
  
  \item \textbf{Nan\c{c}ay Coherent Dedispersion Pulsar Machine}: Don Backer, Ismael Cognard, Paul Demorest, Joeri van Leeuwen and Dan Werthimer were responsible for the project idea. I was solely responsible for the development of the prototype gateware for the FPGA. I relied on help from the project originators regarding the overall design of the instrument, and about the interface with the telescope. \\
  
  \item \textbf{Real-time Coherent Dedispersion Processor}: Glenn Jones and Joeri van Leeuwen were responsible for the project idea\footnote{The use of FPGAs for performing coherent dedispersion is certainly not unique to us; I simply mean here that Glenn and Joeri conceived this particular project. We have collaborated with a team at NRAO Green Bank that is also investigating the use of FPGAs for real time coherent dedispersion, and there are surely other groups elsewhere that are compelled by advances in FPGA technology to look at this application.}. Glenn Jones provided the original architecture for the instrument, and Joeri van Leeuwen provided advice during the design phase. I was responsible for the more detailed design, and for conducting tests to determine what FFT lengths are possible. \\
\end{enumerate}

\newpage

\chapter{Background}

This chapter provides a brief introduction to radio astronomy, instrumentation for radio astronomy, and instrumentation for pulsar science in particular. The work presented in this thesis made extensive use of the CASPER hardware, tools and software, and we provide an overview of the CASPER technology and approach in this chapter. \\

\section{An Engineer's View of Radio Astronomy and Instrumentation}


Radio astronomy is the field of astronomy whose observation technique is based on the capture and analysis of radio frequency electromagnetic radiation using {\it radio telescopes}. Traditional astronomy is based on observations at optical (i.e. visible) wavelengths, but astronomical sources emit radiation not only as visible light, but across the electromagnetic spectrum -- from gamma-rays ($f > 10^{20}$GHz) through radio ($f$ in range 3Hz -- 300GHz). \\

In this thesis we are less concerned with the science that radio astronomy enables than with the engineering challenges that radio astronomers face to build ever-more sensitive instruments. Many excellent introductory textbooks on radio astronomy exist, which the interested reader may wish to review for details on the radiative processes\footnote{As it turns out (\cite{ASTR534}, lecture 2), blackbody radiation from stars, which astronomers in the early 20th century expected would be the main source of radio EM radiation, is not a source that is normally observed, even with modern instruments. The blackbody radiation $B_\nu(T) = \frac{2h\nu^3}{c^3}(1/(e^{\frac{h\nu}{kT}}-1))$ ($h$ is Planck's constant; $c$ is the speed of light in a vacuum, $k$ is Boltzmann's constant, $T$ is the body's temperature and $\nu$ is the frequency) from stars (where $T$ is of order $10^4$K) at frequencies achievable in the 1930's ($\nu < 1$GHz) results in a radio flux on earth that is too small to be detected. As a result, astronomers did not consider trying to observe radio emissions. Radio astronomy was started by accident when a radio engineer for Bell Telephone, Karl Jansky, discovered during investigations into natural causes of radio static (which interfered with Bell's radio transmissions) a steady radio noise at 20.5MHz. He deduced from its periodic changes in strength that the source of the noise was outside the solar system. This was the first detection of non-blackbody radiation at radio wavelengths from an extraterrestrial source. Many other sources, with varied emission mechanisms, have subsequently been discovered. Ironically, it is now the radio astronomers that are dramatically affected by radio frequency interference (RFI) from terrestrial emissions caused by telecommunications companies, and who spend much of their time devising techniques to mitigate this RFI!} that result in electromagnetic emissions that can be detected on earth, and what it is possible to deduce about the physics of astronomical objects using the collected data. For example, see \cite{RadioAstroIntro} or the excellent course notes from Condon and Ransom \cite{ASTR534}. \\

\section{Single Antenna Radio Astronomy}

From the engineer's perspective, radio astronomy is concerned with the detection of radio waves that are continually arriving at the earth, and the subsequent processing of the data that is collected. The observational tool of the radio astronomer is the radio telescope, and in the simplest case this is a (typically directional) antenna with large ``effective area\footnote{The effective area $A_e$ of an antenna is related to its gain $G$ via the reciprocity theorem (which can be understood as a consequence of the time-reversibility of solutions to Maxwell's equations). Specifically, $A_e = \frac{\lambda^2 G}{4\pi}$ (where $\lambda$ is the wavelength).}'' (\cite{ASTR534}, lecture 8) whose electrical output is digitized\footnote{Before the advent of modern digital electronics, a considerable amount of processing was done using analogue circuits, but current radio telescopes now typically digitize the received signals as soon as possible to avoid contamination with noise, and for further error-free digital processing.} and recorded. Broadly speaking, astronomers are interested in measuring the power across some bandwidth $B$ at some centre frequency $f_{sky}$ from some particular location\footnote{Astronomers have a coordinate system for the sky observable from earth that uses two values (Right Ascension and Declination) to define any point in the sky, independent of the location on earth that the observation of some subset of the observable sky was viewed from, the time of the observation, and the portion of the sky the observing was viewing. Given a UTC date and time of an observation, the latitude and longitude coordinates of a point on the earth at which the observation took place, and the azimuth and elevation of a ray into the sky, it is possible to determine the point in the universal coordinate system that this observation corresponds to.} in the sky. Astronomical signals tend to be very weak, so a large effective area is usually very important. Indeed much of the focus of new radio telescope developments is in the creation of telescopes that have ever larger collecting areas. However, another key metric that is used to specify radio telescope performance is ``angular resolution''. In this case, smaller values are desirable, since astronomers want to localize phenomena as accurately as possible. The angular resolution of a single-antenna telescope is directly related to the antenna's beam pattern. Ideally the beam should have zero sidelobes and a main beam that has a small angle.\\ 

A simple dipole antenna on its own is not sufficient to meet the dual requirements of large effective area and high angular resolution. Typically a reflector (``dish'') is used to collect and focus the signal onto a feed antenna. The effective area is directly related to the diameter $D$ of the reflector by its projected geometric area $A_e = \pi D^2/4$. The angle between the half power points of the beam, $\theta_{HPBW}$, is related to both the reflector's diameter and the signal wavelength $\lambda$; specifically $\theta_{HPBW} \propto \frac{\lambda}{D}$. This is convenient, since it means that we can simultaneously increase the effective area and angular resolution by increasing the reflector's diamater. However, by satisfying the astronomers's desire for more localization precision in this way (i.e. the ability to better determine the specific locations of sources due to a smaller beam width), we necessarily reduce the ``amount\footnote{Astronomers typically measure sky coverage in square degrees (deg$^2$).}'' of the sky that they can observe at any one time. Astronomers have a third desire for radio telescopes: they would like to be able to view as much of the sky simultaneously as possible, i.e. they would like large {\it sky coverage}. This is important for {\it sky surveys} where the objective is to ``map'' the sky (i.e. to look for and record the locations of sources of radio emissions). Astronomers want to understand the time evolution of these sources, so clearly the ability to view a large portion of the sky simultaneously is advantageous, lest one miss important dynamical events of sources that aren't observed frequently enough. An even more compelling case for large sky coverage is for transient surveys, in which astronomers want to find and record the locations of short (time\footnote{Transient signals may have duration less than 5ms.}) signals, which are possibly the result of once-off events (and hence can't be reobserved). \\

These conflicting requirements have the result that any given single-dish (i.e. single-reflector) radio telescope is not able to provide optimal science capabilities to all its users -- it will be better-suited to some types of experiments than others. However, this only hints at the parameter space that needs to be optimized in modern telescope design: in the 1950's, merely a decade after Reber's groundbreaking galactic survey \cite{Reber40} marked the first radio astronomy science study, astronomers at the University of Cambridge  began working on radio interferometers. Interferometers provide a wide range of additional capabilities, but their design is even more complicated than that of single-dish telescopes, with even more trade-offs called for. \\

\subsection{Radio Telescope Arrays}


Very soon after radio astronomy became an active area of study, astronomers realized that for some experiments it would not be practical to build a sufficiently large single-dish telescope to get the angular resolution they desired. Large single-dish radio telescopes are indeed very large: the Parkes Radio Telescope in Australia is relatively small with $D=64$m. The Lovell Telescope at Jodrell Bank in the UK has $D=76$m. The Green Bank Telescope in West Virginia has $D=100$m. These are fully-steerable (meaning that they can be pointed to nearly arbitrary positions in the sky). Clearly building a steerable telescope is more challenging than building a fixed position telescope, but also more useful. However, the telescope at Arecibo in Puerto Rico is built in a natural depression, which allowed engineers to construct a dish that is far larger than would be feasible otherwise; there, $D=305$m! \\

However, even with a spherical reflector with the enormous diameter of Arecibo's, it is not possible to obtain suitable angular resolution for many observations that astronomers wish to carry out. Due to the cost of materials (primarily steel), it is not practical to build ever-larger single-dish telescopes. For example, to obtain angular resolution of 1 arcsecond (which is a common requirement), a single-dish telescope would need to have $D=40000$m \cite{UNM423}. \\

A set of ingenious techniques have been developed over the past 50 years to construct and use {\it synthesis aperture arrays}\footnote{The invention and use of some of the core ideas in interferometry-based radio astronomy have certainly not been restricted to this field -- interferometric systems in communications, RADAR and SONAR have also been extensively studied.}. The key idea behind this class of telescope, which are often referred to as radio interferometers, phased arrays, or radio telescope arrays (or contemporaneously simply as radio telescopes), is that it is possible to somehow combine the signals from a set of several single-dish antennas to obtain a new signal that results in an observation with far better angular resolution than the individual dishes are capable of. More precisely, if the distance between the two most separate dishes in the set is $l$, then it is possible to synthesize an aperture with an effective diameter of $D=l$. \\

\begin{figure}[htp]
	\centering
		\includegraphics[width=4in]{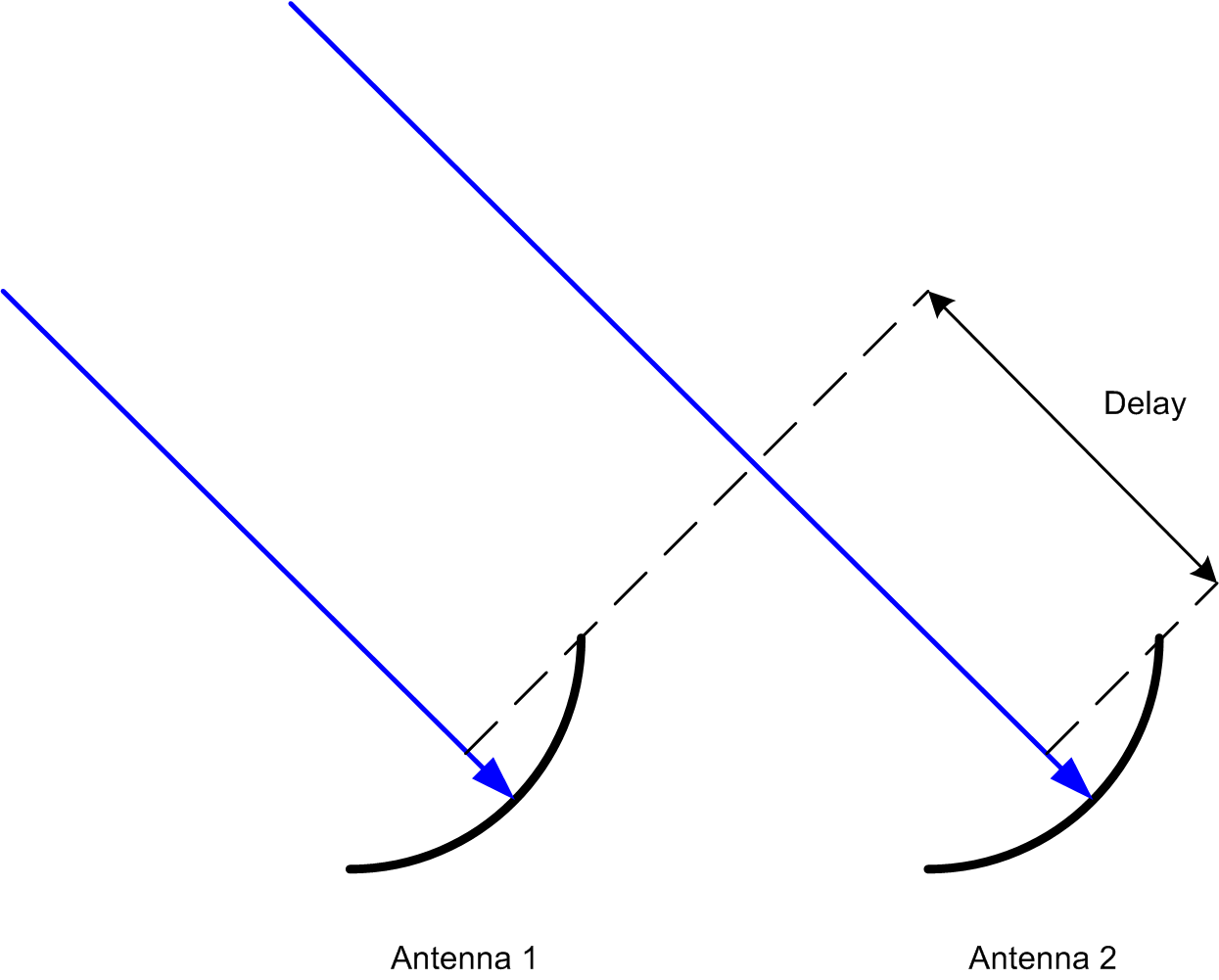}
		\caption[A Two-Element Phased Array.]{A Two-Element Phased Array. The signal from the source (blue) will arrive at antenna 2 after it arrives at antenna 1. This delay can be removed later; if the signals are then summed, an artificial beam is formed.}
	\label{fig:BackgroundBeamformingDelay}
\end{figure}

Figure \ref{fig:BackgroundBeamformingDelay} shows how a signal from a source will arrive at different times at two dish antennas. If the delay between the arrival of the signal at antenna 1 and its arrival at antenna 2 is $\tau$, then we can form an artificial beam by delaying the output of antenna 1's receiver by $\tau$, and adding it to the output from antenna 2. This procedure can be extended to work for $N$ antennas, where one antenna is the reference, and the other $N-1$ antenna outputs are artificially delayed so that they are aligned with the output of the reference antenna. The sum of these signals is then the artificial beam. \\

``Beamformers'' allow astronomers to obtain excellent angular resolution using just two dishes. The primary motivation for having more than two antennas in a phased array is to provide improved sensitivity. If the sensitivity using a single dish is unity, then the sensitivity of a phased array with $N$ such dishes is $N$. The Allen Telescope Array, which currently has 42 dishes, has a digital beamformer that performs the delays and summations in FPGAs: the signals from the antennas are sampled (in practice there are 84 signals, since each dish has a dual polarization antenna) and streamed directly to the beamformer, whose output (the synthesized beam) is then streamed to a set of instruments that consume it\footnote{At the ATA, the primary consumer is that SETI processing system that searches for possible signs of intelligent extraterrestrial life by looking for specific patterns in the data; the other consumer at time of writing is a spectrometer for pulsar science, whose construction is discussed in this thesis.}.  \\

One of the crowning technical achievements of radio astronomy has been the development of the {\it synthesis imaging} technique. This technique allows astronomers to use an array of antennas to form an image of the sky, essentially by manipulating the pairwise correlations of signals from all antennas. A comprehensive coverage of synthesis imaging is provided in \cite{TMS01}. Discussion of the engineering requirements for ``correlators'' to perform these correlation calculations in real-time, to allow for synthesis imaging, is beyond the scope of this introduction, but suffice to say that it is a very demanding computational task. FPGAs provide an excellent platform for processing of arrays with tens to hundreds of antennas, and bandwidths of up to several GHz. \\

\section{Pulsar Science}

Pulsars are neutron stars that are highly magnetized, and which rotate at up to \~700Hz. Pulsars, through a mechanism that is not yet fully understood, periodically emit broadband electromagnetic pulses\footnote{The time duration of each of these pulses is typically less than 1ms.}. The emission period $P$ is thought to be the same as the rotation period. Since the discovery of the first pulsar in 1967, approximately 1800 pulsars have been found. The fastest known pulsars have $P \sim 1$\,ms, and the slowest known pulsars have $P \sim 10$\,s. These rotational velocities are particularly impressive when one considers that pulsars typically have masses larger than our Sun's\footnote{Pulsars are extremely dense objects: they typically have diameters of order ten kilometers. Their density, mass, large magnetic fields and high spin frequencies make them excellent laboratories for studying physical regimes that are impossible to recreate on earth.}. Far more detailed explanations of the science of pulsars, and indeed the techniques for their observation, are provided in the books by Lorimer and Kramer \cite{LK05}, and Lyne and Smith \cite{LS03}. \\

\begin{figure}[htp]
	\centering
		\includegraphics[width=5.5in]{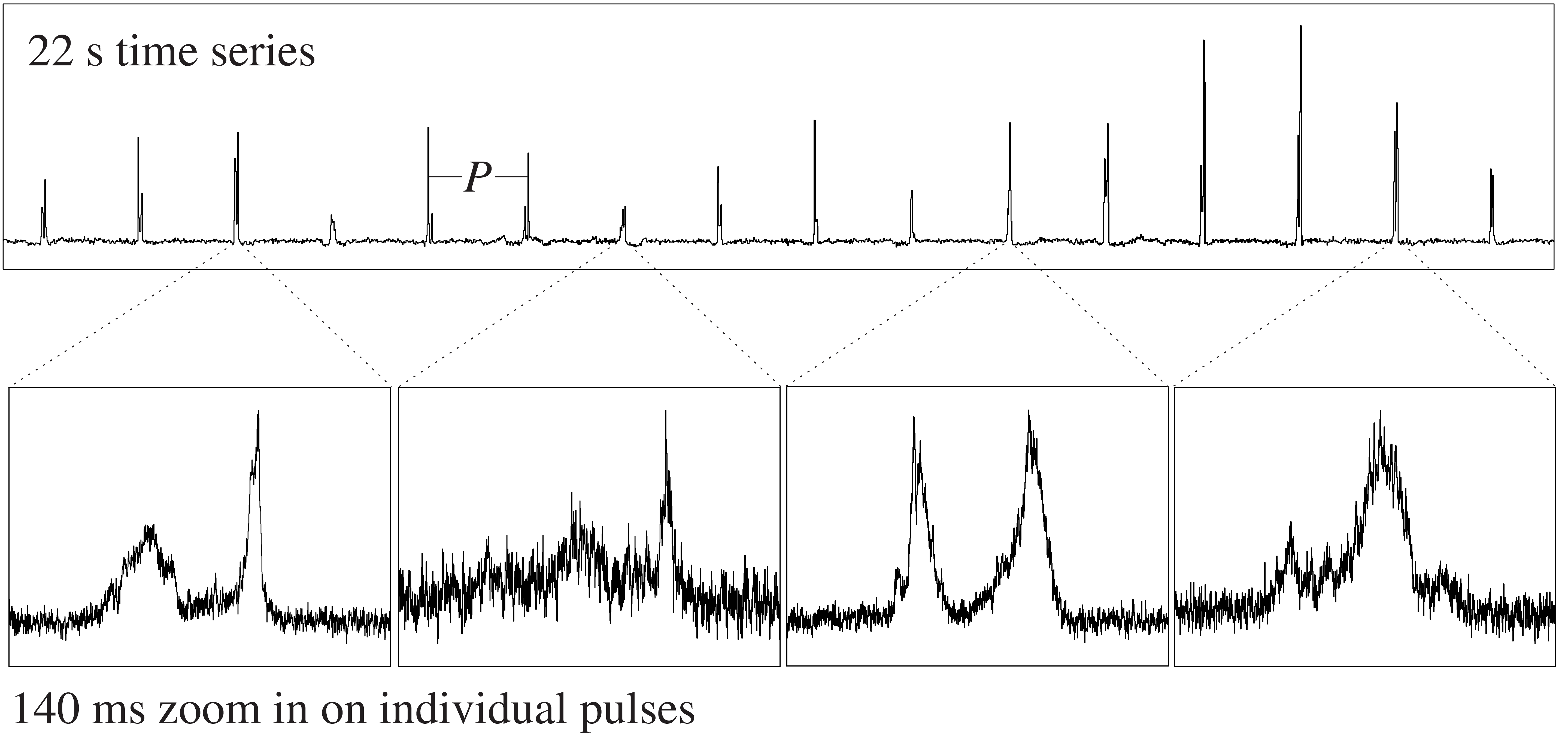}
		\caption[Time series data taken at Arecibo showing individual pulses from PSR B0301+19.]{From \cite{LK05}. Time series data taken at Arecibo showing individual pulses from PSR B0301+19. The horizontal axes in both the main figure and the insets is time, and the vertical axes in all figures is power.}
	\label{fig:BackgroundPulsarPulses}
\end{figure}

Figure \ref{fig:BackgroundPulsarPulses} shows pulses from pulsar\footnote{Pulsars are designated by the moniker ``PSR'' followed by an abbreviation of their their coordinates.} PSR B0301+19 observed at Arecibo. The pulse period is stable, but the individual pulse amplitudes and shapes may change quite dramatically.

Most pulsars emit pulses that are too weak to detect individually on earth, even with a large telescope such as Arecibo -- the pulses are ``hidden'' in the noise\footnote{In this case, we take ``noise'' to mean all the other signal that we obtain that isn't from pulses from the pulsar we're attempting to observe.}. Therefore a large portion of the tools in the pulsar scientist's chest are related to extracting this weak signal from noisy data. \\

Many techniques have been invented to find weak periodic signals in datasets; the periodicity of the signals emitted by pulsars is what has made it possible for astronomers to discover thousands of pulsars, most of which are too weak to yield individually-distinguishable pulses. Details of pulsar searching are available in \cite{LK05}. This periodicity, however, also allows us to relatively easily observe known pulsars. If we know the period $P$ of a particular pulsar\footnote{It's not possible to simply look up a value for $P$ in a table and then use it as-is; it turns out that a number of corrections need to be applied first. The most important is the barycentric correction to account for the earth's motion around the Sun. Here we assume that all the necessary corrections have been carried out.}, and we digitally record a set of sampled time-domain data (i.e. received power as a function of time) of length $T$ on a single antenna telescope pointed at that pulsar's known location, we can determine the pulsar's average pulse shape (its ``pulse profile'') using the following procedure\footnote{The procedure given is more of a general approach, and is missing some details. Most importantly, we have omitted the {\it dedispersion} stage that is usually needed (and is always helpful).}: \\

\begin{enumerate}
  \item Divide the dataset into $T/P$ subsets $A_1, A_2, \cdots, A_{T/P}$, each of length $P$ seconds.
  \item Compute the sum $S = \sum_{i=1}^{T/P} A_i$. In practice the $A_i$ are arrays containing the time samples, so explicitly we must be compute the sums $S\left[j\right] = \sum_{i=1}^{T/P} A_i\left[j\right]$ for all $j=1,2,\ldots,N$. $N$ is the number of samples in a subset $A_i$, and is related to $P$ by $N=2BP$, where $B$ is the sampled bandwidth\footnote{This relation arises because Nyquist's Sampling Theorem requires that the sampling rate for a signal with bandwidth $B$ be at least $2B$.}.
  \item Plot $S[j]$.
\end{enumerate}

This technique is known as ``folding'' (because the data is being repeatedly folded onto itself, at the pulsar's period), and is routinely used by astronomers when they observe known pulsars. Several test results presented in this thesis show folded pulsar data. While individual pulses from pulsars may be highly variable, the integrated (folded) pulse profile is usually very stable, provided that a sufficient number of pulses are integrated\footnote{This number is usually of order hundreds or thousands.}. The pulse profile is frequency-dependent; profiles obtained from observations using different sky frequencies can look considerably different. \\

\begin{figure}[htp]
	\centering
		\includegraphics[width=3.5in]{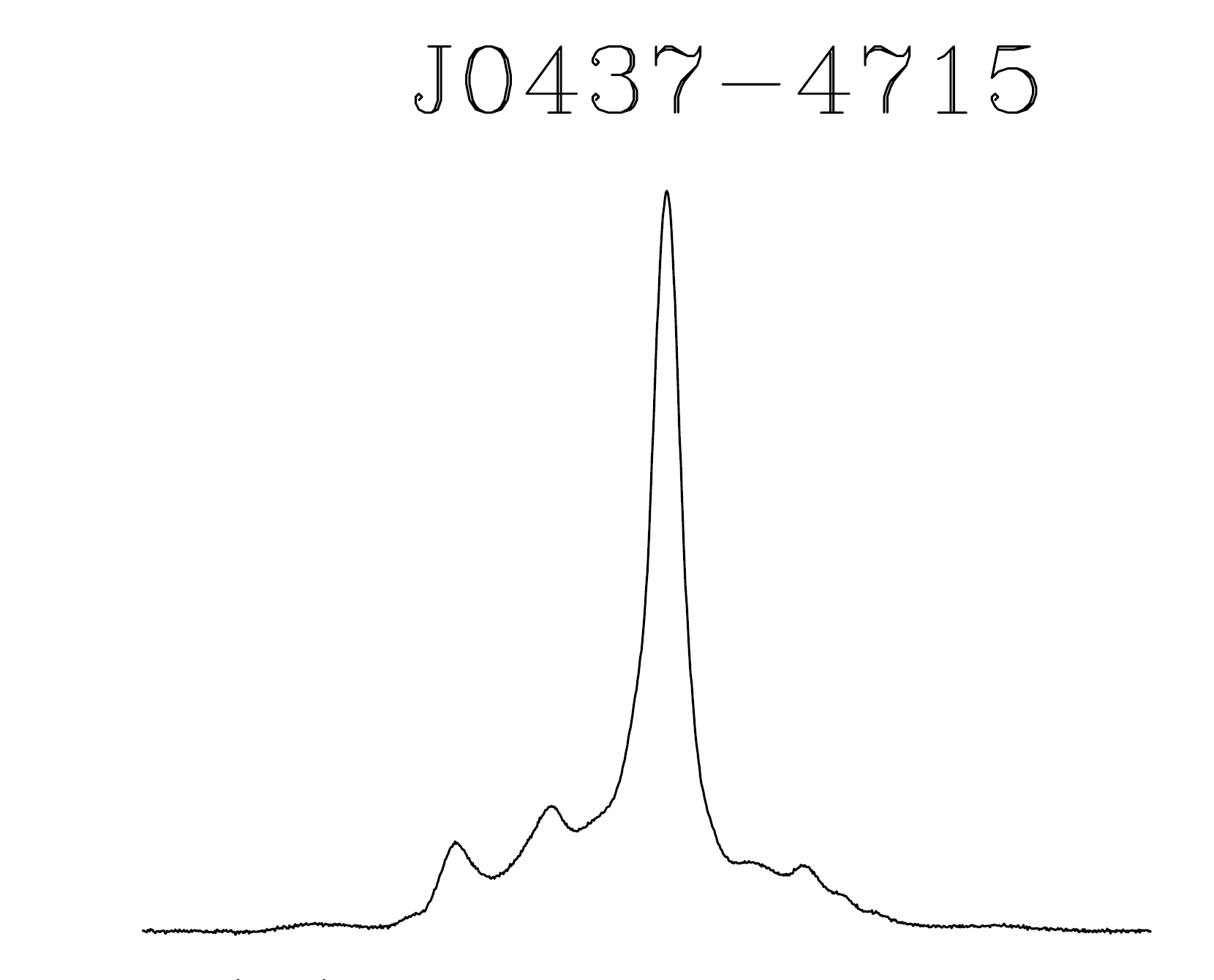}
		\caption[Pulse profile of PSR J0437-4715.]{From \cite{LK05}. Pulse profile of PSR J0437-4715.}
	\label{fig:BackgroundPulseProfile}
\end{figure}

Figure \ref{fig:BackgroundPulseProfile} shows the integrated pulse profile from PSR J0437-4715, from data in the European Pulsar Network database \cite{EPN}. \\

The discussion thus far has ignored the effect that the interstellar medium (ISM) has on the signals from pulsars as the pulses travel towards the earth. The ISM is an ionised plasma that the electromagnetic radiation from pulsars interacts with in a variety of ways. In this thesis, we are most concerned with an effect known as {\it dispersion}, since it has important bearing on how to build effective instruments for performing pulsar observations. \\

Dispersion has the effect of delaying a pulse from a pulsar as a function of frequency. Specifically, the time delay between two frequencies $f_1$ and $f_2$ is given by:

\[
\Delta t \approx 4.15 \times 10^6 \text{ms} \times \left(f_1^{-2} - f_2^{-2}\right) \times \text{DM}
\]

Here DM is the ``dispersion measure'', which is related to how far the pulse traveled\footnote{More precisely, the dispersion measure is the integrated density of the free electrons along the line-of-sight: $\text{DM} = \int_0^d n_e \text{d}l$ where $n_e$ is the electron density \cite{LK05}.}. \\

\begin{figure}[htp]
	\centering
		\includegraphics[width=5in]{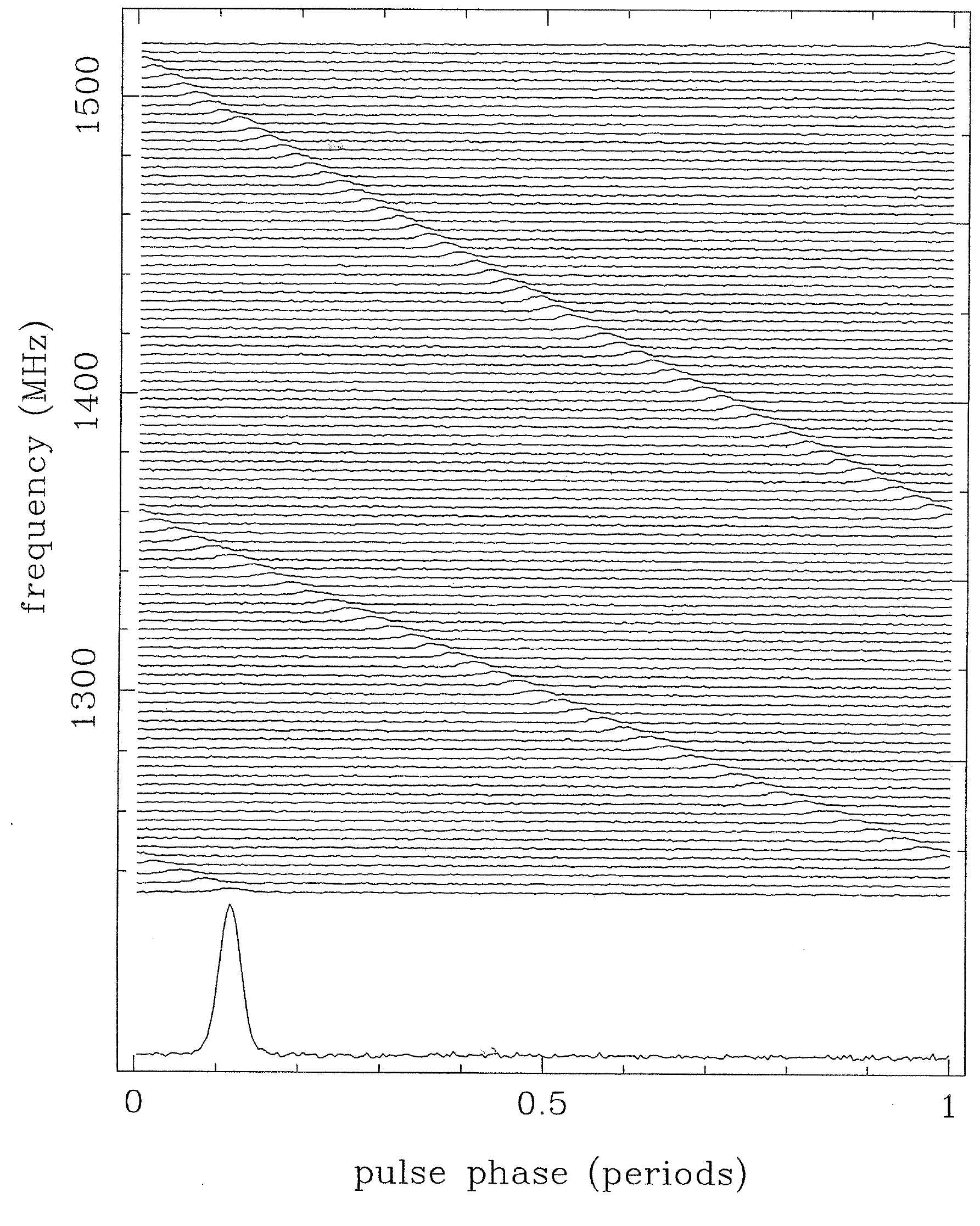}
		\caption[Frequency-Phase Plot of Dispersion of PSR B1356-60.]{From \cite{LK05}, courtesy Andrew Lyne. Dispersion of PSR B1356-60 from an observation at the Parkes Radio Telescope. This pulsar has a period of 128ms, so the horizontal axis can be interpreted as a time period of 0 -- 128ms. The data has been folded using the period to produce this figure. The dispersion is clearly visible in the frequency-time plot. The lower panel shows the pulse profile after dedispersion.}
	\label{fig:BackgroundDispersion}
\end{figure}

If we channelize the data from a pulsar observation, we can easily see the dispersion. Figure \ref{fig:BackgroundDispersion} shows a frequency-time plot of folded data from PSR B1356--60 where the dispersion delay is clearly visible. \\

Counteracting dispersion (using so-called ``dedispersion'' techniques) is a key task in pulsar observations. The instrumentation used to perform pulsar observations should be designed to assist with this, and much of the data analysis work following an observation involves dedispersion. Dedispersion literally attempts to reverse the dispersive effects: the {\it incoherent dedispersion} technique is a computationally inexpensive method that does not completely eliminate dispersion, and the {\it coherent dedispersion} technique is a computationally expensive technique that gives very accurate results by totally reversing dispersive effects (constrained only by numerical precision). \\

The incoherent dedispersion technique works by shifting the different channels in a channelized data set (such as that illustrated in Figure \ref{fig:BackgroundDispersion}) so that the time delays the channels underwent are reversed. The inaccuracy of the method results from the fact that the channelization process necessarily produces finitely many channels, and hence there is dispersion of data within each channel that does not get corrected. Nevertheless, incoherent dedispersion is still widely used due to its efficiency, and the fact that it provides sufficient accuracy for many applications on many telescopes. \\

The coherent dedispersion technique recognizes that the dispersion effect can be considered as a filter $H$, and that hence all that is required to return the received data to its dedispersed form is to apply the inverse of $H$. In practice, it is more efficient to apply the inverse filter in the frequency domain, since a convolution in the time domain is simply a multiplication in the frequency domain. Nevertheless, the size of the discrete Fourier transform required (and its inverse, following the multiplication by the inverse filter) is considerable -- often of order 1 million points -- so coherent dedispersion is an expensive technique to apply. It is widely used in pulsar timing applications, where the accuracy it provides is necessary. \\

\begin{figure}[htp]
	\centering
		\includegraphics[width=3.5in]{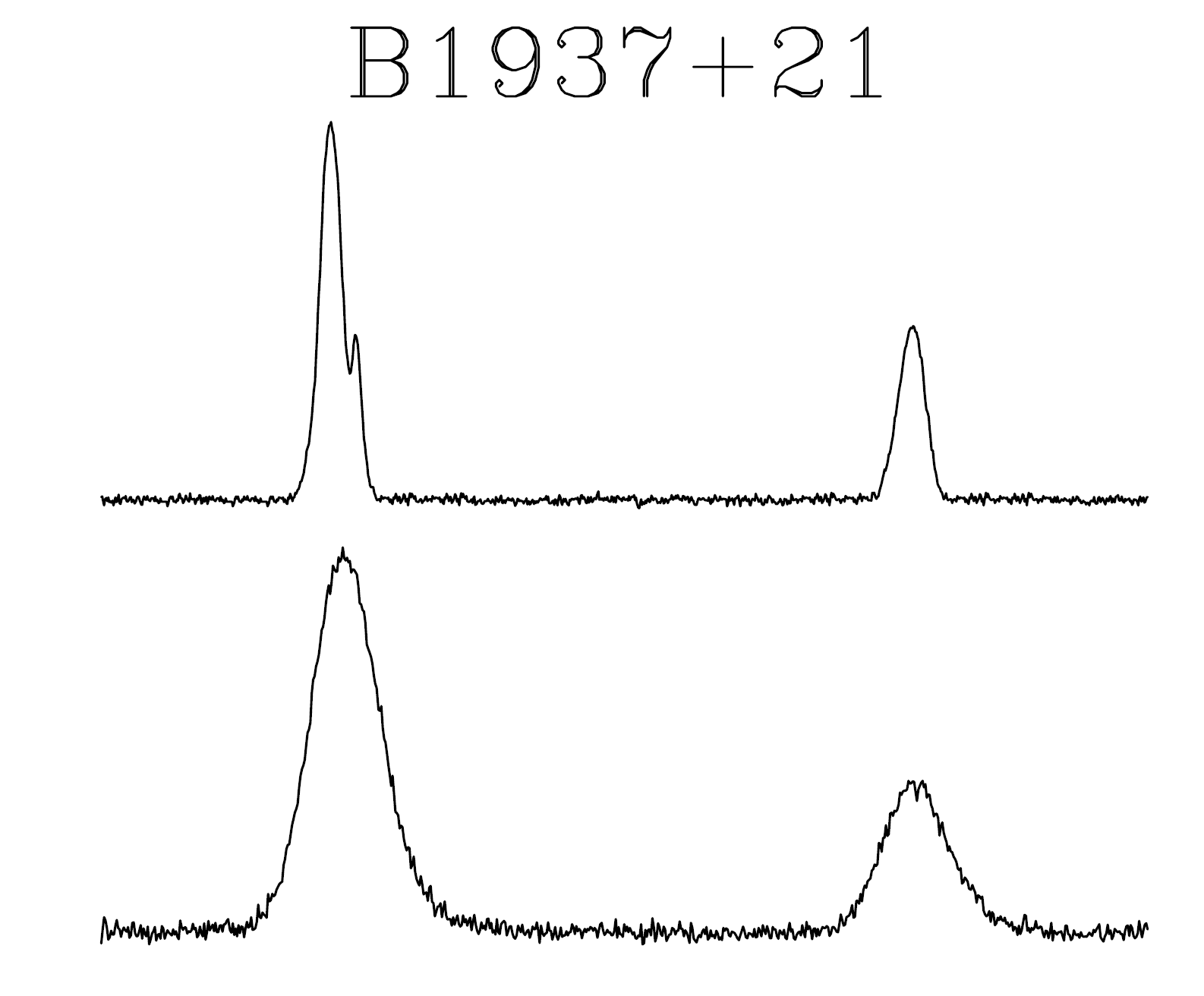}
		\caption[Pulse profiles of PSR B1937+21 showing the effect of coherent versus incoherent dedispersion.]{From \cite{LK05}. Pulse profiles of PSR B1937+21. The upper profile was obtained using coherent dedispersion, and shows the true pulse shape. The lower profile was obtained using incoherent dedispersion, and the reduced time resolution (due to inner-channel smearing) is clearly evident.}
	\label{fig:BackgroundInCoDeDiVsCoDeDiResult}
\end{figure}

Figure \ref{fig:BackgroundInCoDeDiVsCoDeDiResult} shows the effect that coherent versus incoherent dedispersion can have on the observed pulse profile. \\

\section{Instrumentation for Pulsar Science}

Because of the need to perform dedispersion on pulsar data, the data from a telescope needs to be digitized and then channelized. In the case of incoherent dedispersion, the need for channelization is obvious since the technique operates on channelized data; the need for channelization when coherent dedispersion is to be applied is more subtle. In practice it is sometimes not computationally feasible to coherently dedisperse the entire set of data at once. Therefore a coarse channelization may first be performed to produce channels that individually can be coherently dedispersed.\\

We thus see that regardless of dedispersion technique, a {\it spectrometer} (i.e. a digital sampler and channelizer) is necessary. We will discuss shortly the differences between the requirements for instruments intended for incoherent versus coherent dedispersion. The spectrometer is typically implemented in special-purpose hardware (and, in recent times, using FPGAs) as opposed to general-purpose computers because the data rates are prohibitively large for a computer (or cluster) to economically handle. \\

The large data rates required for spectrometers for pulsar science are a direct result of astronomers's desire to observe ever-larger bandwidths. Typical\footnote{This observation is based on the author's interactions with scientists from the Allen Telescope Array, the Parkes Radio Telescope, NRAO Green Bank and GMRT.} current observing bandwidths for pulsar studies range from 50MHz to 1GHz. With $B=1$GHz in a dual polarization system, the data rate from sampling the data is 4GB/s, assuming that each sample is 8 bits\footnote{Most modern radio astronomy instruments typically use 8-bit sampling. The dynamic range afforded by 8-bit samples is useful for mitigating RFI, which might saturate a 1-bit or 2-bit sampler. As ADC technology improves, astronomers will undoubtedly start using even higher sampling bitwidths, leading to even larger data rates.}. FPGA-based hardware can be used to channelize data at high sampling rates more easily than is possible with general-purpose computers. FPGA's are also very well-suited to streaming data applications, whereas general-purpose CPUs are not. \\

A range of FPGA-based spectrometers for radio astronomy have been deployed over the past 5 years with great success (see for example \cite{CASPER}). As FPGA vendors release larger devices due to improved transistor density, there are two main parameters that astronomers wish to have improved by using the new resources: bandwidth and channels. Specifically astronomers would like to observe larger bandwidths, and would (for incoherent dedispersion) like finer channelization (i.e. more channels). \\

Larger bandwidths are desired because sensitivity improves as the square root of the processed bandwidth. Increasing the processed bandwidth is one of the only ways of improving sensitivity on existing telescopes\footnote{This assumes that the available analogue bandwidth is larger than the bandwidth of the present digital processing systems. Clearly if the analogue bandwidth is already all being processed, then the only way of increasing the bandwidth is by upgrading the analogue systems and the digital systems. This type of upgrade does routinely happen, but is far more expensive than just upgrading the pulsar spectrometer digital backend!}. An increase in the number of channels is generally also useful for several reasons. If the bandwidth is increased, then just keeping the same per-channel bandwidth clearly requires an increase in the total number of channels. In the case of spectrometers for incoherent dedispersion applications, decreasing the per-channel bandwidth is usually desirable, since it means that a smaller amount of the total bandwidth will be wasted by the excision of narrowband RFI, and the lower the channel bandwidth, the less dispersion there is within each channel. \\


As we have alluded to, there are two main types of pulsar spectrometer: pulsar spectrometers for studies where incoherent dedispersion will be used to subsequently dedisperse the data, and pulsar spectrometers for studies where coherent dedispersion will be used. \\

                      

\begin{figure}[htp]
	\centering
		\includegraphics[width=8in,angle=90]{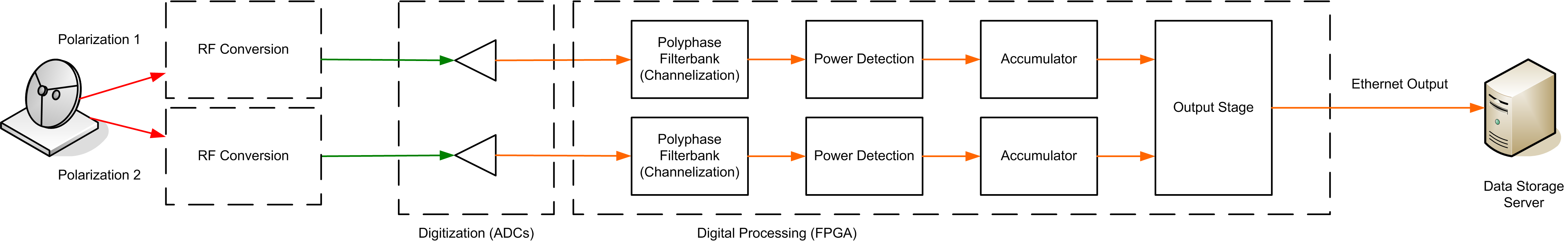}
		\caption{Data Flow in a Spectrometer for Incoherent Dedispersion Applications.}
	\label{fig:BackgroundPulsarSpectrometerDataFlow-InCoDeDi}
\end{figure}

\begin{figure}[htp]
	\centering
		\includegraphics[width=8in,angle=90]{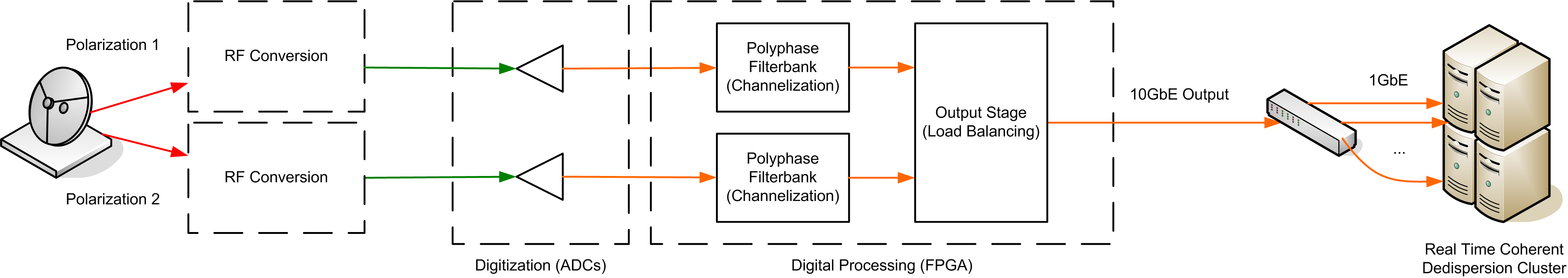}
		\caption{Data Flow in a Spectrometer for Coherent Dedispersion Applications.}
	\label{fig:BackgroundPulsarSpectrometerDataFlow-CoDeDi}
\end{figure}

Figures \ref{fig:BackgroundPulsarSpectrometerDataFlow-InCoDeDi} and \ref{fig:BackgroundPulsarSpectrometerDataFlow-CoDeDi} show the canonical data flow and functionality for spectrometers for incoherent and coherent dedispersion applications respectively. The former design produces a power spectrum (using a polyphase filterbank) that is then accumulated (integrated) before it is outputted. The output in modern instruments is typically via Ethernet. The purpose of the accumulation stage is to reduce the data rate to the point where all the data for an observation can be stored to hard drives. The accumulator functions by summing the power spectra -- it produces an average\footnote{The accumulator does not divide its output by the number of spectra it added together.} power spectrum over the time that it sums over. There is a tradeoff in the choice of accumulation length (i.e. how many spectra are summed in a single accumulation/integration): the longer the accumulation length, the slower the data rate becomes, but the worse the time-resolution of the instrument becomes. Typically spectrometers for incoherent dedispersion applications have between 128 and 4096 channels, and time resolution of between several microseconds and several milliseconds. Bandwidths range from several tens of MHz through to 1GHz. \\

The spectrometer for coherent dedispersion applications is conceptually simpler -- it channelizes the input data (into chunks that typically have bandwidths between 1MHz and 50MHz) and then sends the raw polyphase filterbank output to a cluster, via a switch. Because the data rate is too high for one computer to process, the output is statically load balanced: the channels are divided amongst several machines. For example, in a 32 channel system, there might be 8 processing computers, and so the spectrometer will send 4 adjacent frequency channels to each processing computer in turn. Because power spectra are not formed, it is not possible to accumulate the data\footnote{The reason that the PFB output can't be summed is essentially that pulsar signals look like noise in the voltage domain, so an accumulation of the voltages would tend towards zero.}. The compute cluster performs coherent dedispersion on the output channels from the spectrometer in real time, because the data rate is typically too high to store all the data.

For the purposes of testing a new pulsar observing system it is usually easiest to attempt observations of bright (high flux) pulsars. Once the basic functionality of the system has been verified (via the production of a correct pulse profile) it is then often helpful to attempt an observation of a fast (small period) pulsar. Table \ref{tbl:Pulsars} provides a list of bright and fast pulsars that are useful test sources for Northern hemisphere telescopes.

\begin{table}[h!]
\centering
\caption[Pulsars Useful for Instrument Tests.]{From \cite{GBSpigot}. Pulsars Useful for Instrument Tests.}
\label{tbl:Pulsars}
\vspace{.2 in}
\begin{tabular}{|l|l|l|}
\hline
{\bf Name}	& {\bf Period (s)}		& {\bf Flux Density at 1.4GHz (mJy)} \\
\hline
\hline
\multicolumn{3}{|c|}{Fastest of the Very Bright} \\
\hline
\hline
B0355+54    & 0.1563824177774  		& 23 \\ 
\hline
B1929+10  	& 0.226517634984      & 41 \\
\hline
B0950+08  	& 0.2530651649482     & 84 \\
\hline
B2020+28  	& 0.3434021577859     & 38 \\
\hline
B1933+16  	& 0.3587384107696     & 42 \\
\hline
B0329+54		& 0.7145196822210			& 203 \\
\hline
\hline
\multicolumn{3}{|c|}{Brightest of the Very Fast} \\
\hline
\hline
B1937+21    & 0.0015578064724     & 16 \\
\hline
B0531+21    & 0.03308471603       & 14 \\
\hline
B1855+09    & 0.0053621004540     & 4 \\
\hline
J1713+0747  & 0.0045701365242     & 3 \\
\hline
J1012+5307  & 0.0052557490141     & 2.8 \\
\hline
J1022+1001  & 0.0164529296832     & 2.3 \\
\hline
\end{tabular}
\end{table}

\section{CASPER}

The Center for Astronomy Signal Processing and Electronics Research at the University of California, Berkeley, has developed a set of hardware and software tools for building radio astronomy digital instrumentation. We have made extensive use of their technology to develop the projects described in this thesis. In this section we provide a brief overview of the CASPER technologies.

\subsection{The CASPER Approach}


CASPER was created as a result of the realization that digital signal processing hardware technology, and especially FPGAs, have reached a point where it is possible to build almost all the digital instrumentation required in a radio telescope from a common hardware platform. However, current practice is for observatories to custom-build new hardware whenever they need a new instrument. This is not only extremely wasteful of engineering effort, but also results in lengthened development times for instruments. \\

CASPER therefore advocates an approach to building instrumentation that is focused on {\it reusability}. In most cases the benefits that one obtains from custom-designing a hardware processing board for a particular application are far outweighed by the disadvantages in the additional cost and time that such development takes. CASPER has developed a set of reusable, generic hardware modules that can be used to develop a wide range of instruments, including spectrometers, correlators and beamformers. \\

Another thrust of CASPER's efforts is to encourage the use of industry standard communications protocols. Radio astronomy observatories have a history of developing custom protocols for their instruments, which makes it very difficult for instruments from different observatories to inter-operate, or for observatories to share instruments with each other. CASPER advocates the use of industry-standard Ethernet, XAUI \cite{XAUI} and 10Gb Ethernet \cite{10GbE} for instrument control, streaming data and packetized data respectively. \\

\subsection{CASPER Hardware}


CASPER's hardware offerings are described in detail in \cite{CASPER}. There are two current processing boards, one digitization board, and one future processing board. \\

\begin{figure}[htp]
	\centering
		\includegraphics[width=4in]{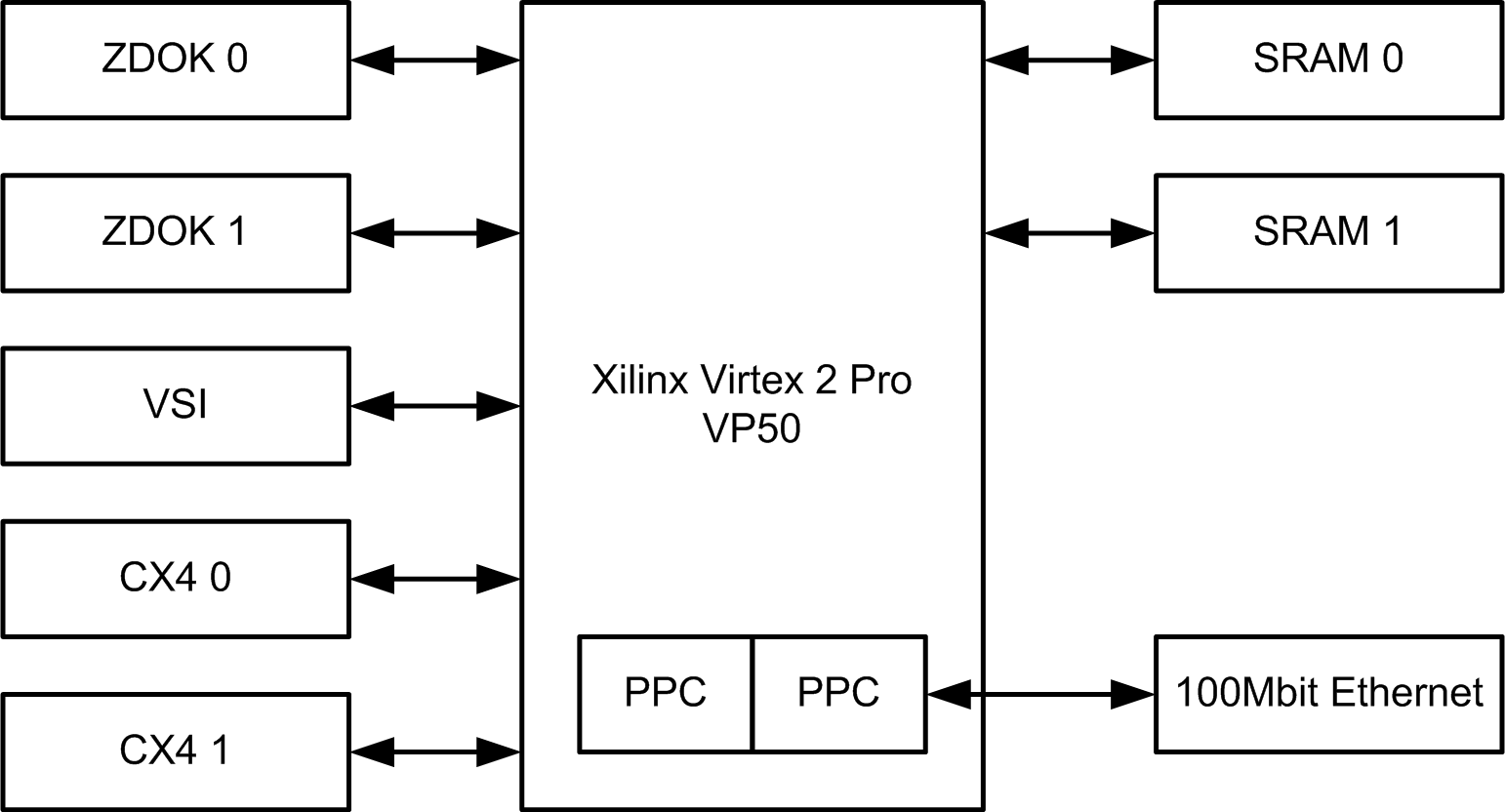}
		\caption{IBOB Block Diagram.}
	\label{fig:BackgroundIBOBBlockDiagram}
\end{figure}

The IBOB (``Internet Break-Out Board'') is a processing board that has the ability to connect to two digitization boards, and has two 10GbE (CX4) ports and one 100MbE port. A basic block diagram for the IBOB is shown in Figure \ref{fig:BackgroundIBOBBlockDiagram}. The IBOB's centrepiece is a Xilinx Virtex 2 Pro VP50 FPGA, which includes two embedded PowerPC cores (of which only one is used). For small instruments, such as the pulsar spectrometers described in this thesis, this board can be used as a standalone device. However, it can also be used as the first stage in larger instruments such as correlators and beamformers, where a set of IBOBs is typically used for digitization and channelization of all the antennas. \\

The iADC is a board that connects to the IBOB via a ZDOK connector. Each iADC board includes two ADCs in a single IC package. The ADCs can be operated individually, at 1GSa/sec each, or can be used in an interleaved mode that effectively gives a sampling rate of 2GSa/sec. \\

CASPER's large processing platform is the BEE2 \cite{CWB05}. This is a hardware platform that includes five Xilinx Virtex 2 Pro VP70 FPGAs, which again each have two embedded PowerPC cores. One FPGA, the ``Control FPGA'', runs a modified version of Linux, BORPH \cite{BORPH}. Each FPGA has at least two 10GbE ports available, and four DDR2 DIMM sockets. This, combined with high-speed IO links between the FPGAs on the board, makes the BEE2 an ideal platform for applications that require large memory bandwidth or IO bandwidth. \\

ROACH (``Reconfigurable Open Architecture Computing Hardware'') is CASPER's next-generation processing board, built in collaboration with the Karoo Array Telescope and NRAO Socorro. ROACH is intended as a replacement for IBOB and BEE2 boards in 2009. It features a single Xilinx Virtex 5 FPGA (either SX95, LX110T or LX155T), four 10GbE ports, an external PowerPC for control and monitoring, 72Mbit QDR SRAM and two DDR2 DIMM slots. It includes the same ZDOK connectors as the IBOB, so that it can also be used with up to two iADC boards. \\

\subsection{CASPER Toolflow and Libraries}

Development for the IBOB, BEE2 and ROACH boards is supported using a toolflow based on MATLAB's Simulink graphical modeling tool, and Xilinx's System Generator product, which allows for development of FPGA designs from within Simulink. Xilinx's EDK and ISE tools are used to build designs with PowerPC support. \\

The Simulink toolflow for the BEE2 was pioneered by Chen Chang, and is described in his Ph.D. thesis \cite{ChangThesis}. Custom libraries for Simulink allow the hardware on the BEE2, and other boards, to be accessed in Simulink designs. For example, there are abstractions for the DRAM and 10GbE interfaces that are available on the BEE2. \\

The Simulink toolflow also provides the concept of ``shared registers'' and ``shared BRAMs''. These resources are physically implemented in the FPGA (just as regular registers and BRAMs are), but are also connected to a bus that allows their values to be accessed and manipulated from the embedded\footnote{The ROACH board does not use an FPGA with an embedded PowerPC, but it supports shared resources by exposing a bus that is connected to an external PowerPC. This PowerPC will also run BORPH.} PowerPC. There is software support to access these shared resources; the IBOB FPGA's PowerPC runs a telnet server that exposes the resources, and the BEE2 control FPGA runs a modified version of Linux, BORPH \cite{BORPH}, that provides a filesystem abstraction for the resources. \\

In addition to hardware support in Simulink, there also exists a CASPER DSP library that has been built specifically to provide the necessary DSP functions that one needs to build radio astronomy instruments using the IBOB, BEE2 and ROACH boards. The CASPER DSP library includes the following core components: \\

\begin{enumerate}
  \item Streaming Parallel FFT
  \item Polyphase Filterbank FIR Filter
  \item Digital Downconverter
  \item Arbitrary Reorder, and Transpose
\end{enumerate}

The DSP library also includes a set of useful ``helper'' blocks, such as counters that freeze instead of wrap, blocks to detect positive and negative edges, and so on. \\

The streaming parallel FFT block is a streaming implementation of the FFT algorithm that allows the designer to input $2^k$ values per clock cycle, for arbitrary $k$. This is crucial for applications where the ADC sampling frequency is higher than the FPGA clock frequency, and a downconverter is not used. Typically the iADC will be clocked at four times the rate of the FPGA on an IBOB, and hence each ADC will present four samples every FPGA clock cycle. Therefore in order to compute a spectrum, for example, it is necessary to have an FFT implementation that accepts four values per clock cycle. \\

The Polyphase Filterbank FIR Filter is used in conjunction with the FFT to produce a polyphase filterbank (PFB). An ordinary FFT exhibits the phenomenon of {\it spectral leakage}, whereby the value in one spectral channel ``leaks'' into adjacent channels. This occurs because the input to the FFT is sampled, not continuous data. The polyphase filterbank significantly reduces this spectral leakage by first convolving the input data with a windowing function. \\

The Digital Downconverter is used to digitally extract a band of interest from the sampled bandwidth. It is implemented in the standard way -- using a lookup for the sine and cosine functions, a ``mixer'' implemented using hardware multipliers, and an FIR filter and decimation stage. \\

The Reorder and Transpose blocks have a variety of uses. Reordering is necessary inside the FFT, but is also a common operation in instruments where it is useful to group frequency components together, for example. An example of where a transpose is necessary is in instruments that have two stages of channelization -- a first FFT does a coarse channelization, and then a second FFT further channelizes each resulting channel. For large matrices, a DRAM-based version of the Transpose function is provided. \\

\newpage

\chapter{The Fly's Eye: Instrumentation for the Detection of Millisecond Radio Pulses}


In this chapter we present the design and implementation of the ``Fly's Eye'' instrument for the Allen Telescope Array. This instrument was purpose-built for an experiment also known as ``Fly's Eye'' to search for bright radio pulses of millisecond duration. We designed and built a system containing 44 independent spectrometers using 11 IBOBs. Each spectrometer processes a bandwidth of 210MHz, and produces a 128-channel power spectrum at a rate of 1600Hz (i.e. 1600 spectra are outputted by each spectrometer per second). Therefore each spectrum represents time domain data of length 1/1600Hz=0.000625s=0.625ms, and hence pulses as short as 0.625ms can be resolved\footnote{Pulses of duration $<$0.625ms can also be detected provided that they are sufficiently bright, but their length cannot be determined with a precision greater than the single spectrum length.}.

\section{Science Motivation}

\begin{figure}[htp]
	\centering
		\includegraphics{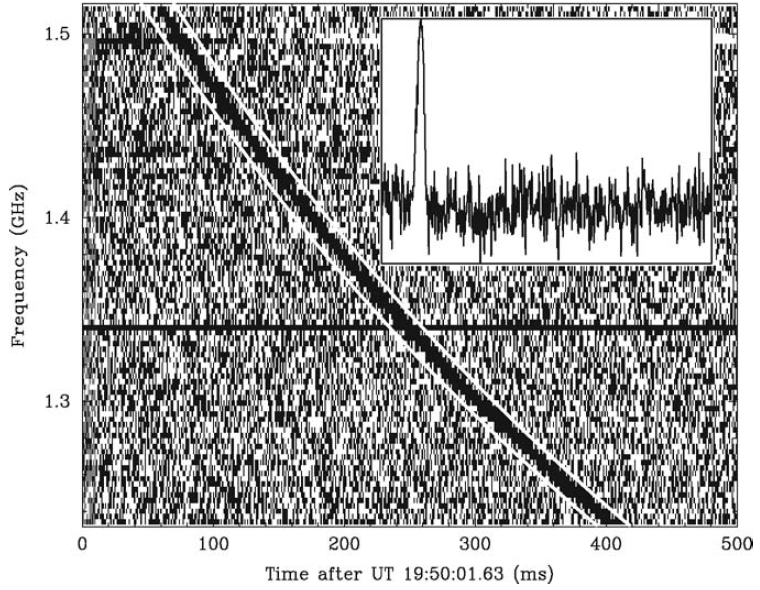}
		\caption[A Single Bright Pulse of Extragalactic Origin.]{From \cite{LorimerPulse07}. The pulse discovered by Lorimer et. al. from data taken on 21 August 2001. The main graph shows the raw captured spectral data, in which the dispersed pulse can clearly be seen. The inset image shows the same data after the pulse has been dedispersed using DM = 375cm$^{-3}$ pc.}
	\label{fig:LorimerPulse}
\end{figure}

In 2007, Lorimer et. al. \cite{LorimerPulse07} announced their discovery of a bright single pulse with duration less than 5ms, which they deduced to be of extragalactic origin. Specifically, the 30Jy burst was dispersed with DM = 375cm$^{-3}$ pc, which based on current models of free electron content in the universe, Lorimer et. al. argue the source of the burst to be up to 1Gpc distant. This, combined with the direction from which the burst came, and other burst characteristics, indicate strongly that the burst is of extragalactic origin. Figure \ref{fig:LorimerPulse} shows the pulse that Lorimer et. al. detected. \\

\begin{figure}[htp]
	\centering
		\includegraphics{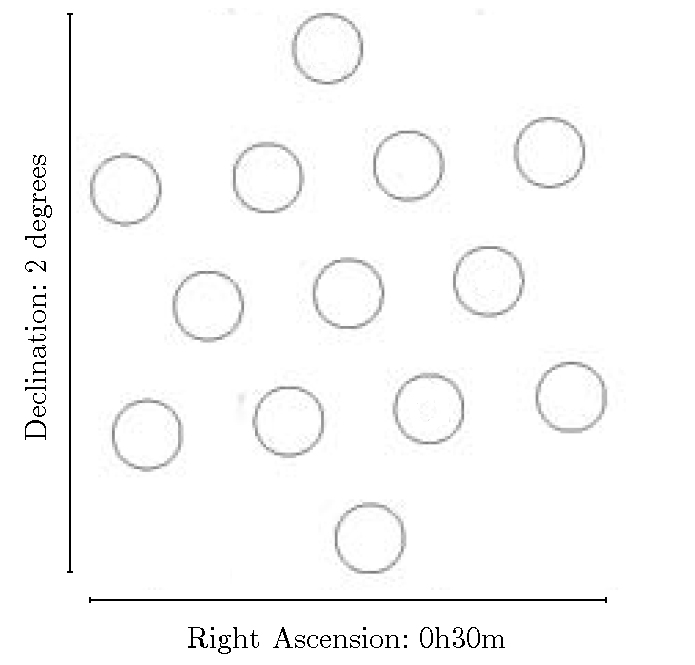}
		\caption[Beam Pattern from the Parkes Pulsar Survey.]{Modified from \cite{LorimerPulse07}. The beam pattern of the 13 beams, with the diameters equal to the half-power width.}
	\label{fig:LorimerBeams}
\end{figure}

The Lorimer et. al. pulse was discovered using the 64m Parkes Radio Telescope with a 13-beam receiver. Each beam had a bandwidth of 288MHz, and the digital backend produced 96-channel power spectra for each beam. Figure \ref{fig:LorimerBeams} shows the beam pattern used. The detected pulse saturated the digitizer of one beam, and also appeared in two adjacent beams, lending further credibility to the conclusion that this was an astronomical signal, not an artifact in the instrumentation. A packed beam pattern such as the one shown also helps to localize sources. \\

Lorimer et. al. found the pulse by dedispersing data from each beam with DMs between 0 and 500 cm$^{-3}$ pc and then searching the dedispersed data for individual pulses with signal-to-noise ratios greater than 4. They processed 90 hours of data from the source of the burst, and over 500 hours of data overall, and did not discover any other pulses. \\

There is now significant interest in follow-up studies to try to detect more similar pulses to further constrain the rate of their occurance, and to hopefully reveal more information about their origin. In \cite{Kulkarni07}, Kulkarni et. al. determine the rate to be $10^{-4} < R(S > 30$Jy$)<0.1$ deg$^{-2}$ day$^{-1}$ based on the discovery of the single pulse, but more data is desperately needed.

\section{Searching for Bright Pulses with the ATA}


The Allen Telescope Array has several advantages over other telescopes worldwide for performing transient searches, particularly when the search is for bright pulses. The ATA has 42 independently-steerable dishes, each 6m in diameter. Since beam size is inversely proportional to dish size, the beam size for individual ATA dishes is considerably larger than that for most other telescopes\footnote{The dishes at the VLA, NRAO Green Bank, Parkes, Arecibo, Westerbork and Effelsberg all have $d > 20$m, with some $d\geq100$m.}. This means that the ATA can instantaneously observe a far larger portion of the sky than is possible with other telescopes. Conversely, when using the ATA dishes independently, the sensitivity of the ATA is far lower than that of other telescopes. However, if the assumption that pulses in the class that Lorimer et. al. discovered are often brighter than the 30Jy pulse from 2001 holds true, then the ATA will be a superior instrument for detecting pulses from this class\footnote{Of course, if it is subsequently discovered that the Lorimer et. al. pulse was an outlier to the upside, and that most pulses have fluxes less than 30Jy, then the ATA will not be as effective an instrument as others at detecting these bursts.}. \\

Another important advantage that the ATA currently enjoys is that it is still under construction, and is not yet in general use. Consequently it is possible to obtain large amounts of observing time on the telescope, whereas this is typically not possible with other telescopes. \\

\begin{figure}[htp]
	\centering
		\includegraphics{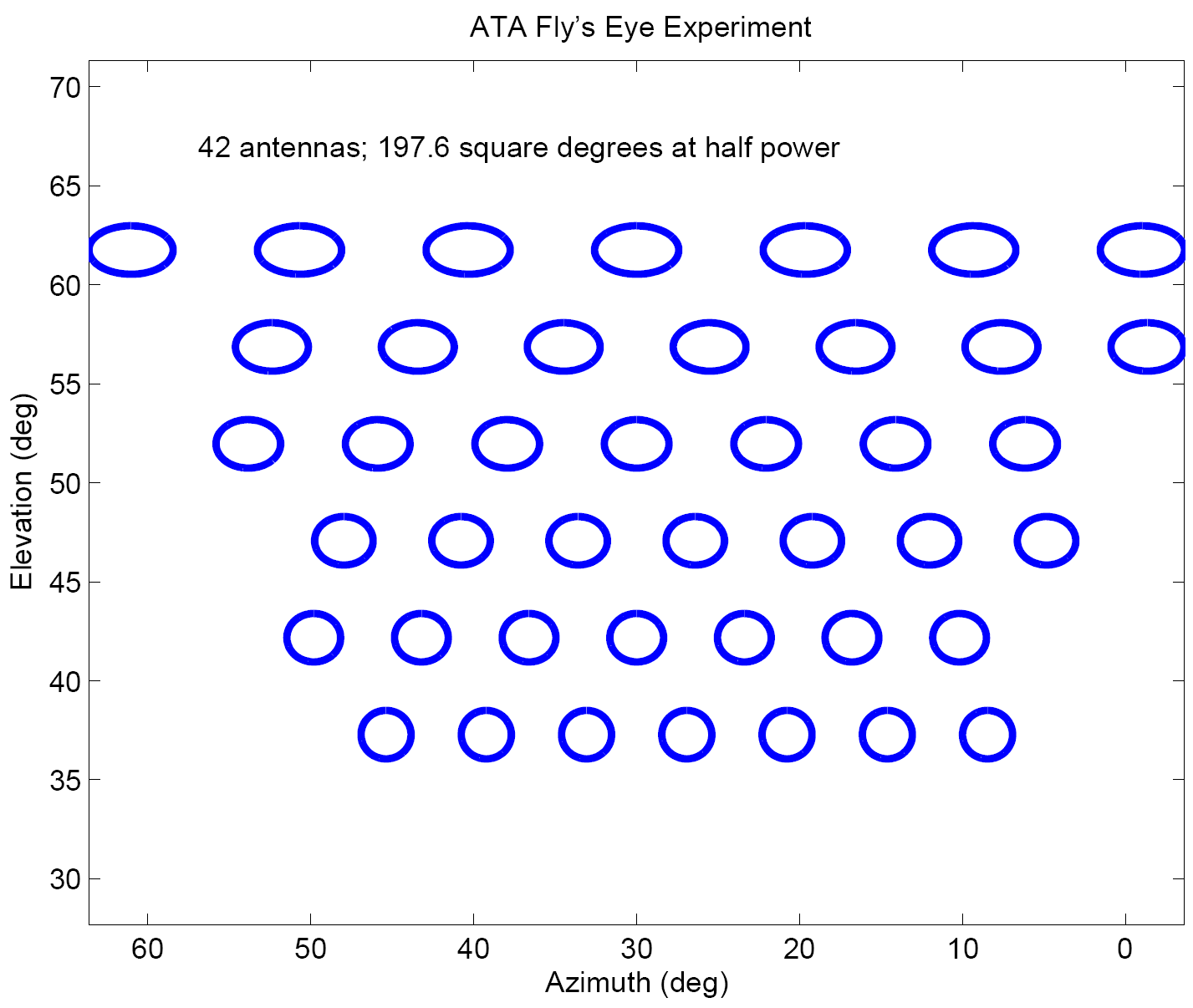}
		\caption[A hexagonal packing beam pattern for 42 antennas at the ATA.]{From \cite{BowerFlysEyeProposal08}. The beam pattern of the 42 beams at ATA, with the diameters equal to the half-power width. This hexagonal packing is pointing north. A south pointing results in poor interference properties, due to the highly populated areas south of the ATA.}
	\label{fig:FlysEyeNorthernPointingBeams}
\end{figure}

\begin{figure}[htp]
	\centering
		\includegraphics{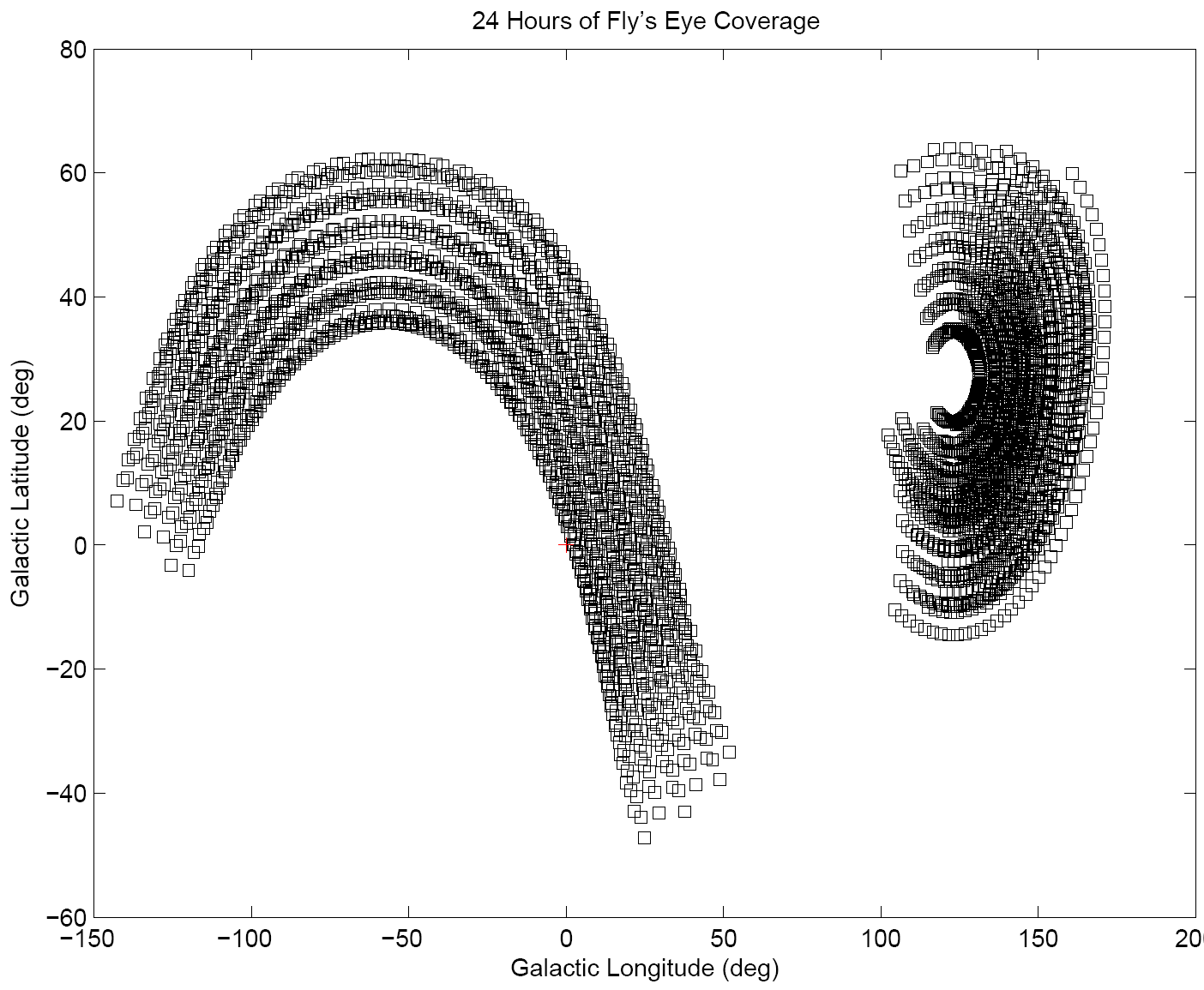}
		\caption[Sky coverage over 24 hours for 42 antennas at the ATA.]{From \cite{BowerFlysEyeProposal08}. The sky coverage of the ATA for an observing period of 24 hours. Both the coverages for southern and northern pointings (corresponding to the respective contiguous regions) are shown.}
	\label{fig:FlysEyeSkyCoverage}
\end{figure}

In February 2008, we submitted a proposal \cite{BowerFlysEyeProposal08} and were awarded 408 hours of observing time on weekends from 15 February 2008 to 4 April 2008. Figures \ref{fig:FlysEyeNorthernPointingBeams} and \ref{fig:FlysEyeSkyCoverage} from \cite{BowerFlysEyeProposal08} show the beam pattern and sky coverage using all 42 antennas.

\section{System Architecture}

\begin{figure}[htp]
	\centering
		\includegraphics[width=4in]{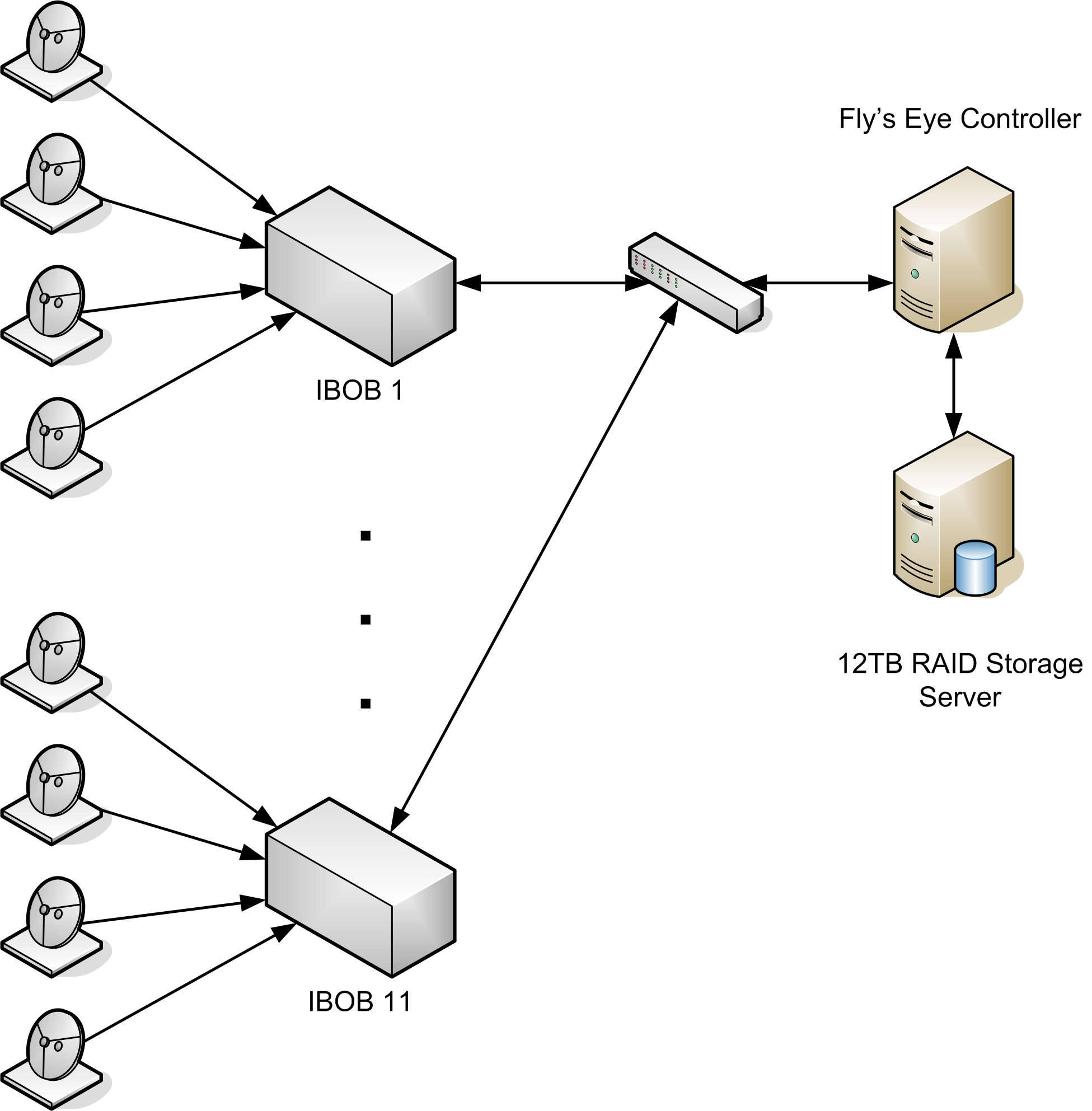}
		\caption[Fly's Eye System Architecture.]{Fly's Eye System Architecture. A selection of 44 analogue signals from 42 dual polarization antennas are connected to 44 independent spectrometers implemented in 11 IBOBs.}
	\label{fig:FlysEyeSystemArchitecture}
\end{figure}

The overall architecture of the Fly's Eye system is shown in Figure \ref{fig:FlysEyeSystemArchitecture}. Each IBOB can digitize four analogue signals, and 11 IBOBs are provided so that 44 signals can be processed. The ATA has 42 antennas, each with two polarization outputs. A selection\footnote{The selection is made with consideration for the goal of maximising field-of-view -- in practice we selected at least one polarization signal from every functioning antenna.} of 44 of the available 84 signals is made, and these are connected to the 44 iADC inputs. The IBOBs are connected to a control computer and a storage computer via a standard Ethernet switch.

\begin{figure}[htp]
	\centering
		\includegraphics[height=6in]{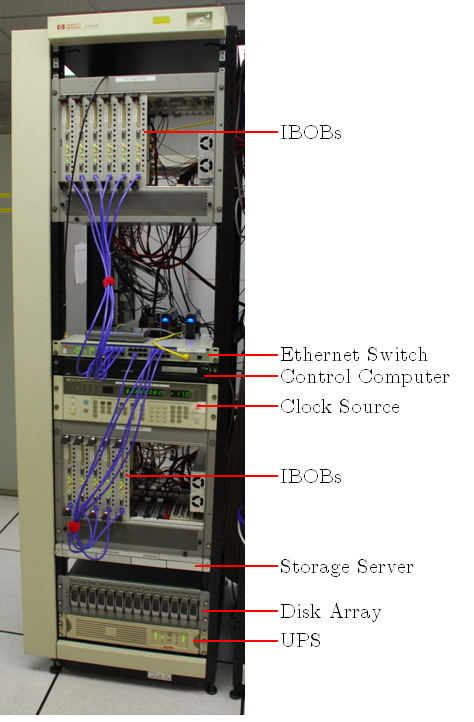}
		\caption[Fly's Eye Rack at the ATA.]{Fly's Eye Rack at the ATA. Two 6U CompactPCI crates (top and bottom) house the 11 IBOBs. The switch, data recorder computer and storage server are all visible.}
	\label{fig:FlysEyeRack}
\end{figure}

\begin{figure}[htp]
	\centering
		\includegraphics[width=8.75in,angle=90]{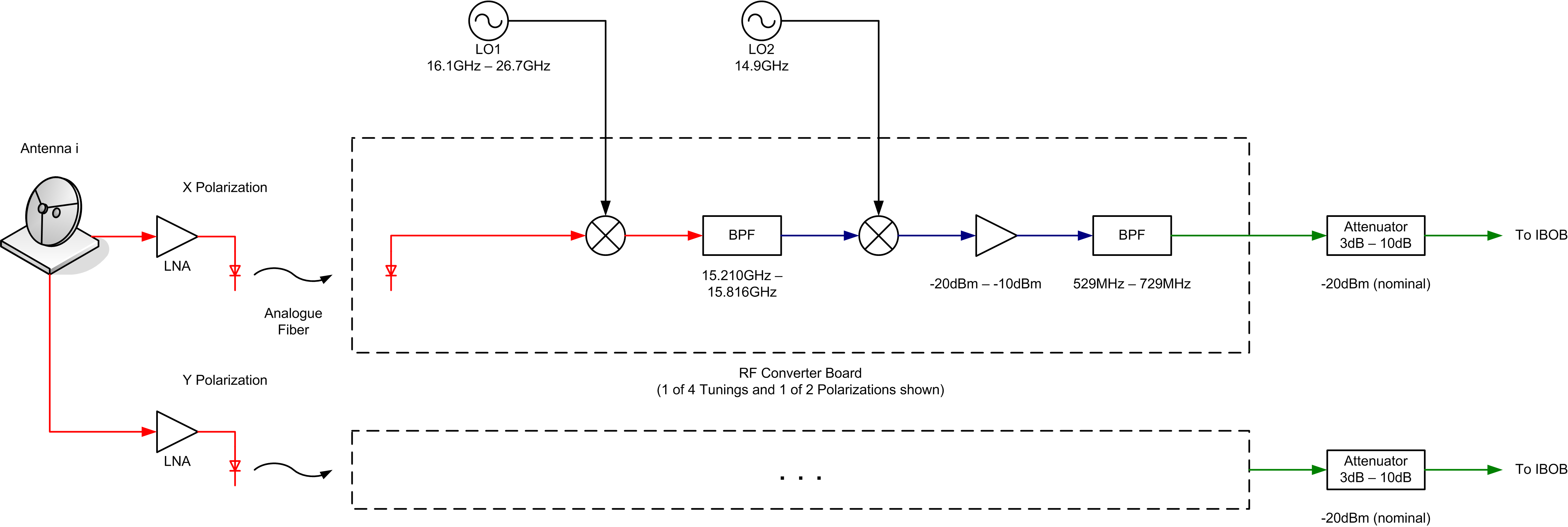}
		\caption{Signal path for the Fly's Eye Experiment at the ATA.}
	\label{fig:FlysEyeATASignals}
\end{figure}

Figure \ref{fig:FlysEyeATASignals} shows the signal path at the ATA for the Fly's Eye experiment. The analogue frontend produces an IF band from 529MHz - 729MHz. The sky frequency is fully controllable due to the presence of the controllable oscillator LO1 in the analogue frontend architecture. We typically used a sky frequency of 1430MHz (i.e. observing the band 1330MHz -- 1530MHz). The IBOB ADCs are clocked at 838.8608MHz, so the IF signal is in the second Nyquist zone. A digital downconverter (DDC) is used to extract a 209.7154MHz band from 524.288MHz -- 734.0032MHz.

\section{A 210MHz Quad Spectrometer}


The quad spectrometer design in each IBOB was based on the ``Pocket Correlator'' (``PoCo'') design by Aaron Parsons \cite{ParsonsPoCo}. The PoCo uses a 1024-channel PFB, whereas the quad spectrometer design required only a 128-channel PFB. This 4-input FX correlator produces power spectra for each input in addition to cross-terms created by multiplying the spectrum from each input with that from every other input. \\

For the Fly's Eye transient detection application, we are only interested in power spectra, not the cross-terms that are calculated by a correlator. We wanted to obtain the maximum dump rate possible with the 100MbE port on the IBOB, which, due to processor bus constraints, is limited to a data rate of 7Mbps. Therefore it was important to eliminate the cross-terms from the output, since their presence would not only be wasteful, but would also greatly reduce the rate at which we could dump power spectra, and hence would increase our integration time. This in turn would degrade our instrument's sensitivity to short bursts. Hence we removed the cross-term calculations from the PoCo design. \\

Two further important modifications were made to the PoCo design. To save resources the PoCo resizes the output data width of the FFT from 18 bits to 4 bits. In our design this is unnecessary, and we opted to maintain the full precision from the FFT\footnote{Since the quad spectrometer only uses a 128-point FFT, there was an ample amount of logic resources available so the full precision could easily be maintained.}. The output buffers in the PoCo were also significantly modified to provide the option of a debug output of all 64 bits from the accumulators, and to maximize the output data rate. \\

\begin{figure}[htp]
	\centering
		\includegraphics[width=8in,angle=90]{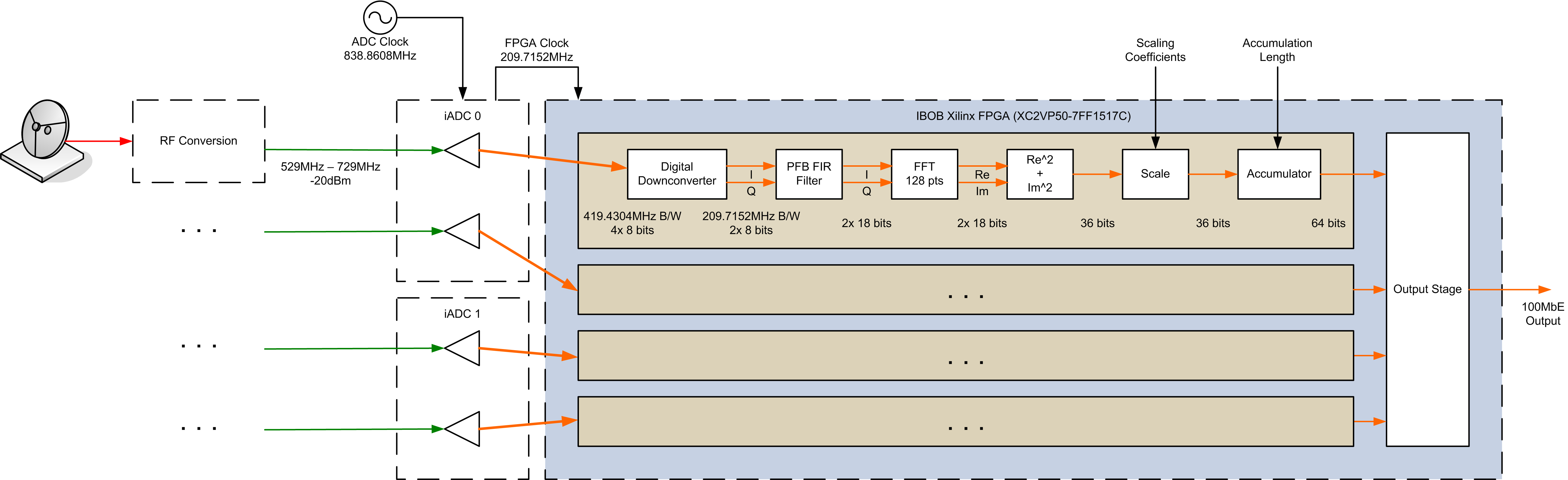}
		\caption[A Quad Spectrometer for the Fly's Eye.]{A Quad Spectrometer for the Fly's Eye. This single IBOB design implements four independent 210MHz spectrometers whose spectra are dumped over the IBOB's single 100MbE connection.}
	\label{fig:FlysEyeQuadSpectrometer}
\end{figure}

\begin{figure}[htp]
	\centering
		\includegraphics[width=8in,angle=90]{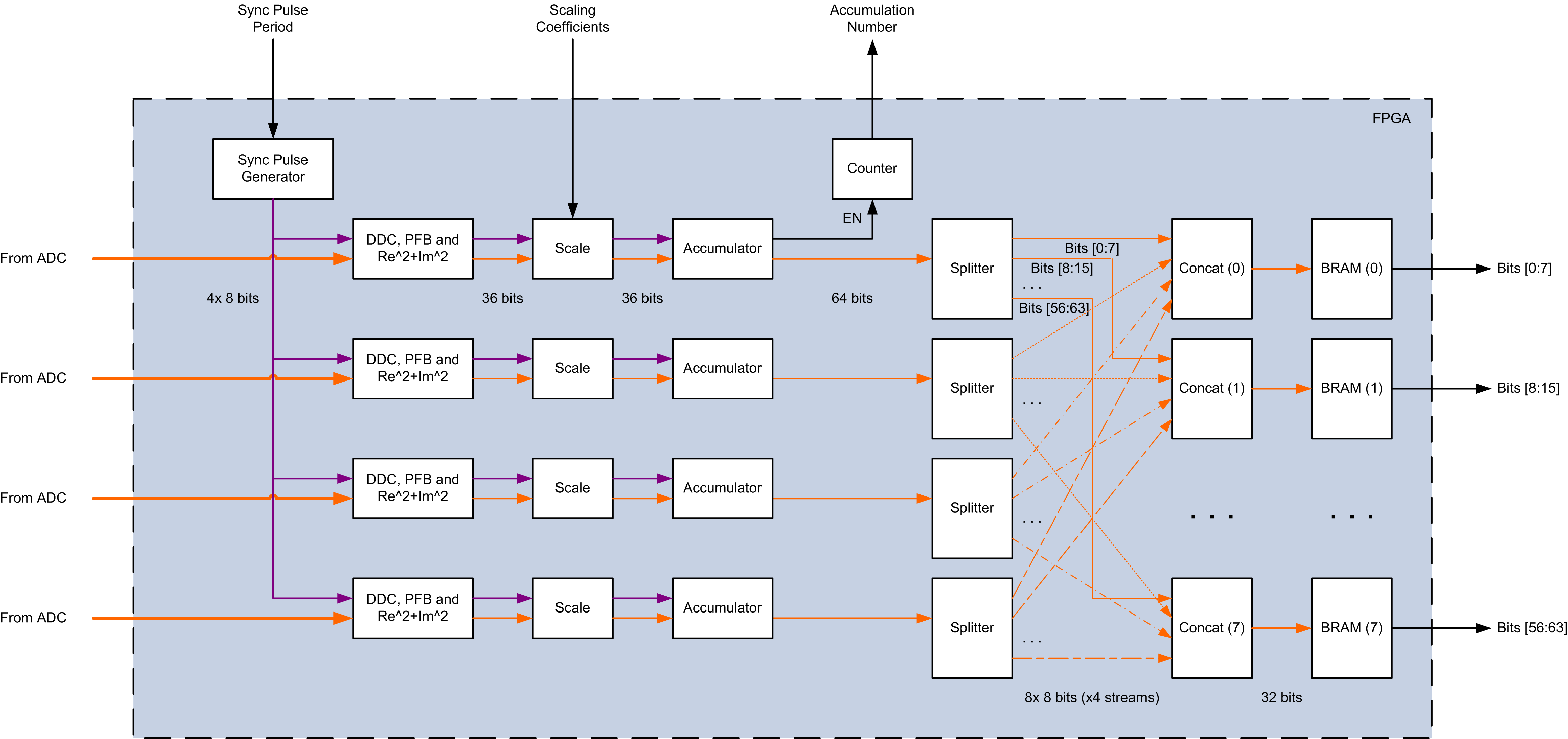}
		\caption[The Output Stage of the Fly's Eye Quad Spectrometer.]{The Output Stage of the Fly's Eye Quad Spectrometer. All 64-bits from each input's accumulator are buffered, and code on the PowerPC determines which set of 8 bits should be dumped.}
	\label{fig:FlysEyeSpectrometerOutputStage}
\end{figure}

Figure \ref{fig:FlysEyeQuadSpectrometer} shows the final design of the modified PoCo design for the Fly's Eye experiment. It is effectively a fast-readout quad spectrometer. Four ADCs attached to an IBOB sample four independent analogue signals at 838.8608MSa/sec. Each digitized data stream is processed in parallel in the FPGA. First each signal is passed through a digital downcoverter, which selects the band 104.8576MHz -- 314.5728MHz from the sampled band 0 -- 419.4304MHz. However, since the Fly's Eye setup operates in the second Nyquist zone, the selected band is actually (838.8608-314.5728)MHz -- (838.8608-104.8576)MHz, which is 524.288MHz -- 734.0032MHz. This includes all of the available signal that is provided from 529MHz -- 729MHz (see the signal flow in Figure \ref{fig:FlysEyeATASignals}). Next, the PFB FIR and FFT channelize each input into 128 bins. The FFT output is passed to an ``equalizer'', which allows the user to scale each frequency bin by a specific constant. In practice we used this feature purely as a scale with no frequency dependence, and instead performed equalization of the band in postprocessing in software. The power spectrum is then computed (by summing the squares of the real and imaginary parts of each FFT coefficient) and accumulated. Finally, the accumulated spectra are written into the output stage BRAM buffers. A counter is incremented when an accumulation is ready, and this signal is used by software running on the PowerPC to determine when to start sending the next set of spectra. Figure \ref{fig:FlysEyeSpectrometerOutputStage} shows this output stage in detail. The 64 bits produced by each accumulator are separated into eight sets of 8 bits. The corresponding sets of 8 bits from all four accumulators are concatenated to produce a 32-bit value that contains the same order 8 bits from all four inputs. \\

\begin{figure}[htp]
	\centering
		\includegraphics[width=4in]{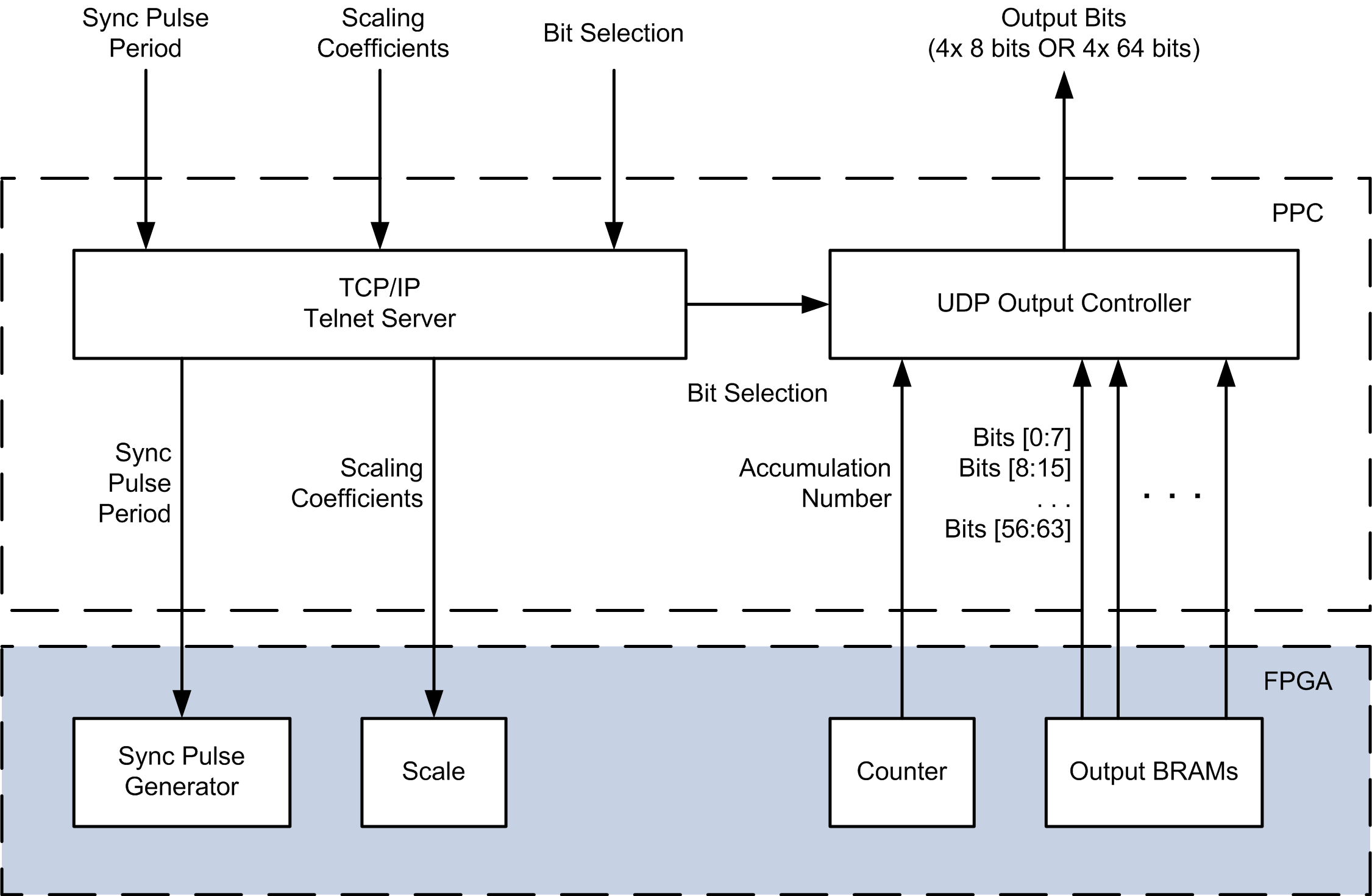}
		\caption[PowerPC-FPGA Interface of the Fly's Eye Quad Spectrometer.]{PowerPC-FPGA Interface of the Fly's Eye Quad Spectrometer. An embedded PowerPC hosts a TCP Telnet server that processes user requests, including the setting of parameters. A UDP output controller runs concurrently, and is responsible for appropriately creating output data packets.}
	\label{fig:FlysEyeSpectrometerPPC}
\end{figure}

During normal operation, only one set of 8 bits is outputted, but this setup allows for a debug mode whereby all eight buffers are outputted, and hence all 64 bits from each of the four inputs are made available. This is particularly useful when the operator is deciding which 8 bits he or she should select to output, and what scale coefficients are necessary. Figure \ref{fig:FlysEyeSpectrometerPPC} shows the interface between the FPGA's embedded PowerPC, the spectrometer design, and the external Ethernet interface. The PowerPC software includes a module, the ``UDP Output Controller'', that is responsible for detecting when a new accumulation is ready and sending it out over the IBOB's 100MbE interface as a UDP packet. If the bit selection parameter is specified to request that all 64 bits be outputted, then this information will be used by the UDP Output Controller, and it will concatenate the bits from the FPGA buffers to recreate the original 64 bit accumulator outputs. \\

\section{Observing Software}


The Fly's Eye system was set up for automatic operation, although at the start of an observing session the operator needs to manually perform some setup and control of the telescope array. We wrote a set of scripts to reset and sequentially configure the IBOBs (with parameters such as accumulation length\footnote{The accumulation length is determined by the ``sync'' pulse period parameter.}, scaling coefficients and bit selection). We wrote a script to set up the telescopes appropriately (it sets the correct pointings and oscillator frequencies). Most importantly, we automated the data recording and monitoring process: we created a script that runs on the control computer that will manage an observation for a given period of time (typically 60 hours). \\

\begin{figure}[htp]
	\centering
		\includegraphics[width=6in]{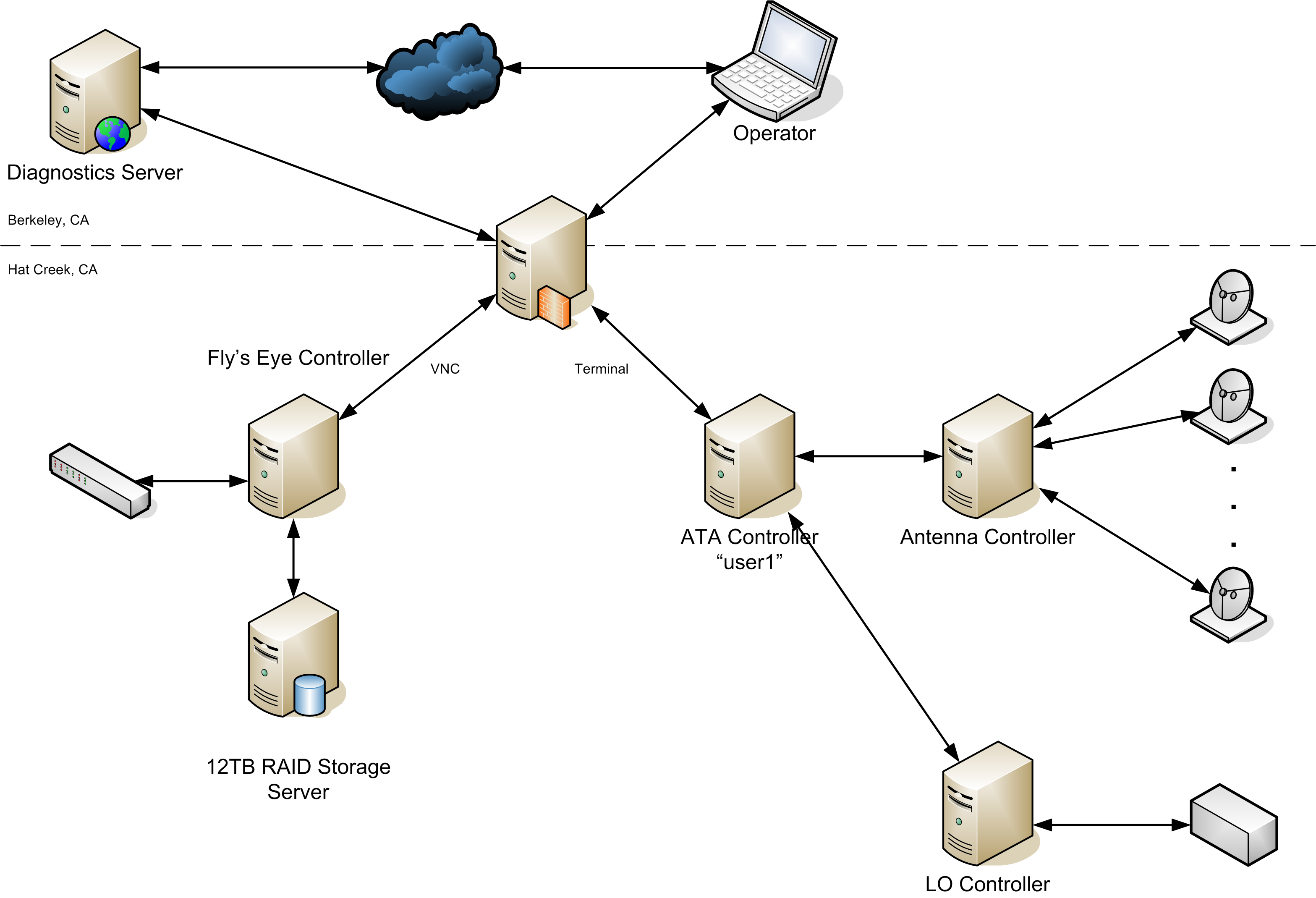}
		\caption[Fly's Eye Control Architecture.]{Fly's Eye Control Architecture. The Fly's Eye equipment is controlled through a computer dedicated to the experiment; this machine is also responsible for managing the data recording. The control of the general ATA infrastructure (antennas and oscillators) is done through a separate, general-use control computer. A remote web server is used to host diagnostic information.}
	\label{fig:FlysEyeArchitectureControl}
\end{figure}

Figure \ref{fig:FlysEyeArchitectureControl} shows the overall architecture of the control and monitoring infrastructure for Fly's Eye. A VNC session to the Fly's Eye control computer is used to run the IBOB configuration and observation scripts. Due to the fact that programs running in a VNC session are by default not killed when the client-server connection fails, we typically used the VNC session for all interactive user control, including with the ATA control computer. The web server shown in this figure is used to host periodically-updated diagnostics from the Fly's Eye control computer. The motivation for this is that during a long automated observation it is useful to have hourly reports that the system is functioning correctly. \\

\begin{figure}[htp]
	\centering
		\includegraphics[width=4in]{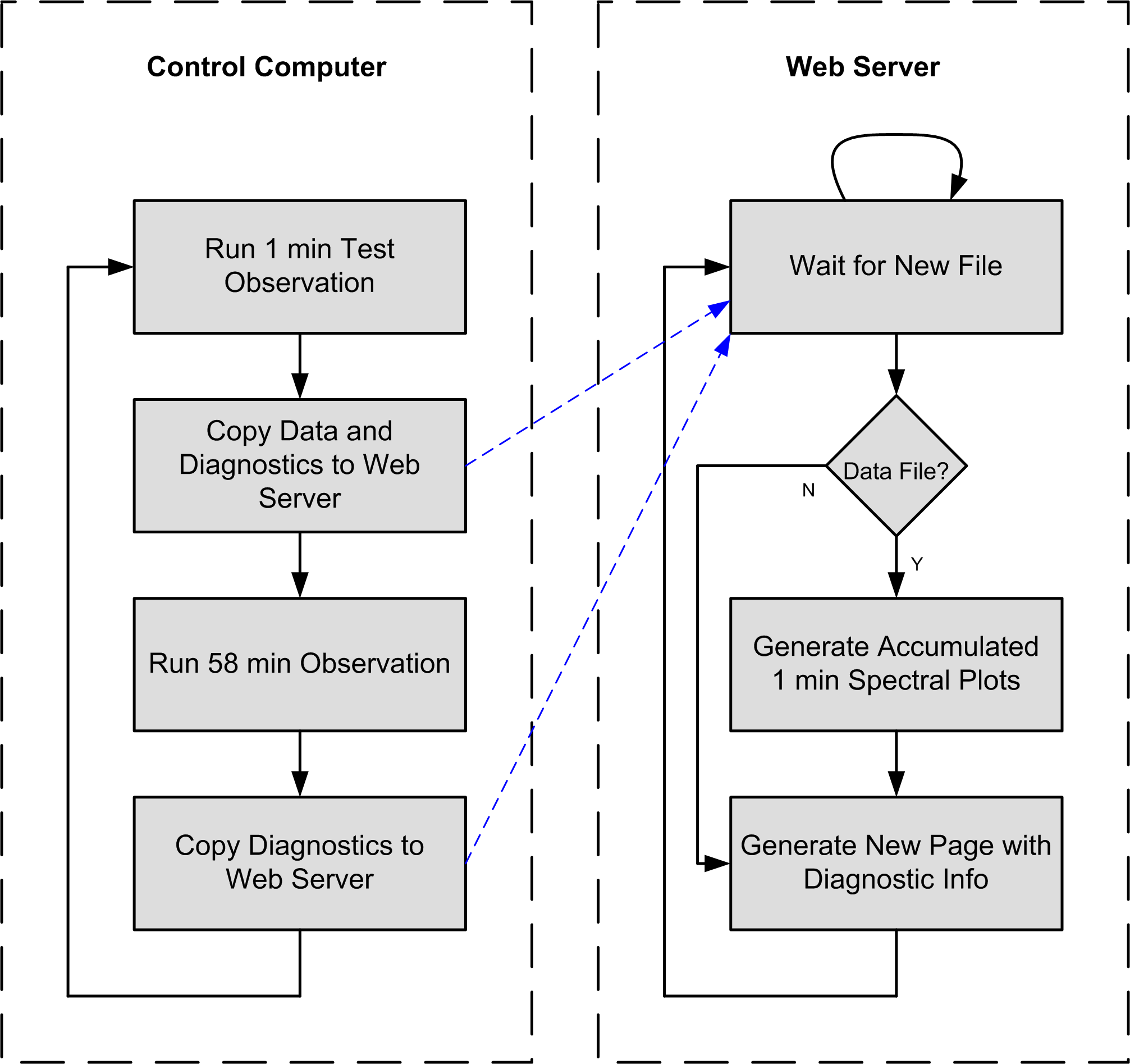}
		\caption[Fly's Eye Control Script Flow.]{Fly's Eye Control Script Flow. The control computer alternates between running a 1 minute test observation and a 58 minute live observation. The test observation data is transferred to the web server, where it is plotted and made available on the web, along with other diagnostic information provided by the control computer.}
	\label{fig:FlysEyeControlScripts}
\end{figure}

Figure \ref{fig:FlysEyeControlScripts} shows the control flow in the scripts that run on the control computer and web server during an observation. The control computer loops in a one hour cycle: it does a 1-minute test observation, and follows it with a 58-minute observation. The web server generates spectral plots from the test observation (44 independent plots) so that the operator can periodically visit a website and have an easy visual check that the system is functioning. After both the 1-minute and 58-minute observations a set of diagnostics about the network connections and disk subsystems is also transferred to the web server for inclusion in the diagnostic web pages.

\section{Offline Processing to Detect Transients}


The analysis required for the Fly's Eye experiment is, in principle, fairly simple -- we wish to search over a wide range of dispersion measures to find large individual pulses. Specifically our processing requires that all the data be dedispersed with dispersion measures ranging from 50 cm$^{-3}$ pc to 2000 cm$^{-3}$ pc. At each dispersion measure the data needs to be searched for `bright' pulses. The practical definition `bright' might change depending on the scientist analysing the data, but essentially the task is to search for pulses whose powers are greater than $\tau \sigma + \mu$ where $\tau$ is some user-specified threshold (for example, $\tau=10$), $\sigma$ is the standard deviation of the signal power, and $\mu$ is the mean signal power. \\

\begin{figure}[htp]
	\centering
		\includegraphics[width=4in]{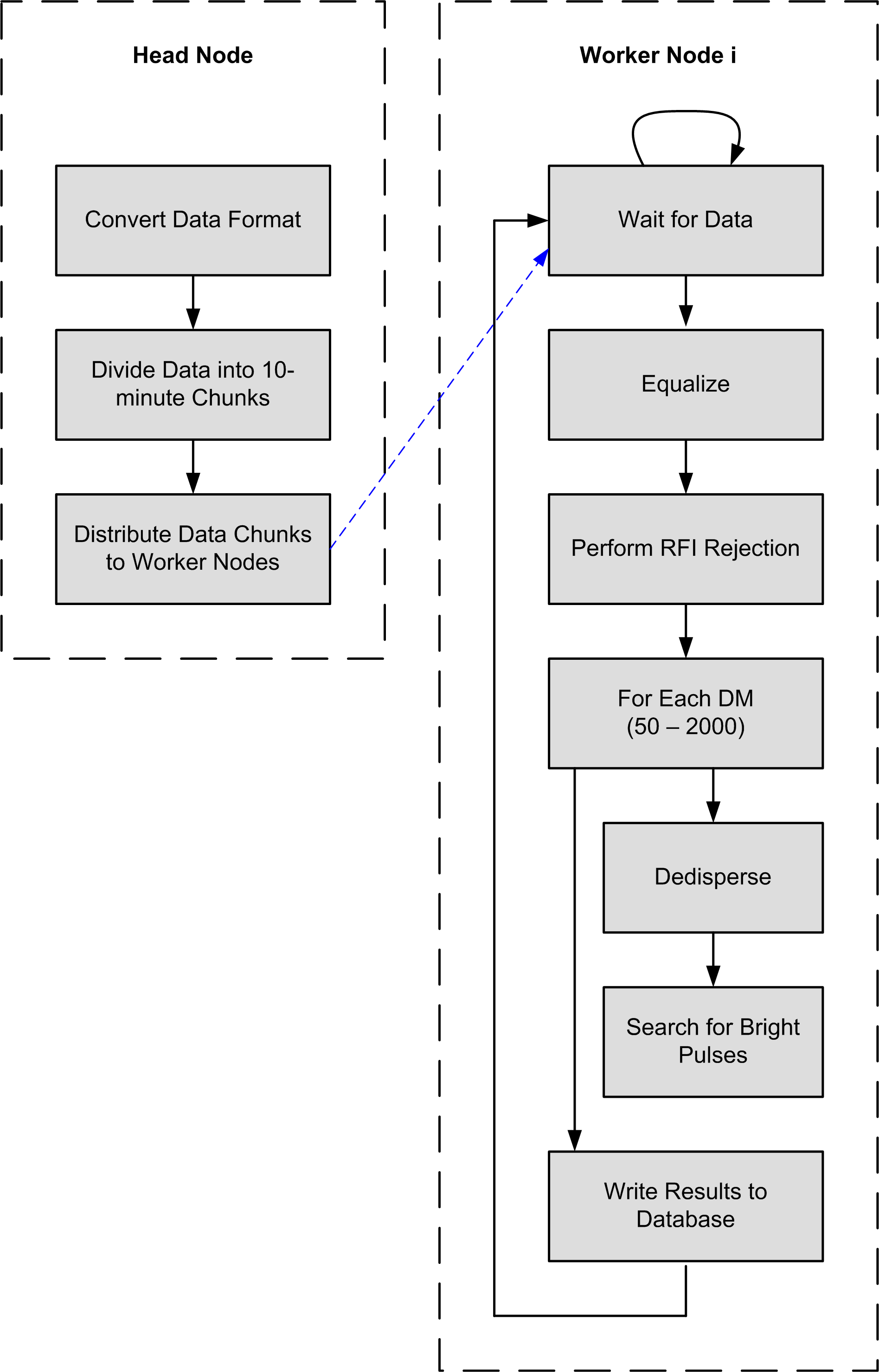}
		\caption[Fly's Eye Cluster Processing Chain.]{Fly's Eye Cluster Processing Chain. The cluster head node is responsible for providing formatted data chunks to the worker nodes on the cluster. The worker nodes search a given chunk of data for bright pulses and write the results to a database for later analysis.}
	\label{fig:FlysEyeProcessing}
\end{figure}

The processing chain is in practice significantly more complicated than this description suggests. Processing is performed on compute clusters. Figure \ref{fig:FlysEyeProcessing} shows the major tasks in the processing chain. The head node's purpose is to format and divide the input data so that it can be assigned to worker nodes. In the worker node flow, the data is equalized, then RFI rejection is performed, and finally a pulse search is performed through the range of dispersion measures. The results are written to a database where they can be subsequently queried. The key feature of the results is a table that lists, in order of decreasing significance, the pulses that were found and the dispersion measures they were located at. \\

\subsection{Equalization}

Equalization is the process of ``equalizing'' or ``normalizing'' the average signal power in each frequency channel to some predetermined value. The power in channel $i$ at (discrete) time $t$ is $P_i(t) \in \left[0,255\right]$, and we compute an average power per channel $\overline{P_i} = \frac{1}{T_0} \sum_{t=0}^{T_0-1} P_i(t)$ over some time period $T_0$ ($T_0$ is typically set to the length of a data chunk, 10 minutes). A particular value $P_i(t)$ is then equalized by dividing (using floating-point arithmetic) by the average $\overline{P_i}$. i.e. the equalized value $P_i'(t) = P_i(t) / \overline{P_i}$. The average power in each channel is thus unity, since $\frac{1}{T_0} \sum_{t=0}^{T_0-1} P_i'(t) = 1$. \\

It is possible to equalize not only the frequency spectrum, but also the average power across all channels. We would like to do this so that gain changes in the system, which result in the average power increasing or decreasing over time (without regard for the source being pointed at), can be mitigated to a certain extent. Sharp gain changes (those that occur in short periods of time) are undesirable, and our equalization technique is only useful for smoothly-varying gain changes, such as those caused by gradual change in temperature. \\

Average power equalization is performed on the frequency spectrum equalized values $P_i'(t)$. We compute the average power over all frequency channels for a single integration (time sample $t$). The power average is defined as $\overline{P'(t)} = \frac{1}{N} \sum_{i=0}^{N-1} P_i'(t)$. $N$ is the number of channels (for Fly's Eye this is always 128). The motivation for why it is possible to normalize the power is that we expect pulses to be dispersed over many time samples, so this procedure should not remove extraterrestrial pulses. \\

With the average powers $\overline{P'(t)}$, we can define the equalization of the powers $P_i'(t)$. The average power equalized values $P_i''(t) = P_i'(t) / \overline{P'(t)}$. This procedure ensures that the average power $P_i''(t)$ is normalized to unity. \\

\subsection{RFI Rejection}

We categorize types of RFI that we can expect as follows:

\begin{enumerate}
  \item Constant (or near-constant) narrowband RFI
  \item Short-timespan wideband RFI
  \item Intermittent RFI (short-timespan and narrowband)
\end{enumerate}

The first two types of RFI can be fairly easily detected using simple methods. However, intermittent RFI (such as that which occurs when an aircraft with an active radar flies through a telescope beam) is difficult to automatically excise. \\

Our strategy for mitigating constant narrowband RFI is simply to identify the channels that are affected, and to exclude them from further processing. This channel rejection is typically performed manually by looking at a set of spectra and identifying obviously infected channels, which are then automatically excluded in subsequent processing runs.\\

Short-timespan wideband RFI is detected automatically. If the RFI timespan is significantly less than $T_0$ (the averaging time for power equalization) then we can easily reject RFI by comparing, for each time $t$, the sum $\sum_{i=0}^{N-1} P_i''(t)$ against some fixed threshold. We expect that this sum will on average be unity, so a value significantly higher than 1 is indicative of RFI. (It is important to set this threshold sufficiently high that a genuine dispersed pulse will not trigger it -- we found that for our data a threshold of 10 appeared to eliminate all short-timespan wideband RFI without accidentally removing bona fide astronomical data.) \\

Intermittent RFI is often quite difficult to automatically distinguish from genuine astronomical pulses, and we followed a conservative approach to try to ensure that we do not accidentally excise dispersed pulses. Our statistic for intermittent RFI is the variance of a single channel over a 10 minute data chunk, $\sigma_i^2 = \left( \frac{1}{T_0} \sum_{t=0}^{T_0-1} \left(P_i''(t)\right)^2 \right) - \left( \frac{1}{T_0} \sum_{t=0}^{T_0-1} \left(P_i''(t)\right) \right)^2$. We empirically found a range for $\sigma_i^2$ outside which it is likely that channel $i$ contains time-varying RFI. Future reprocessing will likely use a more robust method, such as that based on a kurtosis estimator \cite{Nita07}. \\

Our final RFI mitigation technique is manual -- in our results it is easy to see high-$\sigma$ hits that are a result of RFI: these hits appear as simultaneous detections at many dispersion measures. \\

\section{Test Results}

We carried out a series of tests to verify the correctness of the instrument. The results are described in this section. \\

\subsection{Detection of the Hydrogen Spectral Line}

\begin{figure}[htp]
	\centering
		\includegraphics{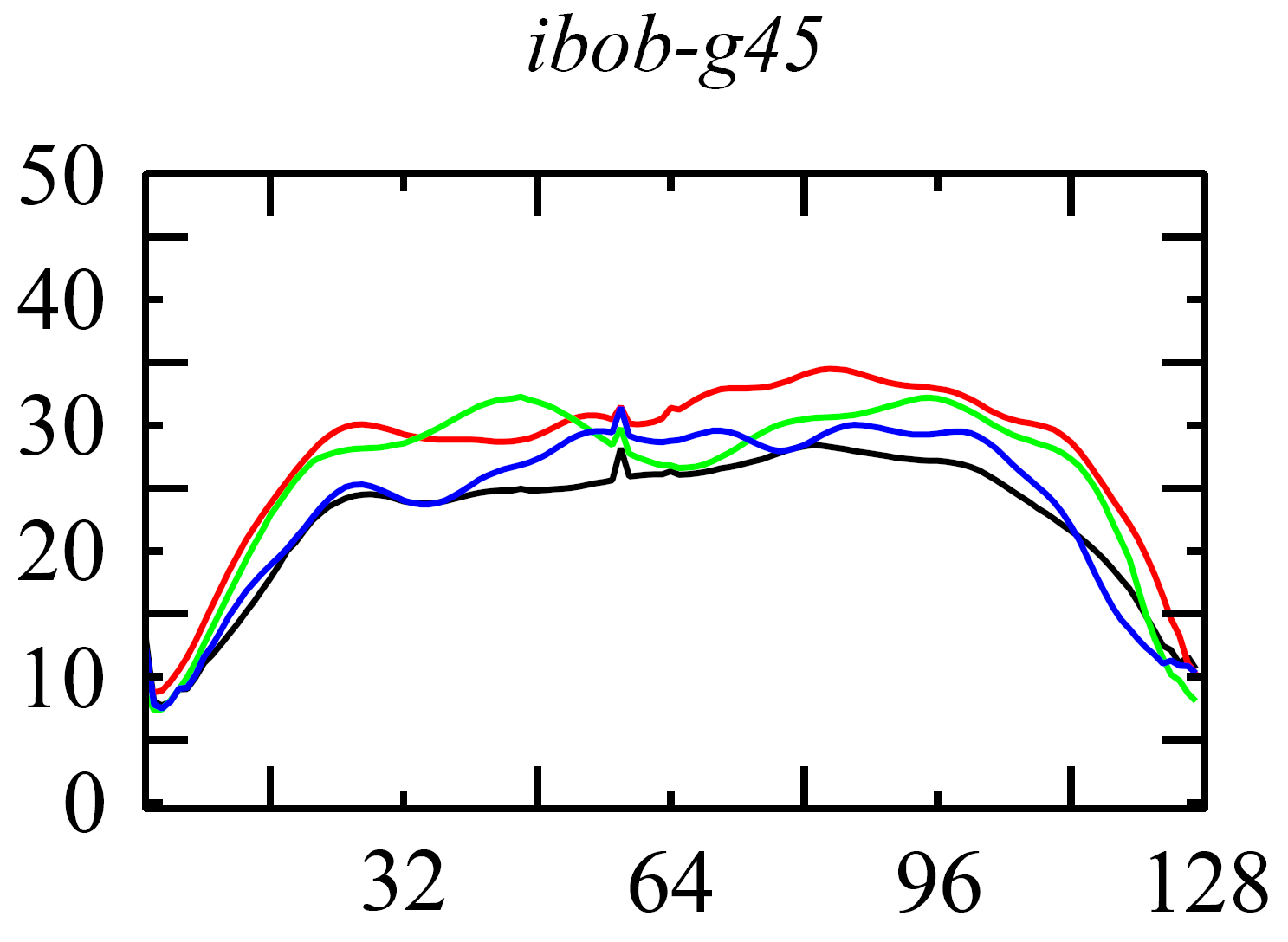}
		\caption[Fly's Eye Spectra from a Single IBOB.]{Fly's Eye Spectra from a Single IBOB. All four spectra from a single IBOB are shown, from a 1-minute test observation conducted on 18 February 2008. The vertical axis is uncalibrated power in arbitrary units, and the horizontal axis is frequency bin (0 - 127). The centre frequency is 1430MHz, and the `bump' clearly visible in all four spectra appearing several bins before bin 64 corresponds to the expected frequency for the Hydrogen spectral line, 1420.40575MHz.}
	\label{fig:FlysEyeHydrogenLine}
\end{figure}

The first test was to verify that the Fly's Eye spectrometers produce correct spectra on a long (multiple seconds) timescale. This is done by adding spectra together from data collected over many seconds, and then viewing the resulting ``integrated'' spectra to inspect them for relevant spectral features. In particular we expect to be able to see the Hydrogen 21cm spectral line when the dishes are pointed at locations that are known to contain much galactic gas. Figure \ref{fig:FlysEyeHydrogenLine} shows one-minute integrated spectra for a single IBOB, in which a bump at $\sim$1420MHz is clearly visible. Similar checks were carried out for all the IBOBs. \\

\subsection{Detection of Pulses from PSR B0329+54}

\begin{figure}[htp]
	\centering
		\includegraphics{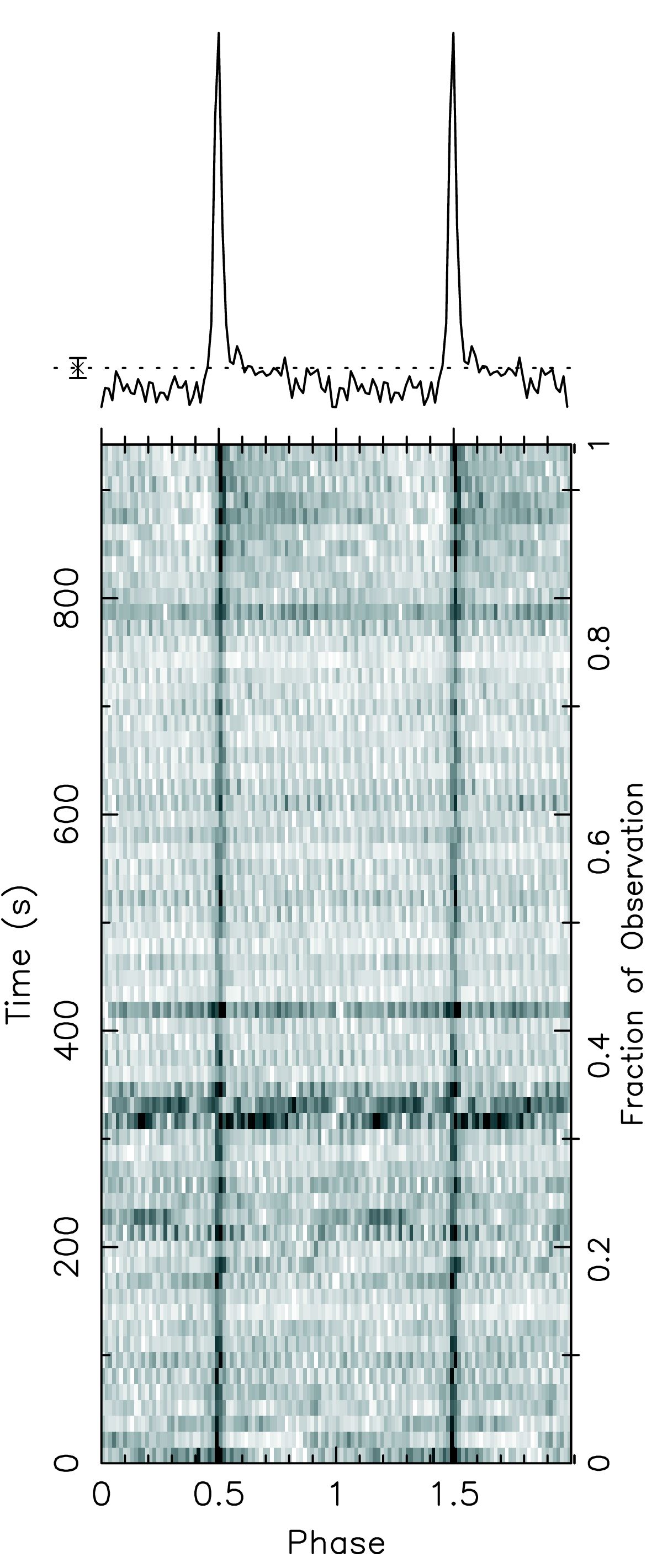}
		\caption[Pulse Profile of PSR B0329+54 obtained using the Fly's Eye.]{Pulse Profile of PSR B0329+54. Data taken from a 15-minute observation from 22 December 2007 were incoherently summed (44 analogue signals), and folded using the known period of the pulsar PSR B0329+54, P0 = 0.7145s. The repeated, folded pulses are visible as dark lines in the time versus phase plot (bottom). The pulse profile from folding several dedispersed pulses on top of each other is shown above. Image courtesy Joeri van Leeuwen.}
	\label{fig:FlysEye0329TimePhase}
\end{figure}

Once the long timescale Hydrogen spectral line test was successfully conducted, a set of tests were carried out to verify that the spectrometers can be used to observe a known pulsar. The first test was to observe a bright pulsar, PSR B0329+54, in the 44 spectra, incoherently summed to improve signal-to-noise. For this test, all the dishes were pointed at PSR B0329+54. In this context incoherent summation means that the power spectra from all 44 spectrometers were simply added together -- this yields a square-root improvement in sensitivity\footnote{In a beamformer the signals are typically added coherently -- that is, the voltages are summed after appropriate delays, not the powers. A beamformer yields a linear improvement in sensitivity, but incoherent summation is a quick, crude technique for improving sensitivity.}. The incoherently summed spectra were dedispersed using the known dispersion measure of PSR B0329+54. The data was then folded at the known pulse period for the pulsar. Figure \ref{fig:FlysEye0329TimePhase} shows the resulting pulse profile from a 15 minute observation. \\

\begin{figure}[htp]
	\centering
		\includegraphics[width=6.5in]{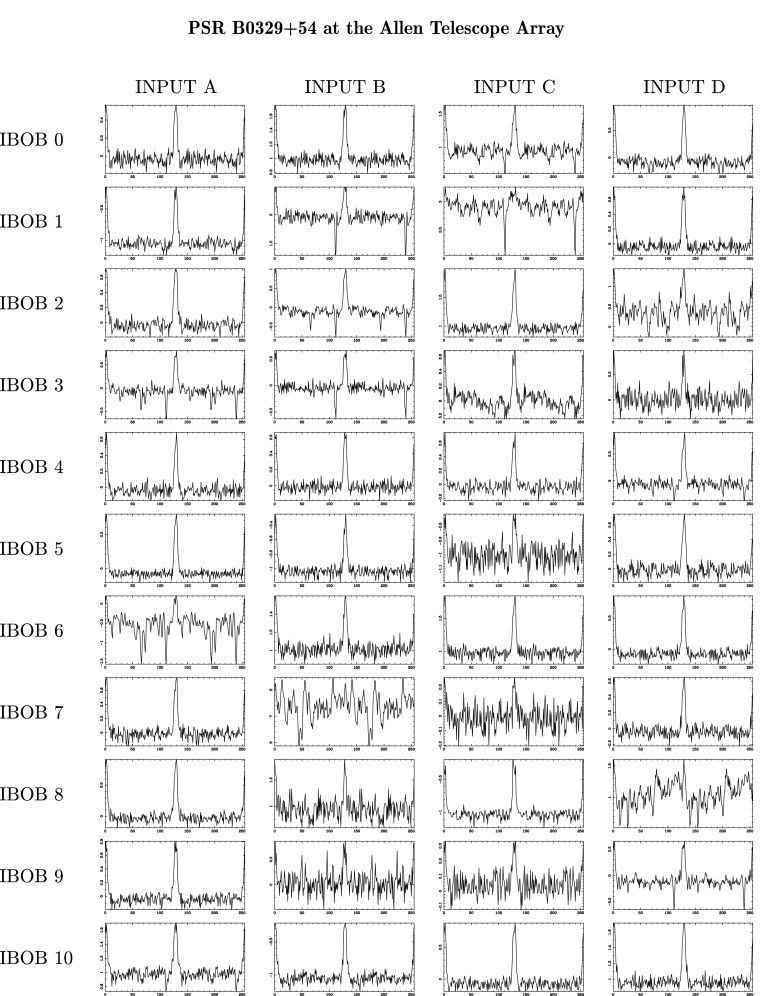}
		\caption[Pulse profiles calculated for all 44 independent spectrometers.]{Pulse profiles calculated for all 44 independent spectrometers. Image courtesy Joeri van Leeuwen.}
	\label{fig:FlysEye0329IndividualSpectrometers}
\end{figure}

After successfully detecting a pulsar in the incoherently summed data, the next test was to detect a known pulsar in individual dishes. The same 15 minute set of data for the summed test was used. However, instead of summing the spectra, each individual stream of spectral data was dedispersed and folded. The resulting pulse profile plots for all the spectrometers are shown in Figure \ref{fig:FlysEye0329IndividualSpectrometers}. A pulse profile is clearly visible in many spectrometers, but not in all of them. This turned out to be due to the fact that the antennas at the ATA have varying signal-to-noise characteristics -- the plots where the pulse profile is clearly visible are connected to antennas that have a better SNR measure than the antennas that yielded data in which the pulse profile cannot be seen. \\

\subsection{Detection of Giant Pulses from the Crab Nebula}

The detection of a pulsar by reconstructing its pulse profile is an excellent test to verify that the short timescale behaviour of the spectrometers is sufficiently accurate to detect dispersed pulses with very short duration. The final test to verify that the Fly's Eye instrument is capable of detecting the transients it was designed to find is to use the instrument to detect an individual pulse from a pulsar. An individual pulse from a pulsar is, to the instrument, indistinguishable from a transient signal, so this test can provide great confidence that the instrument works as required. \\

A fundamental problem exists with attempting to detect a single pulse from PSR B0329+54: despite the fact that this pulsar has the highest average flux of any known pulsar, it is still too weak for individual pulses to be seen in individual ATA antennas, or even in incoherently summed signals from all the dishes. Fortunately Nature provides a solution: the pulsar in the Crab nebula (the ``Crab pulsar'') has a lower average flux than PSR B0329+54, but its pulses are occasionally individually much brighter than the individual pulses from PSR B0329+54. These so-called ``giant pulses'' can be up to 1000 times brighter than the normal pulses \cite{LK05}. Therefore a suitable test of transient detection capability is to observe the Crab pulsar and attempt to detect giant pulses from it. \\

\begin{figure}[htp]
	\centering
		\includegraphics{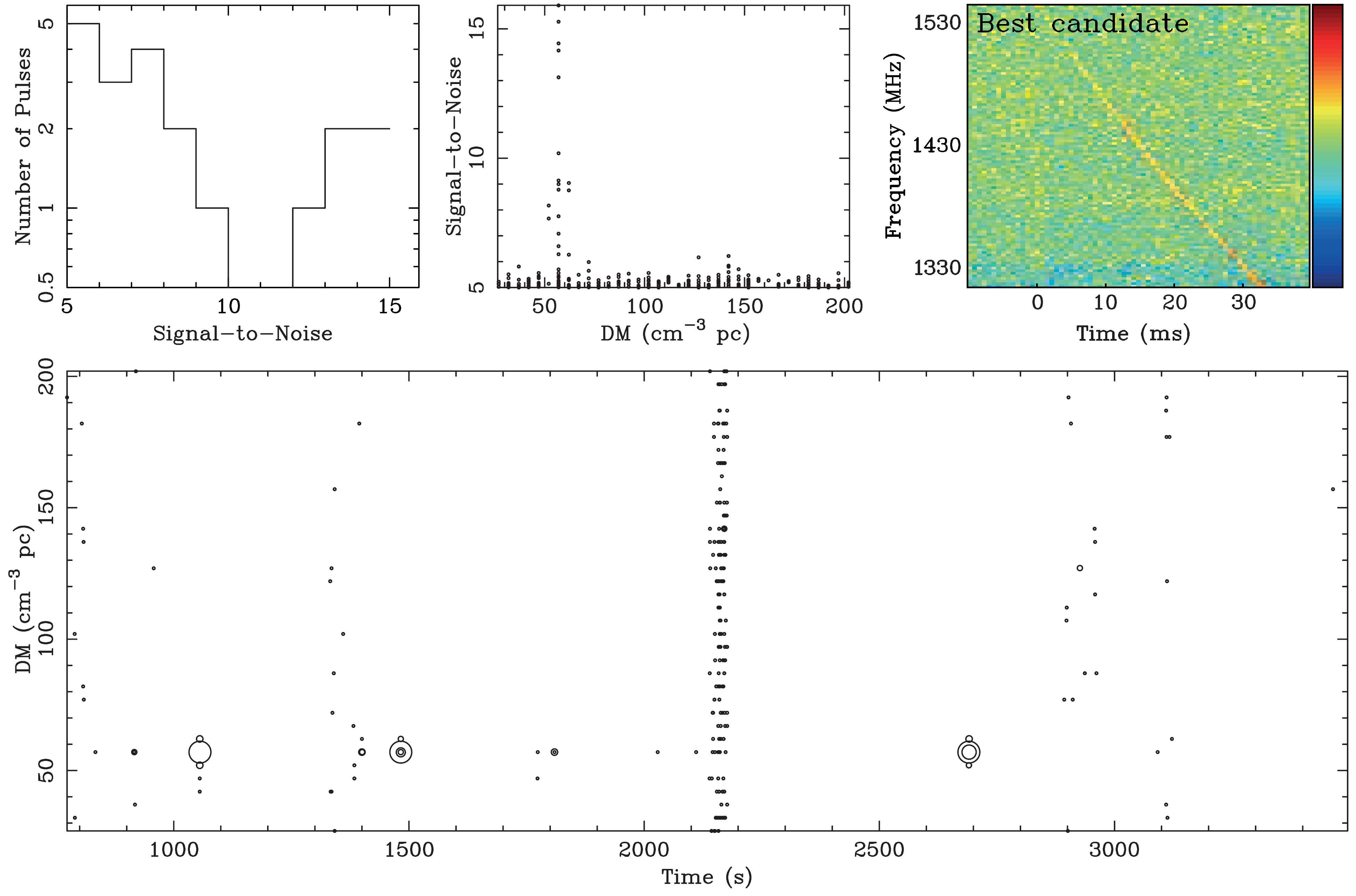}
		\caption[Crab Pulsar Observation Diagnostics.]{Diagnostics on data taken from the Crab pulsar in a 60-minute observation conducted on 22 December 2007. The data was dedispersed using dispersion measures ranging from 0 to 200 cm$^{-3}$ pc. Top-left: single-pulse SNR histogram. Top-centre: noise appears at all DMs, but bright pulses (SNR$>7$) from the Crab correctly appear at DM $\approx 57$ cm$^{-3}$ pc. Top-right: inset of Figure \ref{fig:FlysEyeCrabGiantPulse}. Bottom: pulse detections plotted on the DM versus time plane. Higher SNR detections appear as larger circles. Three giant pulses from the Crab are clearly visible in this plot. Image courtesy Joeri van Leeuwen, Griffin Foster and Andrew Siemion, from \cite{BowerFlysEyeProposal08}.}
	\label{fig:FlysEyeCrabGiantPulseDiagnostics}
\end{figure}

Figure \ref{fig:FlysEyeCrabGiantPulseDiagnostics} shows a diagnostic plot generated from a one-hour Crab observation. The data is an incoherently summed set from the 35 best inputs (the 9 inputs in which PSR B0329+54 could not be detected were discarded). The diagnostic plot was generated after the raw data had been dedispersed using a range of dispersion measures from 5 to 200. The Crab pulsar has dispersion measure $\approx 57$ cm$^{-3}$ pc, so three giant pulses from the Crab pulsar can be easily identified in the lower plot. The giant pulses appear only at the expected dispersion measure, whereas wideband RFI appears across a wide range of dispersion measures. \\

\begin{figure}[htp]
	\centering
		\includegraphics{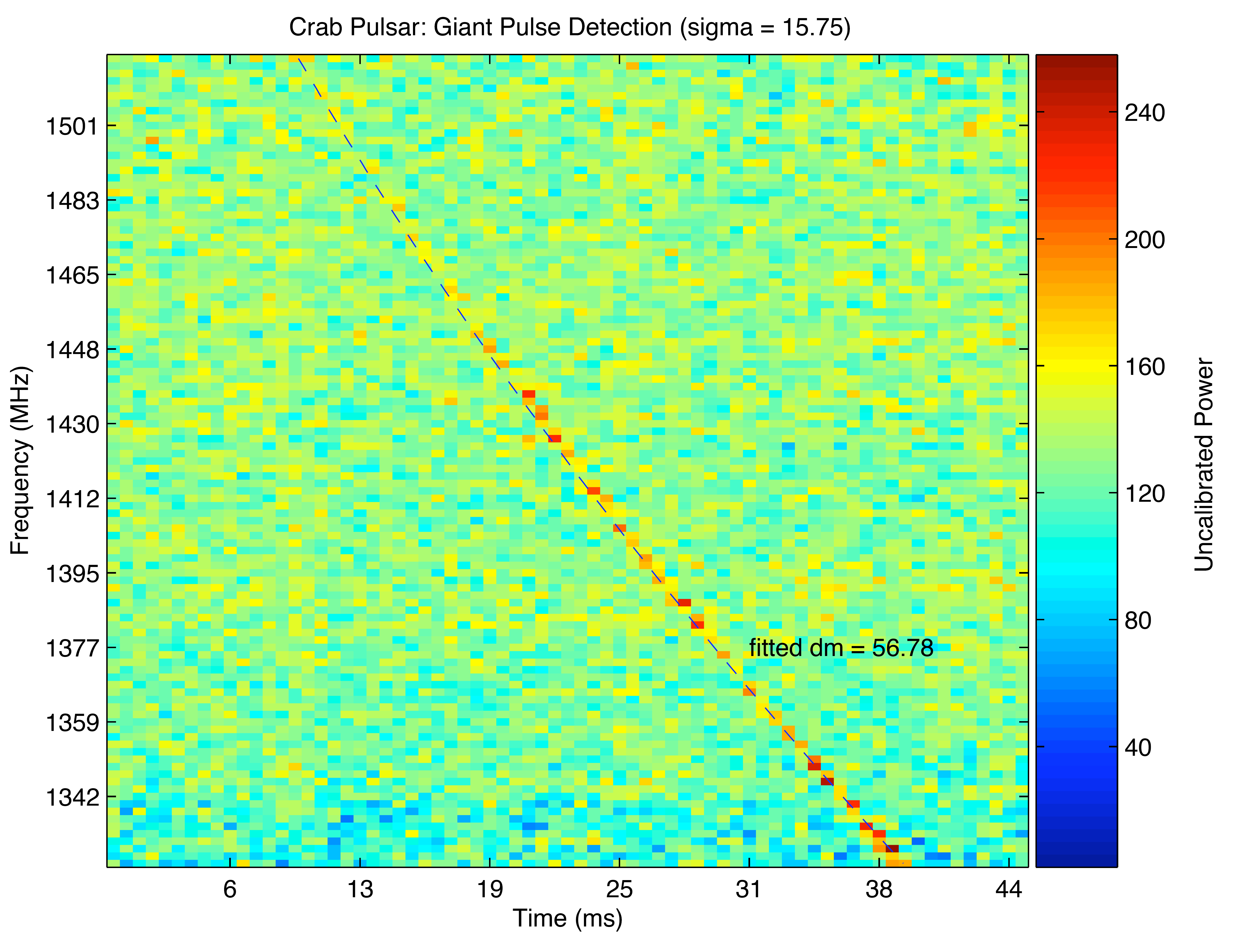}
		\caption[A Giant Pulse from the Pulsar in the Crab Nebula.]{A giant pulse from the pulsar in the Crab nebula. This pulse was detected in a 60-minute dataset taken on 22 December 2007. The dispersion of the pulse, correctly corresponding to DM = 56.78 cm$^{-3}$ pc, is clearly visible. Image courtesy Griffin Foster and Andrew Siemion.}
	\label{fig:FlysEyeCrabGiantPulse}
\end{figure}

A frequency vs. time plot of the raw (summed) data at the time when the brightest giant pulses was detected is shown in Figure \ref{fig:FlysEyeCrabGiantPulse}. The pulse is clearly visible in the data, and a fit to the dispersion measure shows that the pulse is, nearly without doubt, from the Crab (as opposed to RFI). \\

\newpage

\chapter{Fast-readout 10GbE Spectrometers for Incoherent Dedispersion Applications}

The spectrometers that we built for the Fly's Eye experiment, whose design was discussed in the previous chapter, had a data output rate limitation of approximately 7Mbps. This limitation was imposed by the PowerPC 100MbE connection on the IBOB; data is transferred from block RAMs in the FPGA via a bus to the embedded PowerPC, and it is this bus that is the bottleneck. If the output data rate is less than the input data rate (which for a dual polarization power/autocorrelation spectrometer is typically $2\times2\times B \times b$ bits per second, where $B$ is the signal bandwidth and $b$ is the number of bits sampled), then the spectra have to be accumulated to reduce the data rate, and this reduces the {\it time resolution} of the spectrometer. The Fly's Eye spectrometers had a time resolution of 625$\mu$s (calculated as the reciprocal of number of spectra outputted per second). If one wishes to observe pulses that have a duration less than 625$\mu$s, then it is advantageous for the spectrometer to have a smaller (better) time resolution. Pulses from pulsars typically have a power that is small compared to the noise power in the received signal, so the resulting reduction in power that occurs when the spectrometer time resolution is greater than the pulse width can result in the pulsar being harder (or even impossible) to detect. It is primarily for this reason that we want to build spectrometers that have high data output rates. \\

In this chapter we describe our efforts to build several IBOB-based spectrometers that all use a 10GbE connection to output data. This connection bypasses the slow bus that restricts the data rate on the 100MbE connection. We built a dual polarization, real sampling, 400MHz bandwidth power spectrometer for the Parkes Radio Telescope. We modified this design for the Hartebeesthoek Radio Astronomy Observatory, which required a ``Full Stokes'' spectrometer. We also produced a ``Full Stokes'' spectrometer for the Allen Telescope Array that could process 105MHz bandwidth (the currently available bandwidth at ATA) and used a XAUI (digital) connection from the ATA beamformer as its input, as opposed to digitizing analogue inputs.

\section{The Parkes and HartRAO Pulsar Spectrometers}

The Parkes and HartRAO spectrometer share a common architecture and set of features. They were both designed to sample two real input signals at 800MSa/sec, resulting in a 400MHz bandwidth being processed. Both designs use a single 10GbE connection for data output. Both spectrometers use the 100MbE connection for control and testing. The primary differences between the spectrometers are the number of channels (Parkes: 1024 channels, and HartRAO: 512 channels) and the products computed (Parkes: powers only, and HartRAO: powers and cross-terms). The Parkes spectrometer was designed for a new Parkes multibeam pulsar survey, which is an experiment to search for new pulsars. Hence only powers are required. The HartRAO spectrometer, on the other hand, will be used for polarimetry (polarization studies), and this necessitates the output of the cross-correlations between the two polarizations.

\subsection{A 400MHz Fast-readout Dual Power Spectrometer for Parkes}

\begin{figure}[htp]
	\centering
		\includegraphics[width=8.5in,angle=90]{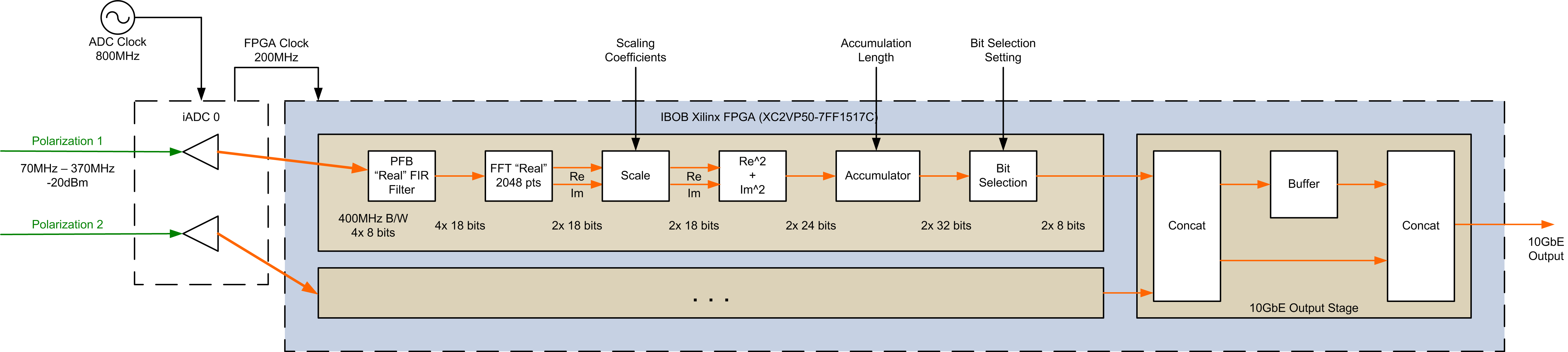}
		\caption{{\it Parspec}: a Dual Power Spectrometer for the Parkes Radio Telescope.}
	\label{fig:ParkesDetailedIBOB}
\end{figure}

Figure \ref{fig:ParkesDetailedIBOB} shows the design of the Parkes spectrometer. The two input polarizations are digitized at 800MSa/sec. The FPGA is clocked off the ADC clock source, divided by 4 on the iADC board; hence the FPGA is clocked at 200MHz. Every FPGA clock cycle, four (8 bit) samples from each polarization are received by the FPGA. Two streaming parallel polyphase filterbank FIR filters and FFTs are used to channelize the data. A 2048 point ``real'' PFB is used; this implementation has optimizations that take advantage of the fact that the imaginary part of the input is known to be zero. Because the output of the FFT on real data is symmetric, the second half of the spectrum is not outputted, hence only 1024 bins are produced. \\

The FFT output is scaled by a user-specified 18-bit coefficient. Each polarization has a separate coefficient. A ``power detector'' then computes the power of this scaled output, and the powers are summed in a vector accumulator. The length of the accumulator is a user-specified parameter. The accumulator has 32 bits of precision, so the succeeding bit selection outputs 8 bits from 32; i.e. there are 4 possible bit selection options -- bits 0 -- 7, bits 8 -- 15, bits 16 -- 23 and bits 24 -- 31. \\

When the accumulator is outputting an accumulation, every clock cycle two sets of 8 bits are produced by each polarization processing chain, and therefore 32 bits are available to be outputted each clock cycle. The 10GbE interface has a data width of 64 bits, so the concatenated data is buffered for one clock cycle every alternate clock cycle; this technique is used to produce a set of 64 bits every second clock cycle during output. \\

\begin{figure}[htp]
	\centering
		\includegraphics[width=4in]{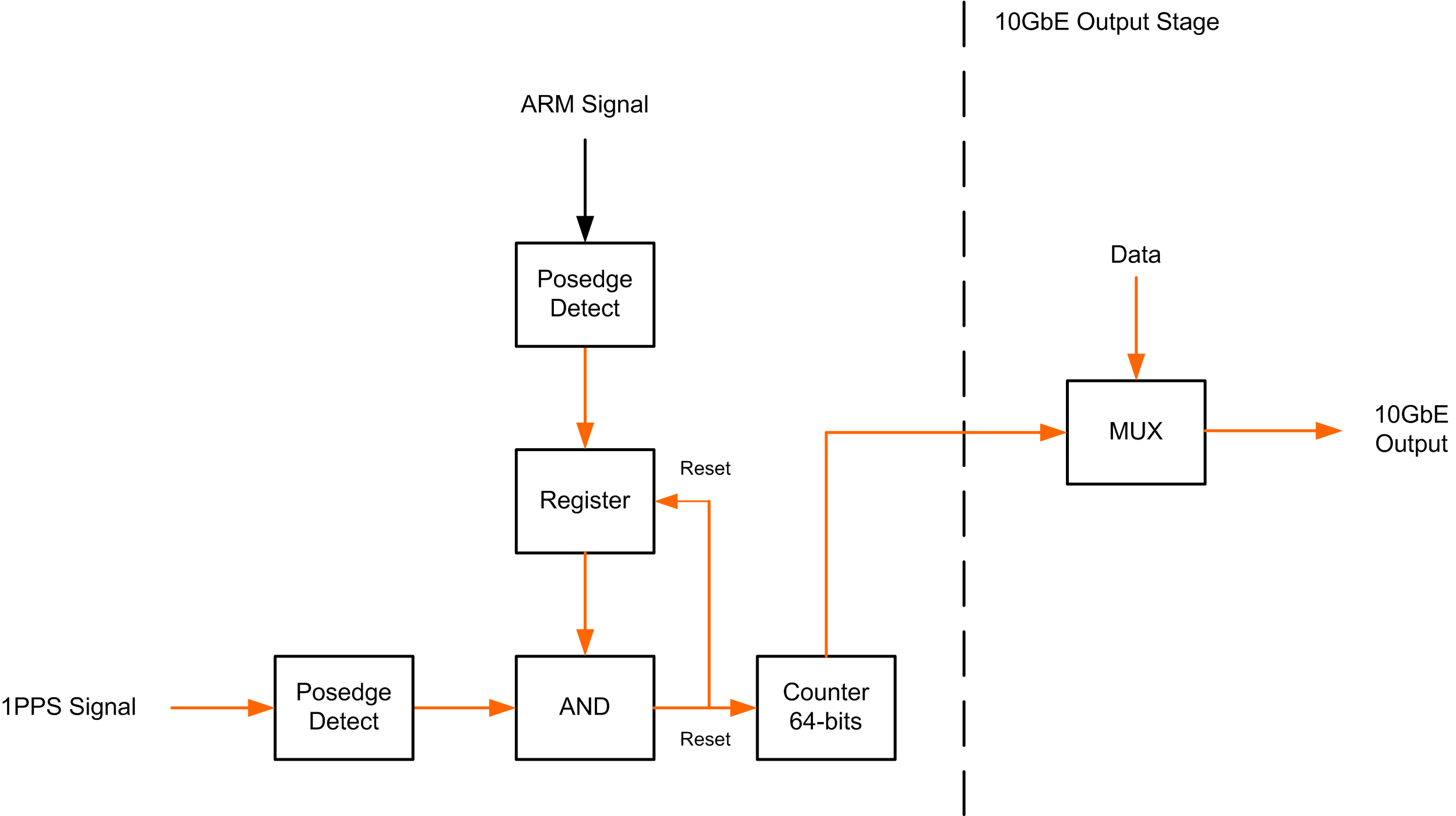}
		\caption[Time Stamping Logic.]{Time Stamping Logic. An ARM signal sent by the user prepares the spectrometer to reset a master 64-bit counter when the next 1PPS pulse arrives. This counter is inserted as the first 64 bits of every packet that is outputted.}
	\label{fig:TimeStamping}
\end{figure}

The output stage has a feature that is not shown in the high-level diagram Figure \ref{fig:ParkesDetailedIBOB}. In pulsar studies it is often necessary to know precisely the time of arrival of pulses. It is also important to know if an accumulation is lost by the inherently unreliable UDP transport used between the IBOB and data recorder computer. Both these requirements are satisfied by the inclusion of a time-stamping mechanism in the design, which is detailed in Figure \ref{fig:TimeStamping}. \\

The 10GbE output stage includes a multiplexer that is used to insert the value from a 64-bit counter at the start of each packet. This counter runs continuously (that is to say, it is incremented every clock cycle), and is only reset to zero in the following circumstance. The user asserts an ARM signal using the 100MbE control link, and this state is stored in a boolean register. When the next 1PPS pulse arrives at the FPGA, it is used to reset to the master counter if and only if the ARM state register has value 1. This reset signal is used to reset the state register to zero. It is also used to reset the sync pulse generator. The counter will not be reset again unless the ARM signal is deasserted and then reasserted, due to the presence of a ``positive edge detection'' function. \\

This scheme can be used for accurate time stamping in the following manner. The control computer should be set to assert the ARM signal on a half-second. If the computer's time is set using NTP, it should be accurate to within several tens of milliseconds. The network and processing latency to enact the assertion should be less than 100ms. The 1PPS signal can be known to an accuracy of micro- or nano-seconds. The control computer can assume that the counter will be reset on the next second (since this is by definition when the 1PPS pulse will occur), hence it is possible to timestamp the data accurately. In practice this requires very careful calibration to determine the latency in the system. One technique is to use an astronomical calibrator -- point at a known source, such as a pulsar, and determine the difference between the expected and measured times of arrival. An alternative, but more difficult, technique is to independently characterize the latency in each part of the entire system, including the analogue frontend, all the electrical and fiber optic cables, the digitizers and the digital processing backend. \\

We implemented the ARM/1PPS scheme in our design, but as at time of writing the Parkes system has not been calibrated to produce an accuracy better than that of NTP. Likewise the HartRAO and BAPP systems have also not yet been calibrated, but the necessary spectrometer features to support accurate time stamping are available to the users.

\subsection{A 400MHz Fast-readout Dual ``Full Stokes'' Spectrometer for HartRAO}

\begin{figure}[htp]
	\centering
		\includegraphics[width=8.5in,angle=90]{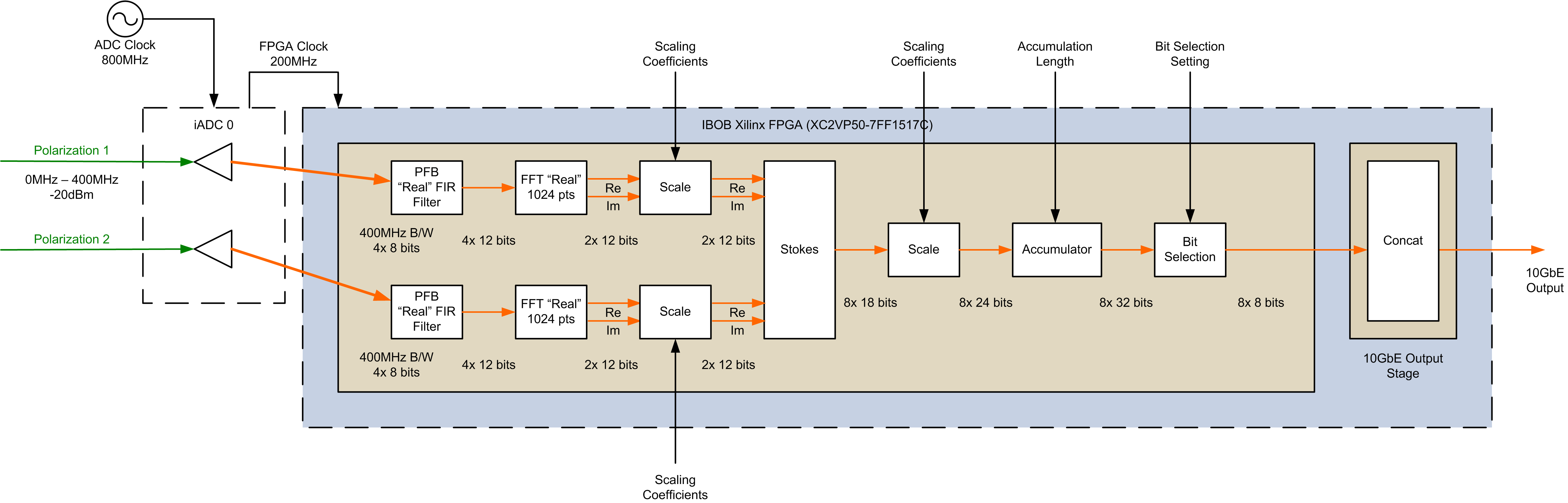}
		\caption{A Dual ``Full Stokes'' Spectrometer for the Hartebeesthoek Radio Astronomy Observatory.}
	\label{fig:HartRAODetailedIBOB}
\end{figure}

Figure \ref{fig:HartRAODetailedIBOB} shows the design of the HartRAO spectrometer. This design is based on that of the Parkes spectrometer. It differs primarily in that it computes both powers and cross-terms. This extra computation necessitated the reduction of the PFB length from 1024 channels to 512 channels, and of the FFT precision from 18 bits to 12 bits. The scaling from the Parkes design, which allows the two polarizations to be independently scaled directly following the FFT, is retained. An additional scaling is performed after the power and cross-term calculations, since the power and cross-term values are expected to differ significantly. Because each set of FFT bins is used to produce four values (the two powers, and two cross-terms), the vector accumulator produces a total of eight 32 bit values per clock cycle while an accumulation is being outputted. After bit selection, eight 8 bit values are available per clock cycle, and hence the concatenated output can be connected directly to the 64-bit 10GbE controller, without the need for the buffering scheme in the Parkes design.

\subsection{Test Results}

\begin{figure}[htp]
	\centering
		\includegraphics{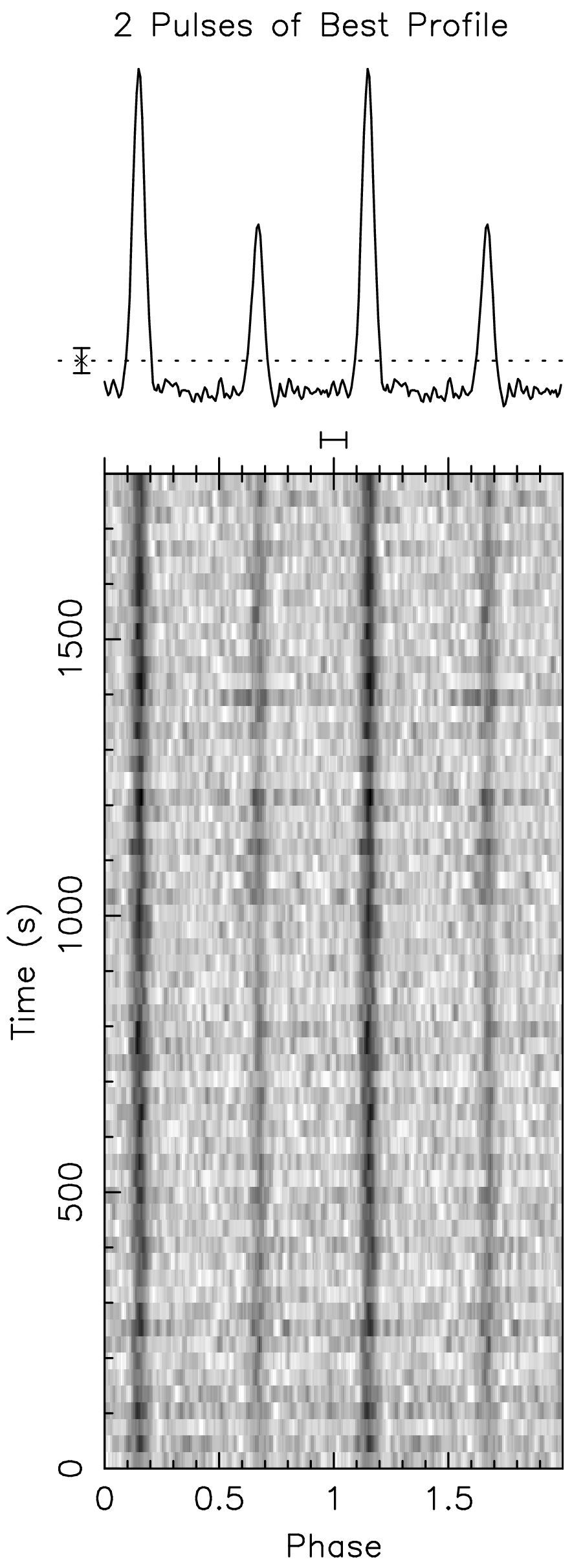}
		\caption[Pulse profile of PSR B1937+21 obtained using the Parkes Spectrometer at NRAO Green Bank.]{Pulse profile of PSR B1937+21 obtained using the Parkes Spectrometer with the NRAO Green Bank 140ft telescope. Image courtesy Glen Langston, Paul Demorest and Scott Ransom (NRAO).}
	\label{fig:Parspec1937}
\end{figure}

The Parkes power spectrometer design was tested both at the NRAO Green Bank 140ft telescope, and at the Parkes Radio Telescope. Figure \ref{fig:Parspec1937} shows the pulse profile for PSR B1937+21 obtained after a 30 minute observation at NRAO by Glen Langston and co-workers. PSR B1937+21 is a fast millisecond pulsar, and has a period of approximately 1.56ms. It is therefore a demanding test of the spectrometer's capabilities. The obtained pulse profile correctly shows the double peak characteristic of this pulsar. \\

\begin{figure}[htp]
	\centering
		\includegraphics[width=4.5in]{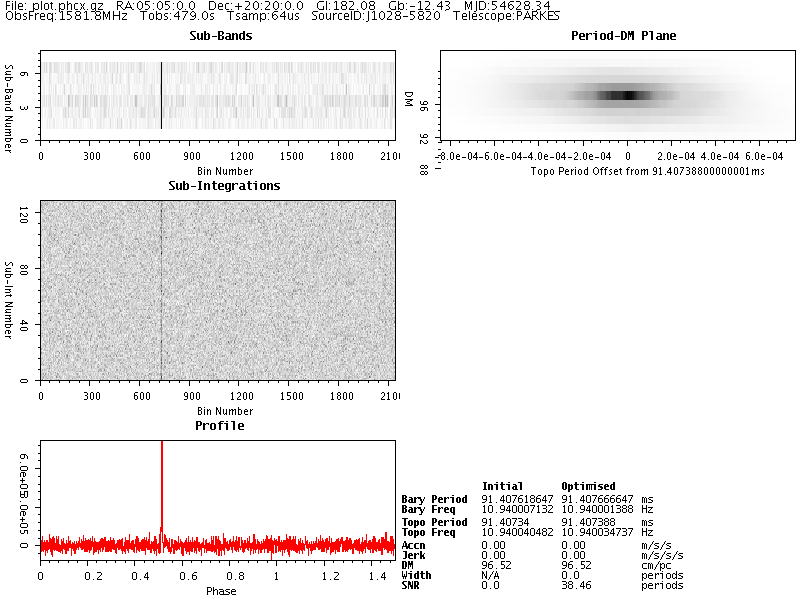}
		\caption[Pulse profile and diagnostics of PSR J1028-5820 obtained using the Parkes Spectrometer with the Parkes Radio Telescope.]{Pulse profile and diagnostics of PSR J1028-5820 obtained using the Parkes Spectrometer with the Parkes Radio Telescope. Image courtesy Mike Keith, Andrew Jameson and Willem van Straten.}
	\label{fig:ParspecJ1028atParkes}
\end{figure}

Figure \ref{fig:ParspecJ1028atParkes} shows the pulse profile (and associated diagnostic plots) obtained from an observation of PSR J1028-5820 conducted using the Parkes Radio Telescope. The produced pulse profile shows excellent time resolution. \\

\begin{figure}[htp]
	\centering
		\includegraphics[width=4in]{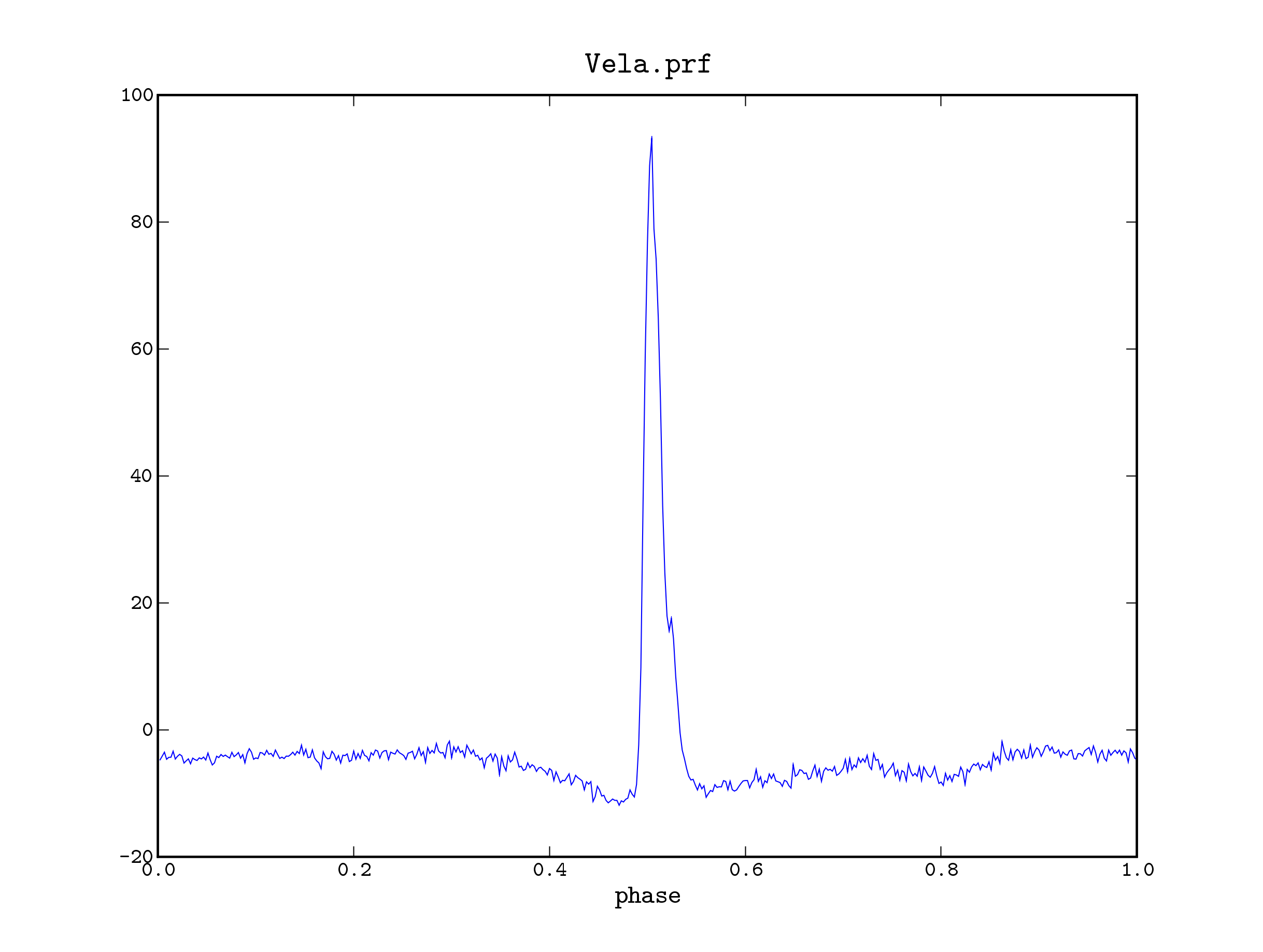}
		\caption[Pulse profile of PSR B0833-45 (Vela) obtained using the HartRAO Spectrometer.]{Pulse profile of PSR B0833-45 (Vela) obtained using the HartRAO Spectrometer with the 26m HartRAO dish. Image courtesy Sarah Buchner (HartRAO).}
	\label{fig:HartRAOVela18cm20s}
\end{figure}

Figure \ref{fig:HartRAOVela18cm20s} shows the pulse profile obtained from an observation of B0833-45 (Vela) conducted using the HartRAO telescope. The observation was performed at a centre frequency of 1668MHz over 20 seconds. The pulse profile produced is very closely matched to the expected profile. \\

\section{The Berkeley ATA Pulsar Processor}

The Berkeley ATA Pulsar Processor (BAPP) is a system consisting of the ATA beamformer, a purpose-built dual spectrometer, and data recording and processing computers. BAPP was built to provide pulsar observing capabilities at the ATA. In our initial deployment, made in March 2008, a bandwidth of 105MHz at a single sky frequency was supported. However, it is intended that when a second dual polarization beamformer is installed, the BAPP spectrometer will be replicated so that concurrent observations of pulsars at different sky frequencies will be possible. This is a possibility due to the ATA's analogue systems, which provide four IF band selection chains per polarization per dish.\\

\subsection{Overall Architecture}

\begin{figure}[htp]
	\centering
		\includegraphics[width=6in]{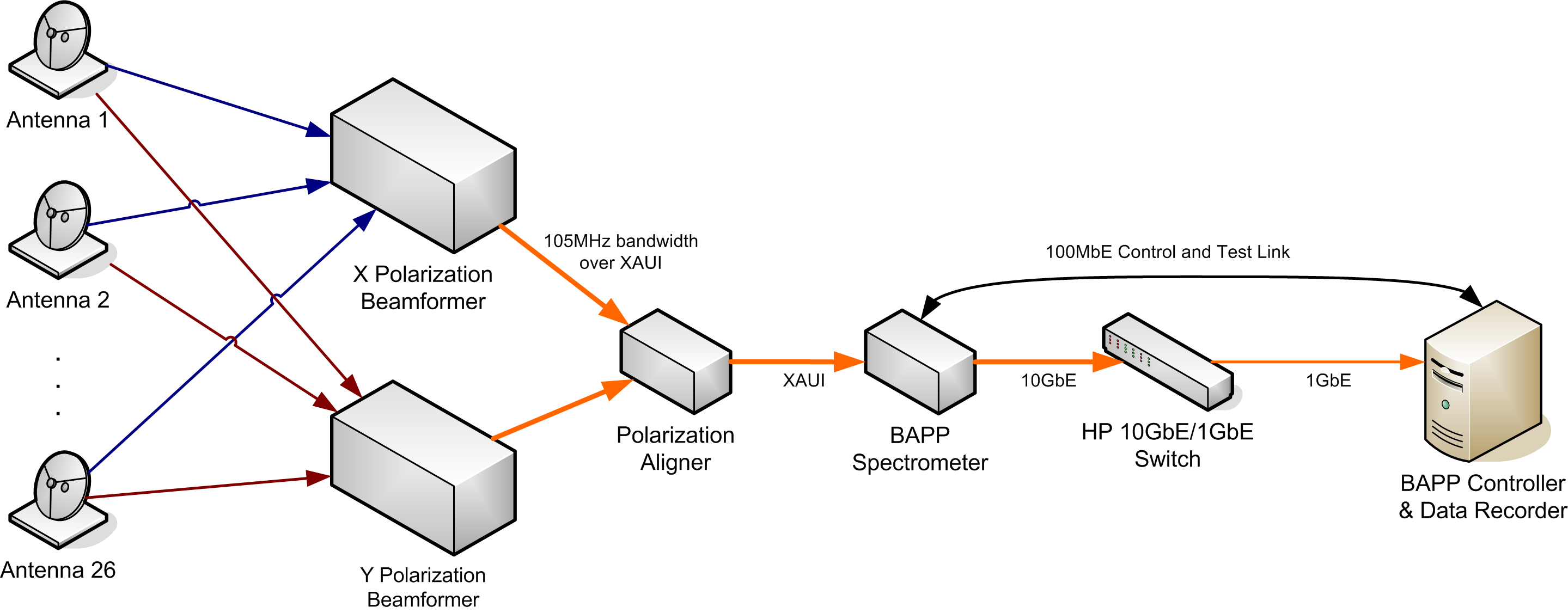}
		\caption[BAPP Architecture.]{BAPP Architecture. The beams from the single polarization ATA beamformers are aligned and sent to the BAPP spectrometer. The spectrometer output is over a 10GbE link to an HP 10GbE/1GbE switch, which in turns sends the data to a control and data recorder computer over a 1GbE link. The control computer configures the spectrometer over a separate 100MbE link.}
	\label{fig:BAPPArchitecture}
\end{figure}

Figure \ref{fig:BAPPArchitecture} shows the overall architecture for the Berkeley ATA Pulsar Processor system, as it was deployed. Two single polarization 26-antenna beamformers send their output to a ``polarization aligner'', which aligns the two different polarization beams and outputs the combined dual polarization beam over a XAUI connection. Recall from Chapter 2 that a beamformer delays the signals from the 26 antenna inputs appropriately so that an effective narrow beam on the sky is formed. The polarization aligner is effectively a final stage in what can be considered a dual polarization beamformer -- it delays the one polarization beam relative to the other polarization beam so that they are aligned in time. This dual polarization signal is sent to the BAPP spectrometer, which is the BAPP subsystem that we are primarily concerned with in this thesis. (For more details on the ATA beamformer systems, see reference \cite{ATABeamformer}.) The data output from the spectrometer is over a 10GbE connection; this is converted to a 1GbE connection by an HP 10GbE/1GbE switch (an HP ProCurve 2900-24G), which is connected to a data recorder computer. The motivation for this wasteful use of a 10GbE connection to carry a data bandwidth of less than 1Gbps is that a data rate of greater than 7Mbps is required to get suitable time resolution, but the IBOB has no Ethernet interfaces other than its 100MbE port and its 10GbE ports. Since the 100MbE port is not capable of outputting data faster than 7Mbps, a 10GbE connection is required. The use of a switch to convert the connection from 10GbE to 1GbE is also not necessary, since the data recorder computer could be designed to use a 10GbE NIC. However, this approach was taken to allow the future creation of ``load balancing'' systems, where the signal is divided amongst several data recorder or processing computers by the switch. An example of such a system is described in the following chapter.\\

\begin{figure}[htp]
	\centering
		\includegraphics[width=8in,angle=90]{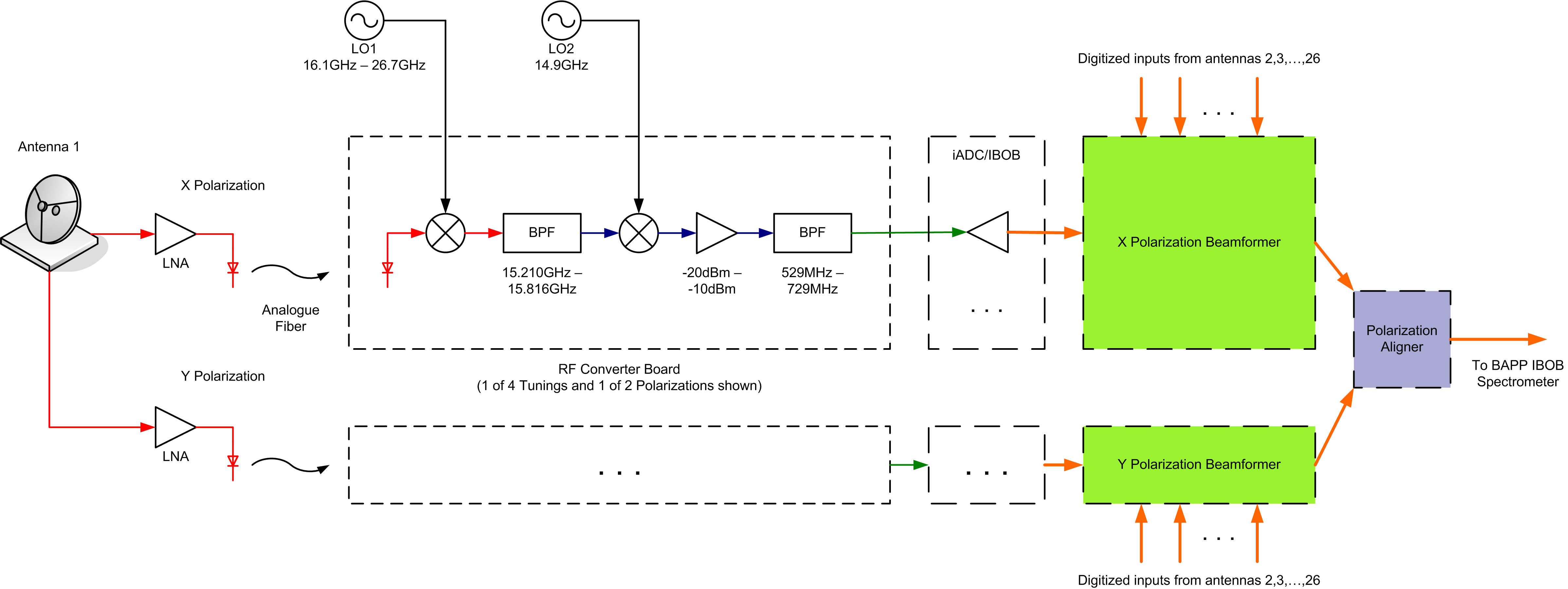}
		\caption[Signal Flow for BAPP at the ATA.]{Signal Flow for BAPP at the ATA. Two separate beamformers are used to each create a single beam for the two polarizations from 26 antennas. The polarization beams are then aligned, and the combined dual polarization beam is sent to the BAPP spectrometer.}
	\label{fig:BAPPATASignals}
\end{figure}

Figure \ref{fig:BAPPATASignals} shows the signal flow at ATA from the antennas through to the spectrometer. The analogue subsystem provides 210MHz bandwidth to the beamformers, but a bandwidth of 105MHz is extracted from this. The analogue subsystem is identical to that described in the previous chapter for the Fly's Eye experiment. (As noted in the signal flow diagrams for Fly's Eye and for BAPP, the ATA has four ``tunings'' per polarization per antenna; one is used for the correlator, one is used for the beamformers, one is used for Fly's Eye, and one is unused.)\\

\begin{figure}[htp]
	\centering
		\includegraphics[width=3in]{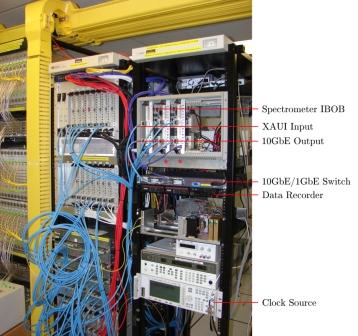}
		\caption[BAPP Installation.]{BAPP Installation. The rack in which BAPP is installed at the Allen Telescope Array is shown. The leftmost (partially) visible rack contains the RF conversion subsystem. The middle rack contains the beamformer (with digitization at the top, using IBOBs, and processing at the bottom, using BEE2s). The rightmost rack contains the BAPP hardware, in addition to several other unrelated pieces of equipment. (Figure compressed to meet arXiv file size requirements.)}
	\label{fig:BAPPRack}
\end{figure}

Figure \ref{fig:BAPPRack} shows the installation of BAPP at the ATA. The combined beamformer output XAUI connection provided to the BAPP spectrometer is visible (although where it connects to at the beamformer end is not -- the beamformer output is from the back of one of the BEE2 boards). The other CX4 port on the BAPP spectrometer IBOB is attached to the HP ProCurve 2900 10GbE/1GbE switch. The 100MbE control and test connection cable to the IBOB (purple) is also visible. This is connected to the Data Recorder computer via the switch, although this is not visible, since the switch is installed with the front panel facing backwards\footnote{This was done to reduce the length of the cable for the 10GbE connection from the IBOB to the switch, since the 10GbE ports on the switch are at the back.}.

\subsection{A 105MHz Fast-readout Dual Spectrometer with Digital Beamformer Interface}

\begin{figure}[htp]
	\centering
		\includegraphics[width=8.5in,angle=90]{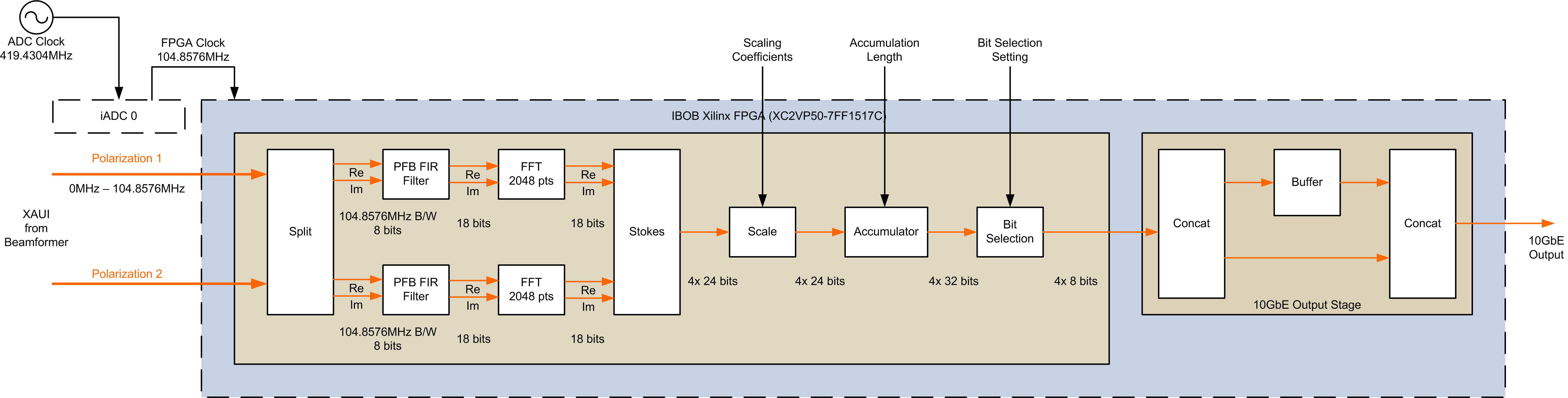}
		\caption[{\it The Berkeley ATA Pulsar Processor} Spectrometer: a 105MHz Fast-readout Dual Spectrometer.]{{\it The Berkeley ATA Pulsar Processor} Spectrometer: a 105MHz Fast-readout Dual Spectrometer with Digital Beamformer Interface}
	\label{fig:BAPPDetailedIBOB}
\end{figure}

The spectrometer design for BAPP is based on the Parkes and HartRAO designs. Because the bandwidth requirement for BAPP was only 105MHz, versus 400MHz for the Parkes and HartRAO spectrometers, it was possible to fit a Full Stokes 2048-channel design onto the IBOB FPGA. Instead of processing four (real) parallel data streams, the BAPP design runs at 105MHz and processes only one (complex) data stream per polarization. As discussed, the input is a dual polarization beam that is streamed over a XAUI connection from the ATA beamformer system. Figure \ref{fig:BAPPDetailedIBOB} shows the design of the BAPP spectrometer. An implementation detail that appears in this diagram is that an iADC board is still used, even though the input is digital -- it was necessary to use an iADC to clock the IBOB externally, with a clock frequency identical to that of the beamformer. The CASPER DSP blocks are designed to be continuously streaming, so the received XAUI data in the spectrometer had to produce one sample of data every FPGA clock cycle, and the only way to ensure this is to have the receiver be clocked at an identical\footnote{As is the standard practice at observatories, the clock sources at the ATA are synchronized via a distributed 10MHz signal, so all the clock sources are in phase. Therefore the sending BEE2 and the receiving IBOB in the BAPP system are clocked using sources that are in phase.} rate to the sender.

\subsection{Test Results}

The BAPP system underwent a set of tests to verify its functionality. The first test was an elementary one designed to verify that the data from the beamformer was being received correctly: the beamformer was configured to output a tone at a specific frequency, and the BAPP power spectra outputs were plotted to check that peaks at the appropriate locations in the spectra were visible. The tone was then moved and scaled, and the BAPP output was again inspected to make sure that the peaks had moved and been scaled as expected. With the basic functionality of the system verified, we proceeded with two astronomical tests.

\subsubsection{Detection of the Hydrogen Spectral Line}

The first astronomical test was designed as the simplest possible end-to-end verification of the BAPP system. The beamformer was calibrated, and then the array was pointed at a source with a significant Hydrogen content. The power spectra outputs were integrated for a one-minute observation, and we verified the presence of the Hydrogen spectral line in the spectra.

\subsubsection{Detection of Pulses from PSR B0329+54}

\begin{figure}[htp]
	\centering
		\includegraphics[width=3in]{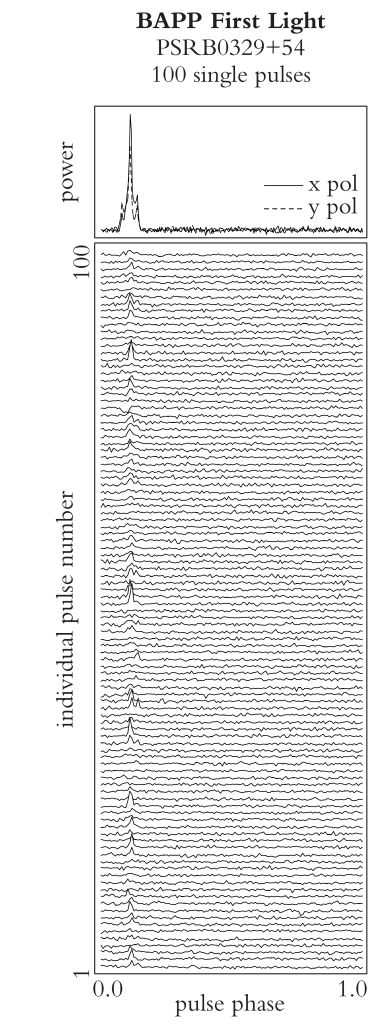}
		\caption{Detection of Individual Pulses from PSR B0329+54. Image courtesy Joeri van Leeuwen. (Figure compressed to meet arXiv file size requirements.)}
	\label{fig:BAPPFirstLight}
\end{figure}

The second astronomical test was to observe a pulsar, and form a pulse profile from that observation. We observed PSR B0329+54, which is sufficiently bright that individual pulses from the pulsar can be detected in the beamformed signal. Figure \ref{fig:BAPPFirstLight} shows a set of 100 individual pulses from PSR B0329+54, and the resulting integrated pulse profile. These pulses are shown after the signal has been dedispersed (with the appropriate, known, dispersion measure) on the BAPP processing computer.

\newpage

\chapter{A Load Balanced Spectrometer for Coherent Dedispersion Applications}

Coherent dedispersion, as opposed to incoherent dedispersion, is used in experiments where it is important to reverse completely the dispersive effect of the interstellar medium \cite{LK05}. Coherent dedispersion is more computationally intensive than incoherent dedispersion, so typically a cluster of computers will be used to dedisperse a bandwidth $B>50\text{MHz}$. The coherent dedispersion algorithm acts on ``raw'' complex voltages, not powers, so data cannot be accumulated (as is the case with data produced for incoherent dedispersion applications). Therefore the output data rate of a spectrometer for coherent dedispersion applications is directly related to the input bandwidth and the output sample bitwidth. \\

\section{Overall Architecture}

\begin{figure}[htp]
	\centering
		\includegraphics[width=6.5in]{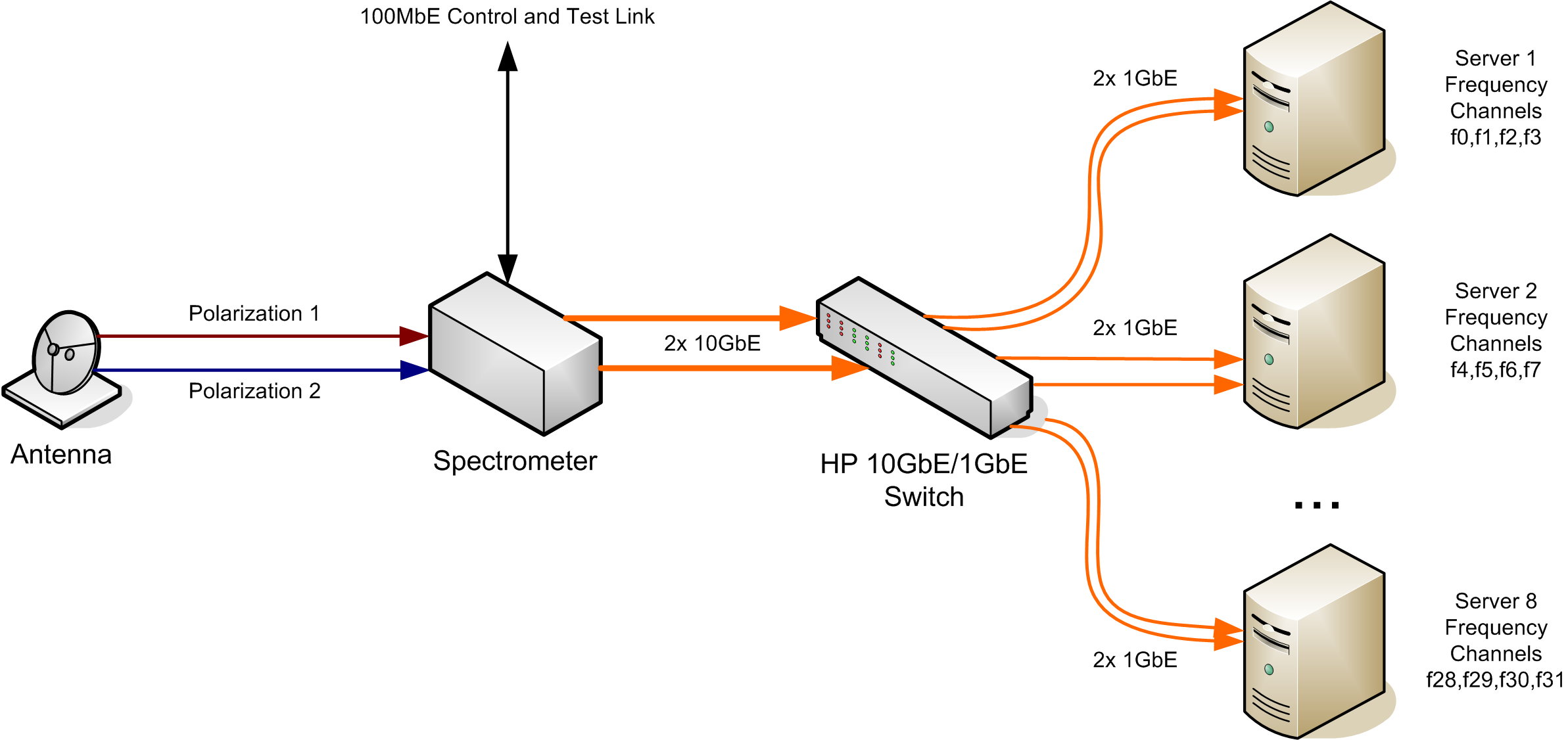}
		\caption{System Architecture for a Coherent Dedispersion Pulsar Study.}
	\label{fig:CoDeDiSpecArchitecture}
\end{figure}

Figure \ref{fig:CoDeDiSpecArchitecture} shows a potential system architecture for an experimental setup to conduct a coherent dedispersed-based pulsar study. Two polarization signals from an antenna are provided as real (i.e. not downconverted) voltages with bandwidth 400MHz. A spectrometer divides the signals into 32 channels each; each polarization's channels are outputted over a separate 10GbE connection. Since a single workstation-class computer is not able to perform coherent dedispersion on a 400MHz bandwidth signal in real time, the workload needs to be spread over several computers. The 10GbE connections are therefore terminated at a switch, which is connected to 8 computers using 1GbE connections, and each computer processes $1/8$th of the bandwidth. \\

Recent investigations into using graphics processing units (GPUs) to accelerate coherent dedispersion calculations \cite{Demorest07,Cognard08} suggest that it will be possible to dedisperse coherently a dual polarization signal with a bandwidth of 50MHz using a single nVidia GT280 GPU. Four computers each equipped with two such GPUs could then coherently dedisperse a 400MHz bandwidth. In this case each computer needs to receive a 100MHz bandwidth; an 8-bit sample bitwidth implies a data rate of $2\text{~polarizations} \times 200\text{MSa/sec} \times 1\text{byte/sample} = 400\text{Mbytes/sec} = 3.2\text{Gb/sec}$. Either four 1GbE connections could be supplied to each computer, or a single 10GbE connection from a 10GbE switch could be used. Due to the use of standard Ethernet output in the design presented in this chapter, and the availability of both 10GbE/1GbE and many-port 10GbE switches, both options are possible without any modification to the spectrometer design. \\

\section{A Static Load Balancing 400MHz Dual Spectrometer}



\begin{figure}[htp]
	\centering
		\includegraphics[width=8.5in,angle=90]{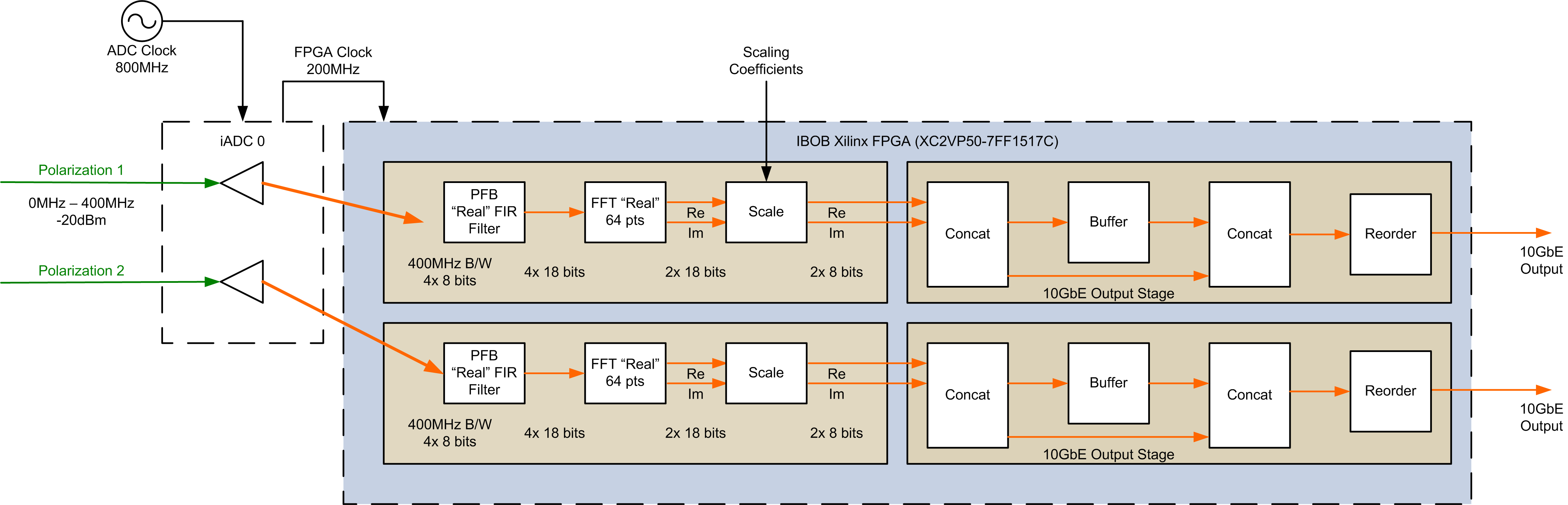}
		\caption{A Fast-readout Dual Spectrometer with ``raw complex voltage'' output.}
	\label{fig:CoDeDiSpecDetailedIBOB-Overview}
\end{figure}

Figure \ref{fig:CoDeDiSpecDetailedIBOB-Overview} shows the overall design of an IBOB-based dual polarization spectrometer that outputs complex Fourier coefficients (i.e. channelized complex voltages) over two 10GbE ports (one for each polarization). The spectrometer was built for the Nan\c{c}ay Radio Telescope, which supplies an IF at baseband. Hence the spectrometer accepts real baseband signals from 0 -- 400MHz as input. \\

Conceptually the design is fairly simple: an input signal is digitized and channelized using a 64-point polyphase filterbank to produce 32 channels. The PFB uses 18 bits of precision, and this is requantized to 8 bits. A scaling function is provided so that the user can shift the bits to select the most significant toggling ones. This stream of 8-bit numbers is sent to the 10GbE output stage. The output stage groups data values by channel (using a reorder block) before transmitting them. The output is statically load-balanced -- the user can set which frequency channels are sent to which IP address. \\

\begin{figure}[htp]
	\centering
		\includegraphics[width=8.5in,angle=90]{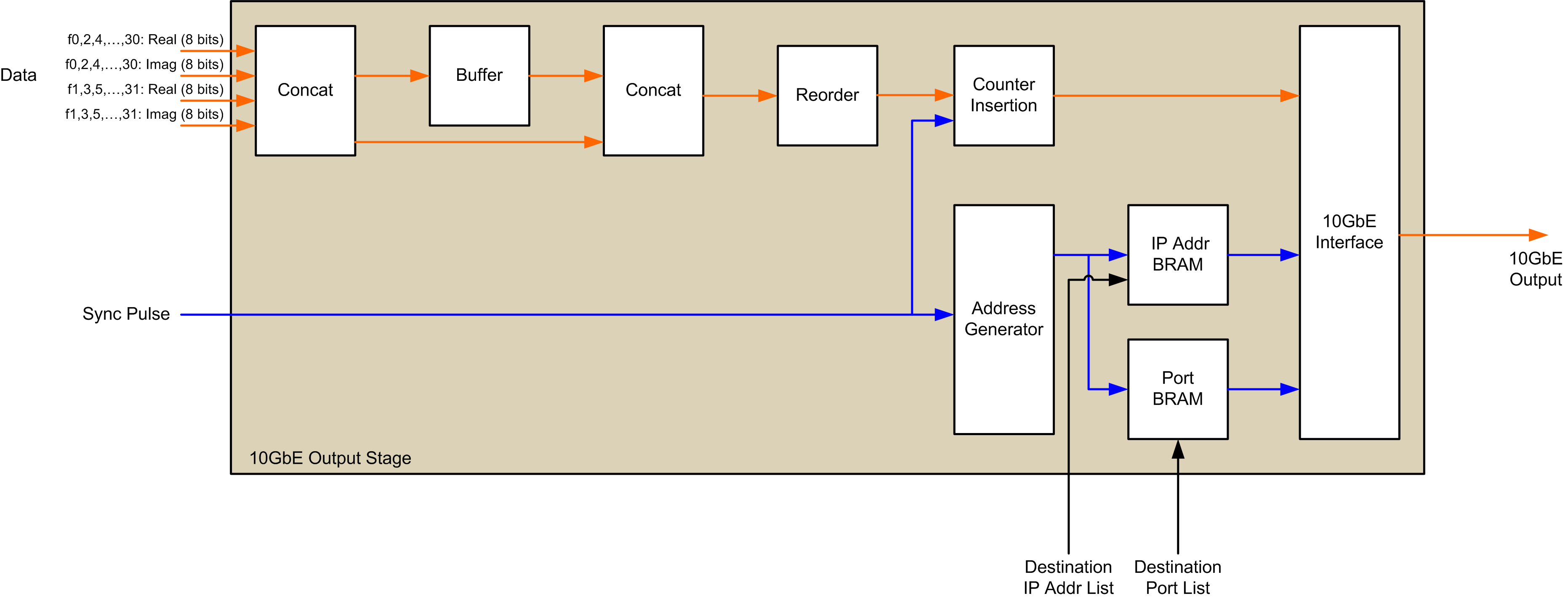}
		\caption{The 10GbE Output Stage.}
	\label{fig:CoDeDiSpecDetailedIBOB-OutputStage}
\end{figure}

The channelization part of this instrument's design is very similar to that of the designs described in the previous chapter. The output stages\footnote{The design includes two identical output stages, one for each polarization.}, however, have significant differences. Figure \ref{fig:CoDeDiSpecDetailedIBOB-OutputStage} shows a high-level view of the output stage in our prototype design. Every clock cycle the output stage receives four 8-bit values from the channelization stage: the real and imaginary parts of two consecutive channels. As is the case with the output stages in our other instruments, we need to buffer every second set of data because the data bus width of the 10GbE core is 64 bits. \\

One new feature of this output stage is its support of static load balancing by allowing the user to specify lists of destination IP addresses and ports. These lists are each stored in a single shared BRAM. Previously two shared registers, allowing the user to specify only one IP address and one port, were provided. A fairly simple design involving a wrapping counter is needed in an ``Address Generator'' unit to enable the design to change destination details at appropriate intervals. \\

A simple approach to implementing static load balancing is to simply send out packets containing four adjacent spectral bins. Every clock cycle the output stage receives two new bin values, so every two clock cycles it has four bins (comprising 64 bits) that can be sent over 10GbE. This is a very simple scheme to implement, since it simply requires that every second clock cycle the Address Generator should increment to select a new address from the IP and port BRAMs. However, there is a practical consideration that causes this approach to fail. We performed a set of network tests on a typical compute server with a 1GbE NIC and found that even with relatively low data rates ($<50$MB/sec) the server would drop a high percentage of packets ($>50\%$) if the packet payload size was sufficiently small (due to the high interrupt rate, which the CPU was apparently unable to satisfactorally service). Packet sizes larger than 100 bytes were found to yield vastly improved packet loss statistics. \\

This finding about packet sizes suggests that a successful approach to static load balancing of spectral data will require that each sent packet have a payload of at least 100 bytes. We chose the value 256 bytes, since this is sufficiently large that packet loss becomes negligible, but not so large that it is not possible to fit the required buffer into the FPGA design. \\

It is desirable to perform the load balancing in such a way that the compute servers receive packets with adjacent spectral bins. It is not desirable to have the spectrometer output different time samples containing all spectral bins to a single compute server. Therefore the easiest means of increasing the packet payload size -- simply increasing the number of clock cycles that the output stage spends on each destination before moving to the next destination in the list -- is not acceptable. What needs to happen instead is for the output stage to build large (256 byte) packets containing many time samples of some small set of bins. \\

If we define the input to the output stage at any one clock cycle as $f_t(k) \in \mathbb{C}$ and $f_t(k+1) \in \mathbb{C}$ (where $t$ is a time index denoting the spectrum, and $k \in \left\{0,2,\ldots,30\right\}$ are the channel/bin numbers), what we desire is to form eight packets $P_i$ ($i \in \left\{0,1,\ldots,7\right\}$) whose contents consists of four adjacent spectral bins from 32 spectra. For example, $P_0 = \left\{ f_t(0), f_t(1), f_t(2), f_t(3), f_{t+1}(0), \ldots, f_{t+1}(3), \ldots, f_{t+31}(0), \ldots, f_{t+31}(3) \right\}$. More generally,

\[
P_i = \left\{ f_t(4 \cdot i), f_t(4 \cdot i + 1), f_t(4 \cdot i + 2), f_t(4 \cdot i + 3), \ldots, f_{t+31}(4 \cdot i), \ldots, f_{t+31}(4 \cdot i + 3) \right\}
\]

The order that the output stage receives data in is:

\[
\left\{f_t(0), f_t(1)\right\}, \ldots, \left\{f_t(30), f_t(31)\right\}, \left\{f_{t+1}(0), f_{t+1}(1)\right\}, \ldots, \left\{f_{t+1}(30), f_{t+1}(31)\right\}, \ldots
\]

Braces here denote values that arrive on the same clock cycle. It is now easy to see that if we wish to create $P_i$ starting with $i=0$ and continuing through $i=7$ before returning to $i=0$, a simple solution is to reorder the data received by the output stage. The reorder definition is as follows:

\[
\left\{0, 1, \ldots, 1024 \right\} \longmapsto \left\{0, 1, 2, 3, 32, 33, 34, 35, \ldots, 992, 993, 994, 995, 4, 5, 6, 7, \ldots \right\}
\]

This reordering can easily be implemented using the CASPER ``Reorder'' block. The only further modification required to support the creation of large packets is to the Address Generator, which needs to only increment the address counter every 64 clock cycles. \\

The final part of the output stage before the 10GbE interface block is a ``header insertion'' stage that, as the name suggests, inserts a header at the start of each packet to be sent. This header is simply a 64-bit counter value that is used both for timing purposes and for detecting packet losses. \\

\section{Test Results}

Time constraints relating to the acquisition of equipment at the Nan\c{c}ay Radio Telescope resulted in us not being able to conduct field tests of our prototype\footnote{{\it Note added in proof}: From 2 -- 5 September 2008, a variant of the instrument described in this chapter was deployed in Nan\c{c}ay. During this week we successfully observed, and produced pulse profiles for, the pulsars PSR B1237+25 and PSR B1133+16.}. However, we conducted basic functional tests in the lab. Specifically we were able to divide the spectrum between two computers through a 10GbE/1GbE switch. (The instrument design allows for 8 different IP addresses (and port numbers) per polarization, so we used each IP address 8 times.) We conducted a ``tone test'' to verify that the spectrometer was working -- this involves directly attaching a sine wave source to an ADC input and verifying that a peak appears in the appropriate bin(s) in the resulting power spectrum. On the receiving computer end we computed the power spectrum by squaring the sums of the real and imaginary parts of the Fourier coefficients that are transmitted by the instrument. Since the channelization portion of the design has been well tested in several other instruments, we are confident that the prototype will produce correct results in a production environment. Our tone tests verified the functionality of the new portion of the design, the load-balancing output stage. \\

\newpage

\chapter{Conclusions}

In this thesis we have described the development of and results from several spectrometer instruments that were purpose-built for transient and pulsar studies. We obtained satisfactory field results from our ATA Fly's Eye, Berkeley ATA Pulsar Processor, Parkes Spectrometer and HartRAO Spectrometer instruments. We also presented work on the design of a prototype spectrometer system for coherent dedispersion applications, and an investigation into the design of an FPGA-based real-time coherent dedispersion system. \\

Throughout we have used the CASPER hardware and tools, as we set out to do, and we have verified the functionality and ease-of-use of many aspects of the hardware and the DSP library. \\

\section{Results Obtained}

We developed four field-tested instruments: the Fly's Eye, BAPP, the Parkes Spectrometer and the HartRAO Spectrometer. Each instrument was tested at at least one operational radio telescope facility and verified to perform adequately for the tasks for which they were design. \\

Specifically the ATA Fly's Eye instrument was verified through observations of PSR B0329+54 and the Crab pulsar, and through observations of giant pulses from the Crab. These giant pulses were detected using the same search mode that is used to find bright transient pulses, and hence verified the end-to-end capability of the system to detect the class of signal the instrument was designed to find. \\

The Berkeley ATA Pulsar Processor spectrometer was successfully integrated with the ATA beamformer. An observation of individual pulses from PSR B0329+54 using 24 dishes forming one beam verified the end-to-end functionality of the system. The pulse profile obtained using BAPP showed excellent agreement with existing profile data. \\

The Parkes spectrometer was tested and verified at both NRAO Green Bank and at the Parkes Radio Telescope. An observation of the fast millisecond pulsar B1937+21 yielded a correct pulse profile and verified the spectrometer's functionality for small-period pulsars. An observation of PSR J1028-5820 with the Parkes Radio Telescope verified the spectrometer's functioning in the Parkes Radio Telescope setup (for which it was designed), and showed excellent time resolution. \\

The HartRAO spectrometer was tested at HartRAO with observations of Vela and several other pulsars. The Vela pulse profile was in excellent agreement with existing profile data. \\

\section{Future Work}

The CASPER hardware successor to the BEE2 and IBOB, the ROACH board, is to be released from beta testing in early 2009. We expect that it will be possible to implement 700MHz dual polarization spectrometer designs on the ROACH board without considerable effort. The ports of the Parkes and HartRAO spectrometers to ROACH should be relatively easy, and will not only increase the available bandwidth, but should allow for improved spectral resolution. \\

Both the ATA Fly's Eye and BAPP instruments are bandwidth limited by the supplied signal, so an upgrade to ROACH would not yield improvements to the bandwidth. However, improved spectral resolution (and in the case of Fly's Eye, time resolution) should be fairly simple to achieve. \\

Once ROACH is available, it will be interesting to investigate its capabilities with the development of a real-time coherent dedispersion system in mind. An implementation of the BEE2-based design presented in Appendix A would allow for such an investigation -- the remaining work would be to discover what values of $N$ (number of channels) and $M$ (FFT length) are achievable in practice. Real-time coherent dedispersion is a significant computational challenge, and a successful implementation using ROACH could have a significant impact on the pulsar observing community due to potential improvements in price/performance by switching from cluster computing to an FPGA processing approach. \\

\newpage

\appendix

\chapter{Investigation into Building a FPGA-Based Real-Time Coherent Dedispersion System}

Coherent dedispersion is used when the best possible time resolution is required. The coherent dedispersion algorithm operates on ``complex voltages'' (i.e. complex Fourier coefficients), and hence the data rate prior to dedispersion cannot be reduced by accumulation. Therefore observing large bandwidths will result in large data rates to the coherent dedispersion system. Typically coherent dedispersion is done in pseudo-real-time using compute clusters, because it is not feasible to store the complex voltages for later processing. However, the processing demands for current typical observable bandwidths ($100\text{MHz} < B < 1\text{GHz}$) are significant. In order to dedisperse coherently a dual polarization 1GHz signal, approximately 15 modern dual quad-core CPU compute servers are necessary \cite{Bailes07}. Even next-generation GPUs are not expected to provide a simple solution: approximately 20 nVidia GT280 GPUs will be needed to coherently dedisperse 1GHz \cite{Cognard08}. \\

One potential alternative to using compute clusters is to use an FPGA-based system. Coherent dedispersion is a relatively simple algorithm that involves two discrete Fourier transforms (the forward and inverse transforms), and multiplication. The data supplied to the procedure can be streaming, so FPGA implementations offer much promise. \\

In this chapter we present a high-level investigation into the design of a real-time coherent dedispersion system using FPGAs. We did not implement the ideas presented to the point of producing a prototype. \\

\section{Coherent Dedispersion on Compute Clusters}

Before we can discuss FPGA implementations of coherent dedispersion, we first need to provide some detail on how this task is performed on conventional compute clusters. \\

Consider the case where the input bandwidth $B$ is split into $N$ channels using a polyphase filterbank, and these $N$ channels are sent to $N$ compute servers (one frequency channel per server). The data $S_i$ arriving at each server $i$ can then be considered as complex samples of a $B/N$ bandwidth signal. \\

The coherent dedispersion of a discrete signal $S_i\left[t\right]$ (where $t$ is discrete time) is calculated as the convolution of $S_i\left[t\right]$ with a ``chirp'' function that is designed to reverse the dispersion in the signal. Convolution in the time domain is equivalent to multiplication in the frequency domain -- because the discrete Fourier transform (and its inverse) can be efficiently calculated using the FFT algorithm, it is more efficient to apply the dedispersion filter in the frequency domain. The filter's definition \cite{LK05} is: \\

\[
H(f_0 + f) = e^{ i \frac{ 2 \pi \mathcal{D} }{ (f+f_0)f^2_0 } \text{DM} f^2 }
\] \\

Here $f_0$ is the centre frequency, $\mathcal{D}$ is the dispersion constant and DM is the dispersion measure. The procedure then is to obtain the discrete Fourier transform, $S_i\left[f\right]$, of the signal $S_i\left[t\right]$, to multiply this with a discrete version of $H(f_0 + f)$, and to then apply the inverse DFT to this result to produce a dedispersed time series of complex samples. \\

There is an important practical consideration in this procedure that can greatly affect the implementation of the algorithm: the length of the chirp function to be used (i.e. how many values are in its domain). This length determines the number of points required in the FFTs \cite{Cognard08}; specifically there needs to be an overlap between FFTs of at least the chirp length, so the FFT length needs to be the nearest power-of-two that is at least twice as large as the chirp length. \\

The chirp length is dependent on the observing frequency $f_0$, the bandwidth of the channel $B/N$, and the dispersion measure, DM, to be applied. It can differ quite significantly, as Table \ref{tbl:CoDeDiFFTLengths} shows. \\

\begin{table}[h!]
\centering
\caption[Chirp Function and FFT Lengths Required for Coherent Dedispersion.]{Table courtesy Ismael Cognard. Chirp Function and FFT Lengths Required for Coherent Dedispersion. $f_0$ is the centre frequency, and $B/N$ is the channel bandwidth.}
\label{tbl:CoDeDiFFTLengths}
\vspace{.2 in}
\begin{tabular}{|l|l|l|}
\hline
{\bf DM (pc cm$^{-3}$)}	& {\bf Chirp Length}		& {\bf FFT Length} \\
\hline
\hline
\multicolumn{3}{|c|}{$f_0=1.4\text{GHz}$, $B/N=16\text{MHz}$} \\
\hline
\hline
250    & 193586   & 512k \\ 
\hline
500    & 387172   & 1M \\ 
\hline
1000   & 774344   & 2M \\ 
\hline
\hline
\multicolumn{3}{|c|}{$f_0=1.4\text{GHz}$, $B/N=8\text{MHz}$} \\
\hline
\hline
250    & 48396    & 128k \\ 
\hline
500    & 96793    & 256k \\ 
\hline
1000   & 193586   & 512k \\ 
\hline
\hline
\multicolumn{3}{|c|}{$f_0=1.4\text{GHz}$, $B/N=4\text{MHz}$} \\
\hline
\hline
250    & 12099    & 32k \\ 
\hline
500    & 24198    & 64k \\ 
\hline
1000   & 48396    & 128k \\ 
\hline
\end{tabular}
\end{table}

In CPU and GPU implementations, the primary limitation that prevents the efficient execution of arbitrarily large FFTs is memory. Since the dedispersion procedure needs to be performed in real-time, a limit on the coherent dedispersion capabilities of a particular cluster system is defined partially by the largest FFT the system can perform and still meet the real-time requirement. We will see that a similar limitation exists, albeit for a different reason, in FPGA-based systems. \\

\section{Coherent Dedispersion on CASPER Reconfigurable Computing Platforms}

A real-time coherent dedispersion system implemented using FPGAs needs to first sample and appropriately channelize the full bandwidth $B$. Let us again denote by $N$ the number of channels, and hence each channel has bandwidth $B/N$. We now need to consider that the system needs to perform coherent dedispersion on all $N$ channels. We cannot perform these all in parallel (if, for example, $N=1024$ we would need to implement thousands of large FFTs in parallel!), but fortunately it is not necessary to do so. Coherent dedispersion is amenable to a ``streaming'' implementation, where it is not necessary to have all the data available at the beginning of the computation -- instead the data is streamed through the design as it becomes available. Since the input data rate to the coherent dedispersion system is equal to the output data rate (the dedispersed output signals have the same bandwidth as the input signals), it is clear that so long as a streaming implementation is possible, it is not necessary to perform the dedispersion on all channels simultaneously. \\

In the system described in the previous chapter, the FPGA board (an IBOB, in that case) channelized the data, grouped adjacent channels and sent them as UDP packets to a set of compute servers. To perform coherent dedispersion, the servers would then each need to buffer $M$ complex samples from a single channel before proceeding, where $M$ is the length of the FFT to be performed. In the case of a streaming FFT implementation, a similar buffering scheme is necessary. However, whereas the $N$ servers each had a length $M$ buffer and operated independently, an FPGA implementation needs to buffer $M$ values from all $N$ channels. This implies the need to use DRAM to store the buffer, since $M \times N$ is typically large for practical applications (see Table \ref{tbl:CoDeDiFFTLengths}). \\

An important consideration when assessing the capability of FPGA platforms to perform coherent dedispersion in real-time is determining the maximum length of FFT that can be implemented. A fully pipelined $M$-point FFT implementation has time complexity $\mathcal{O}$($M$) and FPGA resource complexity $\mathcal{O}$(log $M$). In other words, the amount of resources an FFT implementation uses on an FPGA is proportional to the log of its length. Because FPGA resources are limited, we can't produce arbitrarily large FFTs. Therefore our coherent dedispersion capabilities are limited by maximum length of a single FFT that we can fit on a single FPGA. \\

There are other potential bottlenecks in FPGA platforms that may limit performance (such as DRAM availability and memory bandwidth), but in CASPER's BEE2 and ROACH boards, FFT resource usage sets the limit. \\

\subsection{BEE2}

\begin{figure}[htp]
	\centering
		\includegraphics[width=7.5in,angle=90]{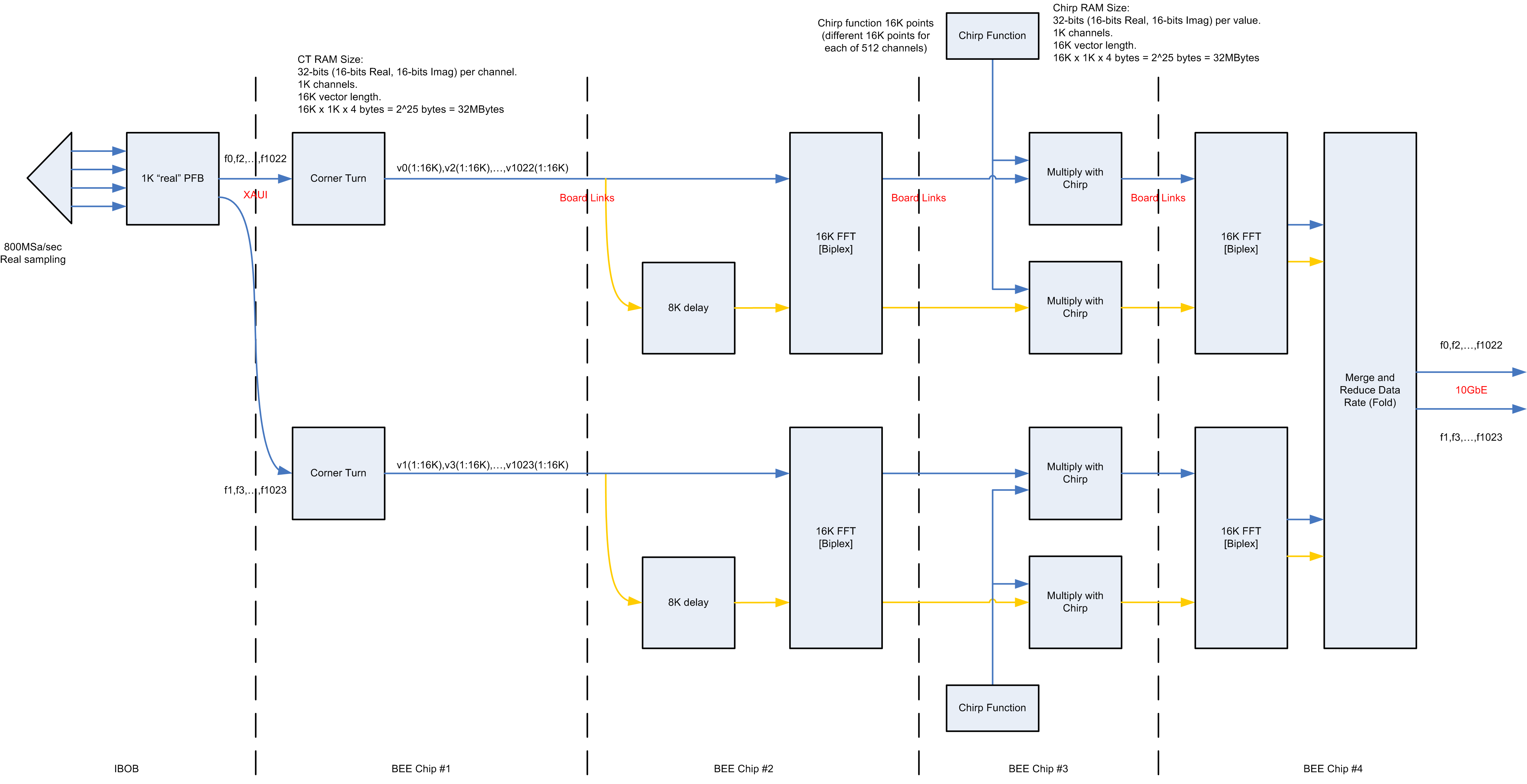}
		\caption[A High-Level BEE2 Architecture for a Real-Time Coherent Dedispersion System.]{A High-Level BEE2 Architecture for a Real-Time Coherent Dedispersion System. This proposed design could process a single polarization with 400MHz bandwidth -- an IBOB samples the signal at 800MSa/sec and channelizes it into two streams (even and odd frequency channels) that are sent to a BEE2 board for coherent dedispersion.}
	\label{fig:RTCoDeDi-BEE2}
\end{figure}

A possible high-level system architecture for a real-time coherent dedispersion system implemented using a BEE2 board is shown in Figure \ref{fig:RTCoDeDi-BEE2}. This system could process a single polarization with 400MHz bandwidth. Since the BEE2 does not have any direct ADC capabilities, an IBOB with iADC is needed to sample the input signal. The IBOB includes an FPGA that is sufficiently large to perform a $N=4096$ channelization, so this capability isn't wasted. In the presented design, a channelization of $N=1024$ is used, giving $B/N = 400\text{MHz}/1024 \approx 0.39\text{MHz}$ bandwidth per channel. \\

The IBOB and BEE2 FPGAs are clocked at 200MHz, so given a real input of 800MSa/sec, the CASPER streaming FFT implementation will supply two channels every clock cycle. This requires that two channels be coherently dedispersed in parallel. If we reduce the bandwidth to 200MHz (from 400MHz), then only one coherent dedispersion calculation need be implemented. \\

The implementation of the dedispersion on the BEE2 is split amongst four FPGAs: the first FPGA implements two $M \times \frac{N}{2}$ buffers in DRAM (called ``Corner Turners'' due to their mode of operation, which is explained shortly); the second FPGA implements four FFTs (two biplex FFTs) that convert the single channel's time series to the frequency domain; the third FPGA performs the multiplication by the Chirp function (whose values are stored in DRAM), and the fourth FPGA performs the inverse FFTs, and merges the data for output. \\

\begin{figure}[htp]
	\centering
		\includegraphics[width=3in]{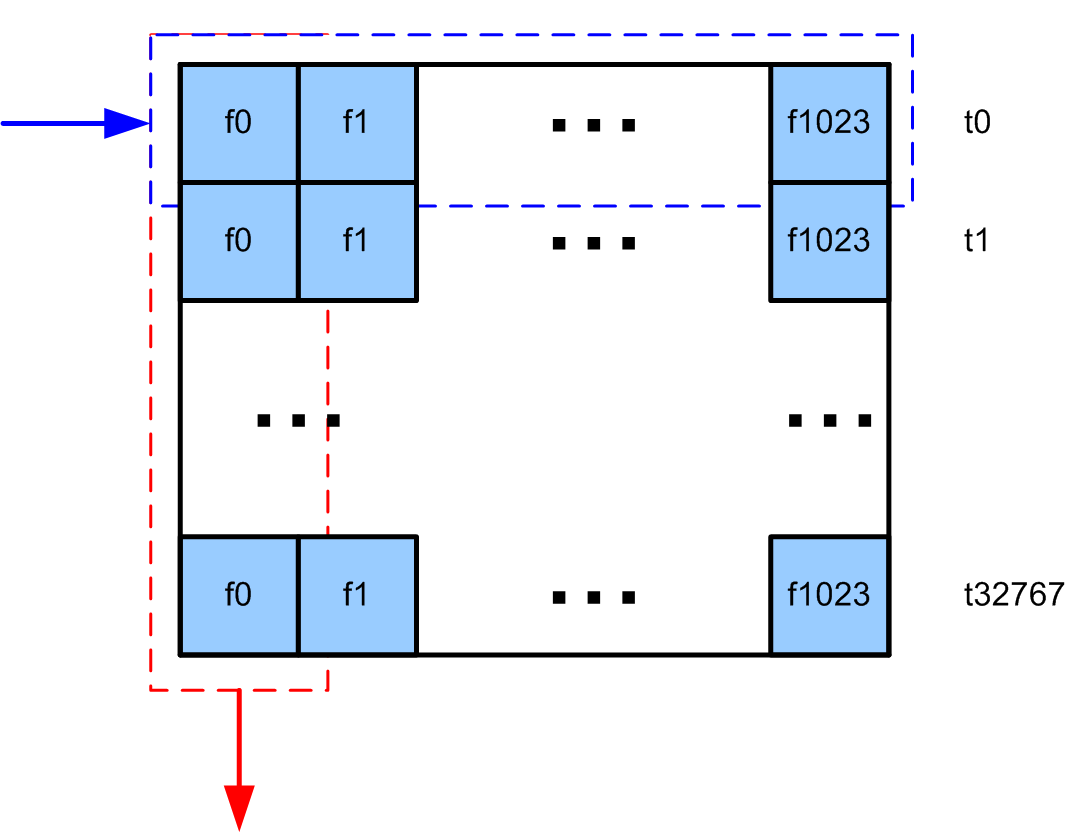}
		\caption[A Corner Turner for a Real-Time Coherent Dedispersion System.]{A Corner Turner for a Real-Time Coherent Dedispersion System. In this example, a 1K channelization PFB may write values in {\it rows} (blue) into a memory that has 32K rows and 1K columns. A ``Corner Turner'' reads out the values in {\it columns} (red).}
	\label{fig:RTCoDeDi-CornerTurner}
\end{figure}

The ``Corner Turn'' (or ``Corner Turner'') on the first FPGA is a special DRAM buffer implementation. It effectively performs a matrix transpose operation: the input to the Corner Turner is written in rows, but the output is read from columns. This is illustrated in Figure \ref{fig:RTCoDeDi-CornerTurner} for an example scenario where an $N=1024$ channelization PFB writes its values to the buffer in preparation for an $M=32768$-point FFT.

In the BEE2 design example, each Corner Turner uses 32MB of DRAM, which is easy to supply. \\

The next FPGA stage, which performs the first FFT in the dedispersion procedure, contains an 8192 clock delay so that an overlap between FFTs of half the points is calculated. This overlapped data is processed in parallel with the non-delayed data. \\

The third BEE2 FPGA performs the multiplication of the data streams with the appropriate chirp function values. Since each of the $N=1024$ channels requires a different 16k chirp values for its dedispersion, the storage requirement for all these values is non-trivial: a total of 64MB is required, which again necessitates the use of DRAM. This FPGA's resources are not heavily taxed, so it may be possible to merge this stage with the previous one. However, since the BEE2 contains 5 FPGAs, and no fewer, a design in practice would probably use a separate FPGA for multiplication to avoid complications. \\

The final BEE2 FPGA performs an inverse FFT on all four data streams. The data rate is at this point quite high, so it may be desirable to reduce the data rate so that the output can be sent to a single computer for storage or further processing. To reduce the data rate, one might compute the power and fold the data. \\

\subsection{ROACH}

We expect that with ROACH the clock speed of the FPGA will be approximately doubled versus that on a BEE2, and we expect to have at least twice as much FPGA resources for implementing large FFTs. The Xilinx Virtex 5 LX155T device (which is compatible with ROACH) has more than double number of slices than the Virtex 2 Pro VP70 used on the BEE2 contains. \\

We therefore expect that it will be possible to implement a dual polarization $B=1\text{GHz}$ real-time coherent dedispersion system using 4 ROACH boards that will be suitable for many combinations of centre frequency and dispersion measure. \\

Appropriate I/O, specifically CX4 ports for XAUI connections, is available on the ROACH boards to support streaming data through a series of 4 boards. \\

\section{Conclusions}

FPGA-based implementations of coherent dedispersion are possible, and may offer benefits over CPU or GPU implementations. With CPU/GPU implementations a static load-balancing spectrometer, such as the one described in the previous chapter, is needed. These systems require a switch. In contrast, a purely FPGA-based system could replace the switch and the compute server cluster with, potentially, just 3 ROACH boards. \\

The capabilities of the BEE2 are too limited to provide a price/performance advantage over current CPU and GPU solutions. A BEE2 and IBOB system can probably be built to have comparable capability to approximately 3 dual quad-core compute servers. However, a system using ROACH boards may very well offer a price/performance and performance/Watt advantage over CPU or GPU implementations. \\

The most significant relative disadvantage to FPGA approaches to coherent dedispersion is that both CPU and GPU approaches offer far more flexibility to the developer. With an FPGA implementation it is more difficult to make changes, add features, and so on. The tradeoff between improved price and worse programmability is one that users will have to consider. \\

One important issue that we have ignored in our high-level investigation in this chapter is the matter of required numerical precision. We have assumed that a 16-bit data path is sufficient, and that we can produce adequate results using fixed-point multiplication. A study conducted by, or with, an experienced practitioner of coherent dedispersion methods is necessary for this.

\newpage

\chapter{Derivation of Expected Noise Correlation Results}

\section{Introduction}

In dual polarization spectrometers we often wish to compute ``cross-terms'' of the two polarizations, which in addition to the powers of the polarizations, can be used to compute the Stokes parameters $I,Q,U,V$. Specifically, if our spectrometer has as inputs polarizations $A$ and $B$, for each frequency channel we compute the powers $\left|A\right|^2$ and $\left|B\right|^2$, in addition to the cross-terms $\Re\left\{AB^*\right\}$ and $\Im\left\{AB^*\right\}$ (where $X^*$ is the complex conjugate of $X\in\mathbb{C}$). If our polarization inputs $A$ and $B$ are from orthogonal linear polarization channels, the Stokes parameters can be computed as: \\

\[
  \begin{bmatrix}
    I \\
    Q \\
    U \\
    V
  \end{bmatrix}
  =
  \begin{bmatrix}
    \left|A\right|^2 + \left|B\right|^2 \\
    \left|A\right|^2 - \left|B\right|^2 \\
    2 \Re\left\{AB^*\right\} \\
    2 \Im\left\{AB^*\right\} \\
  \end{bmatrix}.
\]

Similarly if $A$ and $B$ are the left- and right-handed signals from circular feeds, then the Stokes parameters can be computed as:

\[
  \begin{bmatrix}
    I \\
    Q \\
    U \\
    V
  \end{bmatrix}
  =
  \begin{bmatrix}
    \left|A\right|^2 + \left|B\right|^2 \\
    2 \Re\left\{AB^*\right\} \\
    2 \Im\left\{AB^*\right\} \\
    \left|A\right|^2 - \left|B\right|^2 \\
  \end{bmatrix}.
\]

We need to know how the statistics of the powers relate to those of the cross-terms in order to verify correctness during laboratory tests. In the laboratory we use as inputs independent broadband noise sources with Gaussian statistics; this is done to realistically simulate the signals received by radio telescopes. It is not possible to verify that individual integrations have had their cross-terms computed correctly, but it is possible to verify that the statistics of the polarization powers agree with the statistics of the cross-terms. These statistics-based tests should be sufficient to gain confidence that the cross-terms are being computed correctly\footnote{These tests can only be applied in processing chains that maintain the relative magnitude of signals throughout -- this is not generally the case when the user selects arbitrary sets of 8 bits from the integrators' outputs, or when the output values are independently scaled. In production the user will typically scale and select bit sets for each of the outputs independently in order to maximize the number of ``useful'' bits. When carrying out the statistical tests that we describe, the scaling and bit selection should be set uniformly for all outputs.}. \\

\section{A Mathematical Description of a Full Stokes Spectrometer}

The processing chains in our spectrometers for incoherent dedispersion applications are relatively simple: two input signals, polarizations $A$ and $B$, are digitized and then independently Fourier transformed using an $M$-point Fast Fourier Transform procedure\footnote{More precisely, the signals are channelized using polyphase filter banks, but this does not significantly affect our analysis.}. Effectively every sequential set of $M$ samples are transformed together, and Fourier transforms are carried out continuously on these sets. The resulting Fourier components from each polarization are, independently for each channel, used to compute four values. If we denote the Fourier component of polarization $A$ as $A_i(f)\in\mathbb{C}$ (let $A_i(f)=x_i(f)+jy_i(f)$ where $x_i(f),y_i(f)\in\mathbb{R}$ and $j = \sqrt{-1}$) for the $i$th Fourier transform, and some channel $f$ (where $f \in \left\{0,1,2,\ldots,M-1\right\}$), and similarly define $B_i(f)\in\mathbb{C}$ (let $B_i(f)=z_i(f)+jw_i(f)$ where $z_i(f),w_i(f)\in\mathbb{R}$) for polarization $B$, then the four values that are computed for each frequency channel $f$ and for each subsequent transform $i$ are: $\left|A_i(f)\right|^2 = A_i(f) A_i^*(f)$, $\left|B_i(f)\right|^2 = B_i(f) B_i^*(f)$, $\Re\left\{A_i(f) B_i^*(f)\right\}$ and $\Im\left\{A_i(f) B_i^*(f)\right\}$. Finally these values are ``accumulated'' by frequency channel, which is to say that a sum of $N$ values from subsequent Fourier transforms is computed\footnote{The term ``accumulated'' is often used interchangeably with ``integrated''; likewise ``accumulation'' and ``integration'' are both used to refer to a single sum.}. The value $N$ is known as the {\it accumulation length}. These accumulated values are what the spectrometer outputs. \\

Specifically the following values are computed for a single accumulation (for each channel $f$): \\

\begin{equation}
\Sigma_1 \stackrel{=}{_{\text{def}}} \sum^{N}_{i=1}\left|A_i\right|^2 = \sum^{N}_{i=1} ( {x_i}^2 + {y_i}^2 ) \label{eq:PowA}
\end{equation}

\begin{equation}
\Sigma_2 \stackrel{=}{_{\text{def}}} \sum^{N}_{i=1}\left|B_i\right|^2 = \sum^{N}_{i=1} ( {z_i}^2 + {w_i}^2 ) \label{eq:PowB}
\end{equation}

\begin{equation}
\Sigma_3 \stackrel{=}{_{\text{def}}} \sum^{N}_{i=1}\Re\left\{A_i{B_i}^*\right\} = \sum^{N}_{i=1} ( x_iz_i + y_iw_i ) \label{eq:ReAB}
\end{equation}

\begin{equation}
\Sigma_4 \stackrel{=}{_{\text{def}}} \sum^{N}_{i=1}\Im\left\{A_i{B_i}^*\right\} = \sum^{N}_{i=1} ( x_iw_i + y_iz_i ) \label{eq:ImAB}
\end{equation}

Here we have dropped the dependence on $f$ as a convenience of notation; when the dependence is not shown, it should be taken to be implicit that each value is frequency-dependent. For example, in equation \ref{eq:PowA}, $A_i(f)$ appears as $A_i$; $x_i(f)$ appears as $x_i$, and so on. \\

We wish to know how the statistics of $\Sigma_1$, $\Sigma_2$, $\Sigma_3$ and $\Sigma_4$ relate, in particular when the input signals from polarizations A and B are modeled as independent Gaussian noise sources. \\

\section{The Distribution of the Fourier Coefficients of Real Gaussian-distributed Time-domain Data}

Given that the input signal for polarization A is real, we know that its $M$-point discrete Fourier transform will be symmetric (and hence the second half of the spectrum can be discarded, since it contains no useful information\footnote{This is, in fact, what happens in practice in the FPGA implementation of the FFT.}). The Fourier components $A_i(f)$ of the $i$th transform are computed by transforming the sampled real values $\alpha_i(t)$ where $t \in \left\{0,1,2,\ldots,M-1\right\}$. It is these sampled values $\alpha_i(t)$ that are assumed to be Gaussian distributed, i.e. $\alpha_i(t) \sim N(\mu,\sigma^2)$. The discrete Fourier transform is defined such that: \\

\begin{equation}
A_i(f) = \sum^{M-1}_{t=0} \alpha_i(t) e^{-2\pi jft/M} \label{eq:DFT}
\end{equation}

We would like to know what the distributions of the random variables $A_i(f)$ are. We use the fact that if $X \sim N(\mu_1, \sigma^2_1)$ and $Y \sim N(\mu_2, \sigma^2_2)$, then for arbitrary $a, b \in \mathbb{R}$, the new random variable $Z = aX + bY \sim N(a\mu_1 + b\mu_2, a^2\sigma^2_1 + b^2\sigma^2_2)$. First, we find the real and imaginary components of $A_i(f)$ using the identity $e^{j\theta}=\cos(\theta) + j\sin(\theta)$:

\begin{equation}
\Re\left\{A_i(f)\right\} = \sum^{M-1}_{t=0} \alpha_i(t) \cos(2\pi ft/M) \label{eq:DFTRe}
\end{equation}

\begin{equation}
\Im\left\{A_i(f)\right\} = \sum^{M-1}_{t=0} \alpha_i(t) \sin(-2\pi ft/M) \label{eq:DFTIm}
\end{equation}

Since $\cos(2\pi ft/M) \in \mathbb{R}$ and $\sin(-2\pi ft/M) \in \mathbb{R}$, we can use the addition rule for normal variables given above to write:

\begin{equation}
\Re\left\{A_i(f)\right\} \sim N(\mu \sum^{M-1}_{t=0} \cos(2\pi ft/M), \sigma^2 \sum^{M-1}_{t=0} \cos^2(2\pi ft/M)) \label{eq:DistDFTRe}
\end{equation}

\begin{equation}
\Im\left\{A_i(f)\right\} \sim N(\mu \sum^{M-1}_{t=0} \sin(-2\pi ft/M), \sigma^2 \sum^{M-1}_{t=0} \sin^2(-2\pi ft/M)) \label{eq:DistDFTIm}
\end{equation}

If we assume that the input signal has zero mean, i.e. $\mu = 0$, then equations \ref{eq:DistDFTRe} and \ref{eq:DistDFTIm} simplify to:

\begin{equation}
\Re\left\{A_i(f)\right\} \sim N(0, \sigma^2 \sum^{M-1}_{t=0} \cos^2(2\pi ft/M)) \label{eq:DistDFTReZeroMean}
\end{equation}

\begin{equation}
\Im\left\{A_i(f)\right\} \sim N(0, \sigma^2 \sum^{M-1}_{t=0} \sin^2(-2\pi ft/M)) \label{eq:DistDFTImZeroMean}
\end{equation}

Now we know that if we input a real signal that is normally distributed with zero mean, both the real and imaginary parts of each of the output Fourier coefficients will also be normally distributed, and will have zero mean. We can also see that the variances of the real and imaginary components of the Fourier coefficients $A_i(f)$ have a non-trivial, but computable, dependence on the frequency $f$. For any FFT length $M$ we can easily compute the variance. For example, if we wish to know the variance of the random variable $\Re\left\{A_i(5)\right\}$ (the real part of the 6th Fourier coefficient) when $\sigma = 1$ and $M=1024$, we must compute $\sum^{1023}_{t=0} \cos^2(10\pi t/1024)$.

\section{The Statistics of the Spectrometer Outputs $\Sigma_1$, $\Sigma_2$, $\Sigma_3$ and $\Sigma_4$}

We would like to determine the means and variances of the variables $\Sigma_1$, $\Sigma_2$, $\Sigma_3$ and $\Sigma_4$ given information about the distribution of the input values. Since the definitions of the output variables $\Sigma_1$ and $\Sigma_2$ are defined similarly in equations \ref{eq:PowA}, \ref{eq:PowB}, and likewise $\Sigma_3$ and $\Sigma_4$ are defined similarly in equations \ref{eq:ReAB} and \ref{eq:ImAB}, we will proceed by analyzing $\Sigma_1$ and $\Sigma_3$ and then extend our results to $\Sigma_2$ and $\Sigma_4$. \\

All the proofs in this section were kindly supplied by Dr Rachael Padman \cite{Padman07}. \\

We assume throughout that $x_i, y_i, z_i, w_i$ are independent random variables that are normally-distributed with zero mean and unit variance, i.e. $x_i, y_i, z_i, w_i \sim N(0,1)$. This assumption is justified by the previous section, which shows that it is possible to create Gaussian noise sources to use as inputs that will result in at least one of the spectrometer's Fourier coefficients being a Gaussian-distributed random variable with zero mean and unit variance.

\subsection{The Mean and Variance of $\Sigma_1$ and $\Sigma_2$}

$\Sigma_1$ is defined as $\sum^{N}_{i=1}\left|A_i\right|^2 = \sum^{N}_{i=1} ( {x_i}^2 + {y_i}^2 )$. Calculating its mean and variance from first principles involves first calculating the mean and variance of ${x_i}^2$ and ${y_i}^2$. Define $g_i=x_i^2$. From Calculus, we know:

\[
p(g_i)=\frac{dx_i}{dg_i}p(x_i)
\]

where $p(g_i)$ is the probability distribution function of $g_i$. Thus:

\[
p(g_i)=\frac{1}{2\sqrt{g_i}} \frac{1}{\sigma\sqrt{2\pi}} e^{- \frac{(\sqrt{g_i}-\mu)^2}{2\sigma^2}}
\]

Substituting for zero mean and unit variance, and multiplying by a factor of 2 to account for folding from both sides of the normal distribution to one side of the exponential distribution, we get:

\[
p(g_i)=\frac{1}{\sqrt{2\pi}} g_i^{-1/2} e^{-g_i/2}
\]

We can recognize this as the $\chi^2$ distribution with one degree of freedom, which is defined on the interval $\left[0,\infty\right]$. The mean and variance of $g_i$ are thus:

\[
\mu_{g_i} = \int^{\infty}_{0} g_i p(g_i) dg_i = \frac{1}{\sqrt{2\pi}} \int^{\infty}_{0} g_i^{1/2} e^{-g_i/2} dg_i = 1
\]

\[
\sigma^2_{g_i} = \int^{\infty}_{0} (g_i - \mu_{g_i})^2 p(g_i) dg_i = \frac{1}{\sqrt{2\pi}} \int^{\infty}_{0} (g_i - 1)^2 g_i^{-1/2} e^{-g_i/2} dg_i = 2
\]

The sum $\Sigma_1$ is a sum of $2N$ independent random variables that each have mean $\mu_{g_i} = 1$ and variance $\sigma^2_{g_i} = 2$. Therefore the mean $\mu_{\Sigma_1}$ and variance $\sigma^2_{\Sigma_1}$ of $\Sigma_1$ can be derived as:

\[
\mu_{\Sigma_1} = \text{E}\left[ \sum^{N}_{i=1} ( x^2_i + y^2_i ) \right] = \text{E}\left[ \sum^{2N}_{i=1} x^2_i \right] = \int^{\infty}_{0} \left( \sum^{2N}_{i=1} g_i \right) p(g_i) dg_i = 2N \mu_{g_i} = 2N
\]

\[
\sigma^2_{\Sigma_1} = \text{E} \left[ \left( \sum^{2N}_{i=1} x^2_i - \mu_{\Sigma_1} \right)^2 \right] = \text{E} \left[ \left( \sum^{2N}_{i=1} x^2_i  \right)^2 \right] - \mu^2_{\Sigma_1} = \int^{\infty}_{0} \left( \sum^{2N}_{i=1}g_i \right)^2 p(g_i) dg_i - \mu^2_{\Sigma_1} = 4N
\]

These results can be similarly obtained by considering $\Sigma_1$ as a sum of $2N$ independent squared normal variables, and noting that the distribution of such a sum is just a $\chi^2$ distribution with $2N$ degrees of freedom. \\

Because $x_i$ and $y_i$ are defined in the same way as $z_i$ and $w_i$, we can see that the mean of $\Sigma_2$ is $\mu_{\Sigma_2} = 2N$ and the variance of $\Sigma_2$ is $\sigma^2_{\Sigma_2} = 4N$. \\

\subsection{The Mean and Variance of $\Sigma_3$ and $\Sigma_4$}

We would like to compute the mean and variance of variables of the form $\Sigma_3 = \sum^{N}_{i=1} ( x_iz_i + y_iw_i )$. Given that $x_i, y_i, z_i, w_i$ are independent Gaussian-distributed random variables with zero mean and unit variance, for the purposes of computing the mean and variance of $\Sigma_3$ we can consider this sum instead as:

\begin{equation}
\Sigma_3 = \sum^{2N}_{i=1} ( x_i y_i ) \label{eq:Sigma_3Redefined}
\end{equation}

First we need to find the mean $\mu_{x_i y_i}$ and variance $\sigma^2_{x_i y_i}$ of the product random variable $x_i y_i$. These are defined as:

\[
\mu_{x_i y_i} = \int^{\infty}_{-\infty} \int^{\infty}_{-\infty} x_i y_i p(x_i,y_i) dx_i dy_i
\]

\[
\sigma^2_{x_i y_i} = \int^{\infty}_{-\infty} \int^{\infty}_{-\infty} ( x_i y_i - \mu_{x_i y_i} )^2 p(x_i,y_i) dx_i dy_i
\]

Since $x_i$ and $y_i$ are independent, $p(x_i,y_i)=p(x_i)p(y_i)$. Thus:

\[
\mu_{x_i y_i} = \int^{\infty}_{-\infty} x_i p(x_i) dx_i \int^{\infty}_{-\infty}  y_i p(y_i) dy_i = \mu_{x_i} \mu_{y_i}
\]

Here $\mu_{x_i}$ and $\mu_{y_i}$ are the means of the random variables $x_i$ and $y_i$ respectively, which we defined to be zero. So $\mu_{x_i y_i} = 0$. \\

The computation of the variance can also be completed using the facts $p(x_i,y_i)=p(x_i)p(y_i)$ and $\mu_{x_i y_i} = 0$:

\[
\sigma^2_{x_i y_i} = \int^{\infty}_{-\infty} \int^{\infty}_{-\infty} x^2_i y^2_i p(x_i) p(y_i) dx_i dy_i = \int^{\infty}_{-\infty} (x_i - \mu_{x_i})^2 p(x_i) dx_i \int^{\infty}_{-\infty} (y_i - \mu_{y_i})^2 p(y_i) dy_i = \sigma^2_{x_i} \sigma^2_{y_i}
\]

We defined $\sigma^2_{x_i} = \sigma^2_{y_i} = 1$, so $\sigma^2_{x_i y_i} = 1$. In the redefined equation for $\Sigma_3$, equation \ref{eq:Sigma_3Redefined}, we have a sum of independent variables $x_i y_i$. We can use the fact that the variance of the sum of two independent random variables is the sum of their variances \cite{GrinsteadSnell} to calculate that the variance $\sigma^2_{\Sigma_3}$ of $\Sigma_3$ is $2N$. Similarly the mean $\mu_{\Sigma_3}$ is calculated to be $0$.

\newpage

\chapter{User Guide for the Parkes Spectrometer}

{\it The Parkes Spectrometer User Guide provides a representative set of instructions for setting up and operating an IBOB-based spectrometer. We have written similar guides for the other deployed instruments described in this thesis, but they have been excluded from this document in the interests of brevity, since their contents differs only slightly from that of the Parkes guide reprinted here.}

\section{Introduction}

The ``Parkes Spectrometer'' (henceforth ``{\it Parspec}'') is based on the IBOB hardware platform, and was built using the BWRC/CASPER\footnote{The Simulink toolflow for the BEE2 and IBOB boards was developed by the Berkeley Wireless Research Center, and the DSP libraries were developed by CASPER.} Simulink toolflow and DSP libraries \cite{CASPER}. \\

This User Guide, as the name suggests, does not include detailed descriptions of the internals of the spectrometer design. Merely enough detail is given to allow the user to take advantage of all the features and configuration options that are available. \\

\section{Hardware Setup}

\begin{figure}[htp]
	\centering
		\includegraphics[width=5.5in]{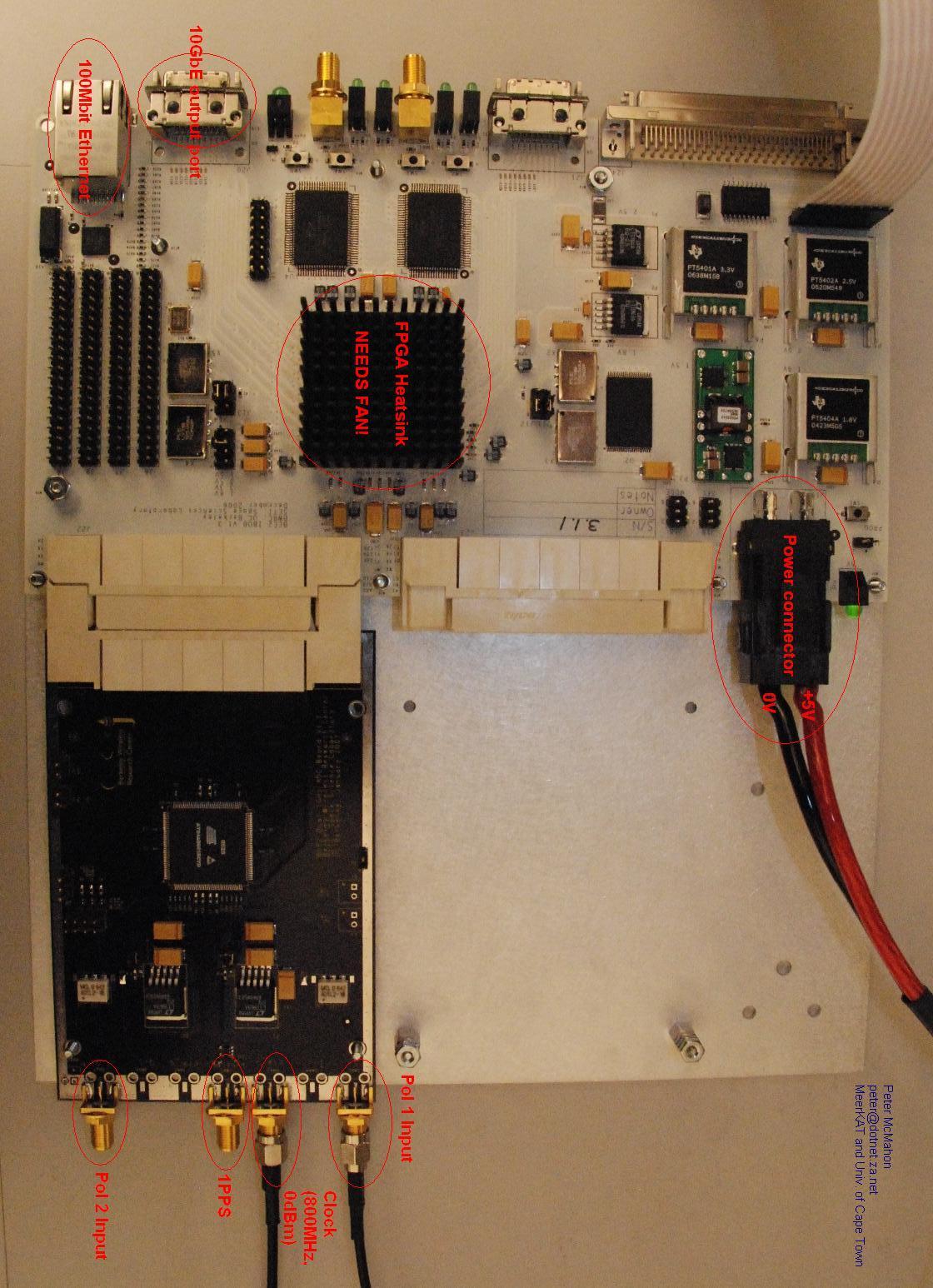}
		\caption{IBOB with labeled connectors and ports.}
  \label{fig:PUGibob}
\end{figure}

Figure \ref{fig:PUGibob} shows an IBOB with labels on the relevant connectors and ports required for setting up the Parspec. To set up the hardware, perform the following steps: \\

\begin{enumerate}

  \item Arrange airflow\footnote{A rule-of-thumb test during operation to verify that the airflow is sufficient is to make sure that the heatsink is not too hot to hold.} for IBOB FPGA. This can take the form of a fan directly mounted on the heatsink, or a fan blowing from bottom-to-top or top-to-bottom over the heatsink (left-to-right, or right-to-left in Figure \ref{fig:PUGibob}).

  \item Connect the clock, 1PPS and analogue polarization signals (``Polarization 1'' and ``Polarization 2'') to the ADC board. The clock should have level 0dBm, and the signals should ideally have power levels below -10dBm\footnote{Power levels above 0dBm may damage the hardware, and signals with power higher than -10dBm may cause overflow in the FFT.}. The 1PPS signal should be 0-2V minimum, 0-3V nominal, and 0-5V maximum, with $50\Omega$ termination. The 1PPS connection is optional.
  
  \item Connect the IBOB's 100Mbit Ethernet port to a control computer (whose IP address is modifiable) using standard Ethernet twisted-pair cable.
  
  \item Connect the IBOB's 10Gbit Ethernet port labeled ``XAUI1'' to the data recorder computer -- either directly (if the computer has a 10Gbit Ethernet NIC), or via a 10GbE/1GbE (CX4/RJ45) switch.
  
  \item Connect the power cable to the IBOB. The red wire should be at +5V, and the black wire should be ground. The power supply should be able to supply at least 10A.
  
  \item Turn on the power supply. If your board came with a pre-programmed PROM, then the design should be loaded from the PROM into the FPGA and the board is ready to be configured. If not, use a Xilinx JTAG programmer to program the FPGA with the Parspec design bitstream.

\end{enumerate}

\section{Software Configuration}

There are two main categories of settings that need to be configured to enable successful operation of Parspec. These are {\it connectivity} settings and {\it spectrometer data} settings. This however is a categorization only for convenience in this guide, and in practice all configuration options are accessible in the same way on the IBOB. \\

All configuration of the design is done over a simple telnet terminal known as {\it TinyShell}. TinyShell features a set of commands that lets you interrogate and modify registers and RAMs in the Parspec design. \\

Once the IBOB is turned on, it should be possible to connect to the board over telnet (port 23) by running a telnet client on the control computer (see ``Hardware Setup'' section). The IBOB runs a telnet server on port 23. Your IBOB's IP address is defined by a set of jumpers on the board; if you don't know your board's IP address, the easiest way to check is to connect to the board using a serial connection\footnote{If you don't know how to do this, please contact your friendly local CASPER representative.} and type {\tt ifconfig} (followed by {\tt <enter>}). \\

It is important to note that it is {\bf necessary to configure all the settings described in this guide} -- none are optional. \\

\subsection{Connectivity Settings}

Convincing your IBOB to send data to your data recorder computer requires that the 10GbE connection be configured appropriately. It also may require some configuration changes on your switch (if applicable) and data recorder computer. \\

\subsubsection{IBOB Configuration}

You need to set up two IP addresses and two UDP ports: the sending IP address and port of the 10GbE interface/connection on the IBOB, and the destination IP address and port (i.e. that of the data recorder computer). You also need to inform the IBOB of the MAC addresses of both the sending and receiving interfaces (the sending interface, on the IBOB, is given an almost-arbitrary MAC address, but the receiving interface MAC address must be correct). \\

All settings are provided as integers, either in decimal or hexadecimal format, rather than as strings. \\

\begin{enumerate}
  \item Set up the destination IP and UDP port using the following commands:
        \texttt{
        regwrite reg\_ip 0x0a000004 \\
        regwrite reg\_10GbE\_destport0 4001 \\
        }
        This results in a destination IP address of 10.0.0.4 and port of 4001 being set (note that $\texttt{0a} = 10$, and it becomes obvious how the IP address is encoded as hex).

  \item Set up the source interface MAC, source IP, source UDP port, and destination MAC in an ARP table using the following commands: \\
        \texttt{
        line  1: write l xd0000000 xffffffff \\
        line  2: setb x40000000 \\
        line  3: writeb l 0  x00000060 \\
        line  4: writeb l 4  xdd47e301 \\
        line  5: writeb l 8  x00000000 \\
        line  6: writeb l 12 x0a000001 \\
        line  7: writeb b x16 x0f \\
        line  8: writeb b x17 xa0 \\
        line  9: writeb l x3020 x00000030 \\
        line 10: writeb l x3024 x486377c1 \\
        line 11: writeb b x15 xff \\
        line 12: write l xd0000000 x0 \\
        } \\
                
        These commands write to memory-mapped registers in the address-space. \\
        
        \texttt{line 1} isn't actually about connectivity at all, but rather the ADC mode. Since this is the section where we're describing the hardware commands, it seemed like an appropriate place to put it. Just enter it and forget about it. \\
        
        \texttt{line 2} sets the base address from which the remainder of the commands operate. \\
        
        \texttt{line 3} and \texttt{line 4} set the source MAC address. This is nearly arbitrary -- just make sure it is a valid MAC address (in particular, the first two hex digits should be zeros). In this example, the MAC address is set to \texttt{00:60:dd:47:e3:01}.
        
        \texttt{line 5} sets the gateway IP address. In this example, the gateway IP address is set to 0.0.0.0. This address does not need to be set correctly unless delivery to an IP on a different subnet is required. \\
        
        \texttt{line 6} sets the source IP address. In this example, the source IP address is set to 10.0.0.1. \\
        
        \texttt{line 7} and \texttt{line 8} set the source UDP port. In this example, this is set to 4000 (\texttt{0fa0} is the hexadecimal representation of 4000). \\
        
        \texttt{line 9} and \texttt{line 10} set the ARP table entry for the destination IP address whose fourth (and final) byte is \texttt{04} (the first three bytes are assumed to be the same as those of the source IP address). In this example \texttt{3020} refers to the address where the first two bytes of the MAC for address 10.0.0.4 are stored, and \texttt{3024} refers to the address where the final four bytes of the MAC are stored. In this case, the MAC of the destination machine (IP 10.0.0.4) was set to \texttt{00:30:48:63:77:c1}. This needs to be set accurately. \\
        
        Since the Parspec design only sends data to the single IP address specified using \texttt{regwrite reg\_ip 0xXXYYZZWW}, you only need to make sure that you have an entry for IP \texttt{XXYYZZWW} in your ARP table. Note that it is assumed in the ARP table that first three bytes of the destination IP address are the same as the first three bytes of the source (10GbE) IP address. e.g. if you set the source 10GbE IP address to be 10.0.0.1, the destination IP is assumed to be of the form 10.0.0.x. It is also possible to send data to a destination IP that is not on the same subnet as the IBOB 10GbE interface. In this case, it is necessary to set the gateway in \texttt{line 5} and to create an appropriate ARP table entry for the gateway IP. \\
        
        Clearly there are scenarios where you may wish to send data to an IP address whose final byte is not 4. This requires both setting the destination IP register accordingly, and creating the appropriate ARP entry. The ARP entries for the even IP addresses can be set as follows: the IP address x.x.x.2 entry is stored at address offsets \texttt{3010} and \texttt{3014}; x.x.x.4 at offsets \texttt{3020} and \texttt{3024}; x.x.x.6 at offsets \texttt{3030} and \texttt{3034}, and so on. The ARP entries for the odd IP addresses can be set as follows: the IP address x.x.x.1 entry is stored at address offsets \texttt{3008} and \texttt{300c}; x.x.x.3 at offsets \texttt{3018} and \texttt{301c}; x.x.x.5 at offsets \texttt{3028} and \texttt{302c}, and so on. The entries from x.x.x.1 to x.x.x.255 are addressable. \\

\end{enumerate}

\subsubsection{10GbE/1GbE Switch Configuration}

The switch needs to be set up to allow {\it jumbo} packets. The UDP payload size of Parspec packets is 2056 bytes. \\

\subsubsection{Data Recorder Computer Configuration}

The data recorder also needs to be set up to allow jumbo packets. In Linux, it is usually possible to allow large packets by setting the ``MTU'' to be sufficiently large with this command: {\tt ifconfig ethX mtu 9000}. This sets the MTU on Ethernet interface ethX (this should be your 10GbE-connected interface) to 9000 bytes. \\

\subsection{Spectrometer Data Settings}

There are three settings that Parsec provides for manipulating the data output: the {\it accumulation length} (which necessarily constrains the output data rate), post-PFB (but pre-accumulator) {\it scaling}, and (post-accumulator) {\it bit selection}. \\

\subsubsection{Accumulation Length}

You need to set the accumulation length, which defines how many filter bank power outputs are added to each other before that output is transmitted. When you modify (including setting for the first time) the accumulation length, you need to modify what is known as the ``sync period'' setting too, according to a given formula. \\

The accumulation length is set using the register \texttt{reg\_acclen}, with the caveat that the register value is one less then actual accumulation length: the command \texttt{regwrite reg\_acclen 12} sets the accumulation length to 13. The minimum accumulation length is 2 (i.e. the smallest value you can set \texttt{reg\_acclen} to is 1). \\

You need to set another register, \texttt{reg\_sync\_period}, to the following value: $n \cdot LCM(1024,k \cdot 512)$, where $k$ is the accumulation length (i.e. $k$ is \texttt{reg\_acclen}+1), $LCM(x,y)$ denotes the {\it least common multiple} of numbers $x$ and $y$, and $n$ is an arbitrary integer (we recommend that you use $n=100$). For example, if $k=13$ and $n=100$, then $100 \cdot LCM(1024,13 \cdot 512) = 100 \cdot LCM(1024,6656) = 100 \cdot 13312 = 1331200$. So to have an accumulation length of 13, you could call \texttt{regwrite reg\_acclen 12} and \texttt{regwrite reg\_sync\_period 1331200}. \\

Incidentally, the accumulation length of 13 results in a spectral dump rate\footnote{These calculations assume a sampling clock rate of 800MHz.} of approximately 30,048 spectra per second, and hence a data rate of approximately 471Mbits/second\footnote{The reader may notice that this data rate is sufficiently low that it can easily be carried by 1GbE. Why then use an expensive and occasionally troublesome 10GbE connection? Simply because the 100Mbit connection on the IBOB is not sufficient to support the output data rate for 30kHz readout, and the next fastest Ethernet port available on the IBOB is the 10GbE port.}. At the minimum accumulation length of 2, the dump rate is approximately 195,313 spectra per second, and hence the data rate is approximately 2.99Gbits/second. \\

The maximum accumulation length is 1024 (which corresponds to a dump rate of approximately 381 spectra per second), but due to the possibility of overflow, it is only advisable to set such a large accumulation length if the input signal power is sufficiently low. The maximum accumulation length irrespective of input signal power is 256 (which corresponds to a dump rate of approximately 1,526 spectra per second). \\

For pulsar studies it is typically important to know the sampling (integration) time $T_{sample}$. This can be calculated directly from the accumulation length. Specifically, assuming that the ADC is clocked at 800MHz, $T_{sample} = \left[((\texttt{reg\_acclen}+1) \times 512) / 200000000\right]$s\footnote{This formula is derived using the fact that with a sampling clock of 800MHz, the FPGA is clocked at 800MHz/4=200MHz and a spectrum is produced every 512 clock cycles.}. \\

\subsubsection{Scaling and Bit Selection}

The ADC samples data with 8-bits of precision, but in the FPGA design, this bitwidth is gradually increased to 32, which is the bitwidth of the data coming out of the accumulator. However, not all 32-bits are outputted over the 10GbE connection -- in the final stage, 8 out of the 32 bits are selected. This selection is user-controlled, but it is not arbitrary: the user must pick one of four bit selection options: bits 0-7, 8-15, 16-23 or 24-31. \\

Which 8 of the 32 bits you select relies on several factors. You typically want to select the most significant bits that are not zero (possibly with some allowance for ``room'' for RFI), and which collection of 8 bits these most significant non-zero bits will be in depends primarily on: a. {\it accumulation length}, b. {\it input signal power}, and c. {\it the scaling parameter}. Since it is difficult to calculate which collection of 8 bits you should choose to output, and it is time-consuming and error-prone to use a trial-and-error approach, Parspec provides a quick way to check: type {\tt bramdump scope\_output1$\backslash$bram} to see the full 32-bits for polarization 1, and {\tt bramdump scope\_output3$\backslash$bram} for polarization 2.\\

The command ``{\tt bramdump}'' displays the entire contents of a single block RAM. Without loss of generality let us discuss polarization 1 and {\tt scope\_output1$\backslash$bram}. The contents of {\tt scope\_output1$\backslash$bram} is the even\footnote{``Why just the even channels?'', you ask. Well, the motivation for this feature is to allow the user to get a sense of the overall power in the spectrum, and hence which bits the user should select. For this purpose looking at just one ``typical'' FFT bin should be sufficient, so being able to look at many bins is actually just a crutch to help you ensure that you don't accidentally output the wrong set of bits because you observed the levels of an atypical bin. We feel that seeing half the bins in the spectrum is sufficient for this purpose. However, it would be nice to allow the user to read out an entire spectrum for other reasons, but due to resource constraints on the FPGA this is not possible, and only the readout of half of the spectra of each of the two inputs is possible.} channels of the accumulated spectrum of polarization 1. With {\tt bramdump}, each channel value will be outputted on a new line -- the first column will be the address (i.e. channel number), and the third will be the binary representation. After the 512 even channels have been outputted, 1536 lines containing zeros will be displayed -- you can disregard these lines. \\

Once you know which bits are toggling, you can set the bit selection parameter. For example, let's say that you note that the largest value in any bin you see in polarization 1 is {\tt 00000000000000000001000100110001}. Clearly if you're forced to output only 8 bits (you are), you would like to output the most significant non-zero bits. Therefore in this case, you would want to output the second set of 8 bits, namely bits 8-15 (which are {\tt 00010001} for this particular value). \\

The command to set the output bit selection is {\tt regwrite reg\_output\_bitselect X} where $X \in \left\{0, 1, 2, 3\right\}$. Specifically if $X=0$, then the bits 0-7 are selected; $X=1$ corresponds to bits 8-15 being selected, and so on. This bit selection setting applies to both polarizations\footnote{You can use scaling to match polarization powers (it's assumed that the powers of the input signals will not be dramatically different).}. \\

What happens if the 32-bit test output shows that your bins have values that are one bit over the boundary of two collections of 8-bits? For example, let's say you have a maximum (and fairly typical) bin value of {\tt 00000001101101010000101100101001}. There is a single bit in the uppermost bit collection (bits 24-31) -- this poses a problem: if you select the uppermost bits, then you will effectively only have 1 bit of precision, but if you select the bits 16-23 you will get invalid data (since the most significant non-zero bit is missing in some, perhaps all, cases!). You clearly wouldn't have this problem if you could select which 8 bits you want without any restrictions, and in this example, you could profitably select the bits 20-27, or some such range. Unfortunately such arbitrary selections aren't possible, but it is possible to effectively shift the bits using {\it scaling} to get the same result. Again using the example presented, suppose you set the bit selection to bits 16-23, and shifted the data to the right by 4 places (i.e. a division by $2^4=16$): you would get the same net result as not performing any shifting, and selecting bits 20-27. \\

Parspec doesn't provide a mechanism to shift bits directly, but it provides an equivalent: there is an option to multiple the FFT output by an arbitrary 18-bit number {\it before} the power detection (c.f. the block diagram in Appendix C). Two scaling parameters are provided: one for polarization 1 and another for polarization 2. To set the scaling factors use the command {\tt regwrite reg\_coeff\_polA B} where $A \in \left\{1,2\right\}$ is the polarization input and $B \in \left\{0,1,2,\dots,2^{18}-1\right\}$ is the scaling value. The 18-bit scaling value is a fixed point number with binary point at 12, so the coefficient $2^{12}=4096$ corresponds to a scaling of 1 (i.e. don't change the bits at all). Because the scaling happens before the power detection, you need to scale by the square-root of what you would otherwise do. To shift by one bit to the right in the output, you need to divide by two, and hence a coefficient of $4096/\sqrt{2}\approx4096/1.414\approx2897$ would be used. Similarly shifting one bit to the left corresponds to multiplication by two, and hence the coefficient would be set to $4096\times\sqrt{2}\approx4096\times1.414\approx5792$. More generally, to shift to the right in the output by $n$ bits, you need to set a scaling coefficient of $4096/\sqrt{2^n}$, and to shift to the left in the output by $n$ bits, you need to set a scaling coefficient of $4096\times\sqrt{2^n}$. \\

Caution should be exercised when setting scaling coefficients, and it is especially important to watch out for overflow and underflow. The scaling multiplication is implemented using saturation logic, so if you set a coefficient that is too high and results in the scaled value being saturated, you should see a saturated output emerge from the accumulator\footnote{This is true if the accumulation length is less than 256 -- if it is more, then there is a possibility of overflow within the accumulator, and the accumulator does not contain saturation logic.}. If you set the scaling coefficient too small, and underflow occurs, you should be able to detect this on the output by the lack of precision. In any case, suffice to say, you should use the scaling factors as fine-grained control over your bits -- to move them between two adjacent bit collections -- and not as a way to shift by more than 4 binary places. Also note that there is more precision available on the low-end than at the high-end, so it is better to divide (i.e. multiply by scaling values less than 1) than it is to increase the values of the data. \\

\subsubsection*{A basic plan for setting the bit selection and scaling coefficient parameters}

Set your scaling coefficients to 1 (i.e. {\tt reg\_coeff\_polX=4096}). Set your accumulation length to what you will use in practice, and then look at the full 32 bits of the accumulator output (for every second bin in a spectrum) using {\tt bramdump scope\_output1$\backslash$bram} for polarization 1 (and {\tt bramdump scope\_output3$\backslash$bram} for polarization 2). Work out which set of 8 bits contains most of the bits you would like to output (typically a couple of bits preceding, and the remainder of the bits following the most sigificant non-zero bit). For example, if a typical output is {\tt 00000001{\bf 01001011}1010011000011101}, then you should pick bits 16-23 (in bold). You can then use the scaling to move the bits 3 places to the right so that 1.) most importantly, the most significant ``1'' bit appears in your selected 8-bits and 2.) your average power is what you want it to be (one might typically aim for an average power of approximately 40 on the scale of 0 to 255). In this example, you might want to shift the bits 3 places to the right, so you'd want to divide by $2^3$, which implies that you should set your scaling coefficient to be $1/\sqrt{2^3}$. This means you should set {\tt reg\_coeff\_polX=}$4096/\sqrt{2^3}\approx4096/2.8284\approx1448$. When you do this, the scope for the same polarization should output something like: {\tt 00000000{\bf 00101011}01011111011000011}. (Obviously the value won't be exactly the same, but the point is that the most significant non-zero bit will have moved to be in the output, and there is room to capture big pulses that are far above the average power.) \\

We recommend that you set the scaling coefficients as close to 1 (i.e. {\tt reg\_coeff\_polX=4096}) as possible to reduce the chance of overflow or underflow. Using the strategy described here should help you to do this. \\

\section{10GbE Packet Format}

The output packet format is very simple. The spectrometer outputs UDP packets whose payloads have the following structure: \\

\begin{table}[ht!]
\begin{tabular}{|l|l|l|l|l|l|l|l|}
\hline
\multicolumn{8}{|c|}{Counter} \\
\hline
$P_1(0)$ & $P_1(1)$ & $P_2(0)$ & $P_2(1)$ & $P_1(2)$ & $P_1(3)$ & $P_2(2)$ & $P_2(3)$ \\
\hline
\multicolumn{8}{|c|}{$\cdots$}\\
\hline
$P_1(1020)$ & $P_1(1021)$ & $P_2(1020)$ & $P_2(1021)$ & $P_1(1022)$ & $P_1(1023)$ & $P_2(1022)$ & $P_2(1023)$ \\
\hline
\end{tabular}
\end{table}

A single packet contains a single 1024-channel spectrum for both polarizations. Here the counter is 64-bits wide, and all the remaining (data) entries are 8-bits wide. Thus the total size of a single packet payload is 2056 bytes. $P_x(y)$ is the power of polarization $x$ in bin/channel $y$. \\

The counter in the packet is the value of an internal counter in the spectrometer that is only reset on an ARM/1PPS event (described in the section below). This counter increments every IBOB clock cycle (i.e. nominally it will increase in value by 200,000,000 every second). This results in the counter value incrementing by $acclen\times512$ (where $acclen$ is the accumulation length, and is equal to ${\tt reg\_acclen}+1$) between each spectrum. Thus it is possible to detect dropped packets by inspecting the counter value of the most recently received packet, and if has incremented by an amount other than $acclen\times512$ compared to the counter value in the next most recently received packet, then packet loss must have occurred. \\

\subsection{Receiving 10GbE/1GbE Packets on the Data Recorder Computer}

The Parspec IBOB will be connected either directly to a computer that has a 10GbE NIC, or to a 10GbE/1GbE switch via 10GbE, which is in turn connected to a computer using 1GbE. Regardless, the computer will receive a stream of UDP packets that have the format described in this section. \\

A quick way to test that the IBOB has been set up correctly and that packets are being outputted is to run a network protocol analyzer, such as Wireshark \cite{Wireshark}, on the data recorder computer. Because the data rate from the IBOB is expected to be high, you should only need to run a capture for approximately one second to save an ample amount of test data. To check that everything is working, you should verify that: \\

\begin{enumerate}
  \item There are packets arriving on the Ethernet interface you expect them to.
  \item The packet size of the packets is 2056 bytes.
  \item The counter in the packet payload is increasing by the correct number (for an accumulation length of 13, the value the counter should be incremented by is $13\times512=6656$).
  \item The payload contains spectral bin values that seem reasonable. To aid in this, it is helpful to have a test tone attached to one polarization input (and no signal on the other). Use the 32-bit test mode described in the {\it Scaling and Bit Selection} section to ensure that the 8-bits being outputted are correct, and then check in the UDP packet payload on the computer that the values appear as you expect them to (i.e. on a tone test, nearly all values should be zero, with non-zero values only at the spectral bin(s) corresponding to the tone frequency).
\end{enumerate}

Once you have established that the packets are arriving at the computer reliably, and contain correct data, the next recommended test step is to capture data with gulp \cite{gulp} (a network capture program that stores packets in pcap format) and process it. The command {\tt gulp -i ethX > datadump.log} will capture packets directly from the interface {\tt ethX} to the file {\tt datadump.log} until the {\tt gulp} process is killed. There is a sample C program that accompanies this User Guide that will process such a dump file and output the received spectral data as text files. (You would likely never actually do this in production, due to the increase in data size, but this program demonstrates nicely how to interpret pcap format packet dumps and the Parspec packet format. The output text files can easily be graphed in MATLAB to verify the correctness of the spectrum.) \\

\section{Precise Timing using ARM and 1PPS}

The purposes of the inclusion of a counter value in output packets are to enable the detection of packet loss, and to enable accurate timing of spectral data. When the IBOB is powered up, an internal counter starts counting with every clock cycle (nominally at 200MHz). Because the startup time is in general not known with precision, the counter value is, on its own, not very useful for time-tagging.\\

However, we provide a means to reset the counter at a precisely known time, and this enables the user to determine the time a spectrum arrived very accurately. The scheme is fairly simple: the control computer (the one connected to the IBOB via the 100MbitE port) must be set up to use NTP so that its clock is accurate to within a few tens of milliseconds. At approximately half-way through a chosen second (e.g. at approximately the time 14:26:53.5) the control computer should toggle the register {\tt reg\_arm} over telnet (i.e. call the command {\tt regwrite reg\_arm 0}, and then immediately follow this with {\tt regwrite reg\_arm 1}; the computer should be connected to the IBOB via telnet before doing this, so that it is guaranteed that this toggle will happen within approximately 0.4 seconds). This will {\it arm} the IBOB, which means that it is set to a state such that when the next 1PPS signal arrives (i.e. on the next second), the internal counter will reset to value 0. The counter will then not be reset until the next time the IBOB is re-armed and another 1PPS signal arrives. In our example, this means that at precisely (with the accuracy of the 1PPS signal and the IBOB clock source) the time 14:26:54.0, the counter will be reset. The data recorder computer can be informed that this is the case, and thus it will be aware of the precise time-stamp of each spectrum from that point on, based on the counter value. \\

\newpage

\section*{\small Acknowledgements}

The development of the Parkes Spectrometer was funded by the MeerKAT project, South Africa's National Research Foundation, and National Science Foundation Grant No. 0619596. Dan Werthimer was responsible for the conception of this project. We thank Glen Langston (with help from John Ford, Scott Ransom and Paul Demorest) at NRAO, and Andrew Jameson and Willem van Straten at Swinburne for beta testing the design and this documentation.

\newpage

\section*{Addendum A: Specifications Sheet for \textbf{Fast-readout Dual Spectrometer}}

\begin{table}[h!]
\caption{Parkes Spectrometer Specifications Sheet}
\begin{tabular}{|l|l|}
\hline
\hline
\multicolumn{2}{|c|}{Each Spectrometer} \\
\hline
\hline
Frequency channels:                            & 1024 (2048 real samples per spectrum) \\
\hline
\multirow{6}{*}{Signal input:}                 & 5MHz - 400MHz {\it or} \\
                                               & 400MHz - 800MHz (2nd Nyquist zone) {\it or} \\
                                               & 800MHz - 1.2GHz (3rd Nyquist zone) \\
                                               & \\
                                               & -20dBm to -10dBm (-15dBm nominal) \\
                                               & 50$\Omega$ SMA \\
\hline
\multirow{2}{*}{Integration time:}             & Minimum: 5.12$\mu s$ (195kHz spectral dump rate) \\
                                               & Maximum: 655$\mu s$ (1.5kHz spectral dump rate) \\
\hline
Polyphase filter:                              & 2 taps, Hamming window \\
\hline
\multirow{2}{*}{Output:}                       & Test mode: 100Mbit Ethernet. 32-bits per spectral bin.\\
                                               & Observing mode: 10Gbit Ethernet. 8-bits per spectral bin. \\
\hline
\hline
\multicolumn{2}{|c|}{Both Spectrometers} \\
\hline
\hline
Clock input:                                   & 800MHz, 0dBm to +4dBm, 50$\Omega$ SMA \\
\hline
\multirow{2}{*}{1PPS input:}                   & 0 to 3V pulse nominal (into 50$\Omega$) \\
                                               & 2V minimum, 5V maximum. Optional. \\
\hline                                               
Power input:                                   & 5V, 7A \\
\hline
Mechanical:                                    & 1x IBOB and 1x iADC board on a 6U, 8HP plate. \\
\hline
\multirow{5}{*}{Control and monitor:}          & Set up accumulation and corresponding sync period. \\
                                               & Set up IP addresses, ports, MAC addresses, and ARP table. \\
                                               & Set scaling: 18-bits, binary point at 12. \\
                                               & Set output bit selection. \\
                                               & Set ARM (optional). \\
\hline
\hline
\end{tabular}
\end{table}

\newpage

\section*{Addendum B: Sample Setup TinyShell Telnet Commands}

\texttt{// Set accumulation length and corresponding sync period\\
regwrite reg\_acclen 12\\
regwrite reg\_sync\_period 1331200\\
\\
// Set bitselection to output bits 8-15\\
regwrite reg\_output\_bitselect 1\\
\\
// Set scaling coefficients to 1\\
regwrite reg\_coeff\_pol1 4096\\
regwrite reg\_coeff\_pol2 4096\\
\\
// Set destination IP to 10.0.0.4 and destination port to 4001\\
regwrite reg\_ip 0x0a000004\\
regwrite reg\_10GbE\_destport0 4001\\
\\
// Move to 10GbE configuration\\
write l xd0000000 xffffffff\\
setb x40000000\\
\\
// Set IBOB 10GbE MAC to 00:60:dd:47:e3:01\\
writeb l 0  x00000060\\
writeb l 4  xdd47e301\\
\\
// Set Gateway IP address to 0.0.0.0\\
writeb l 8  x00000000\\
// Set IBOB 10GbE IP address to 10.0.0.1\\
writeb l 12 x0a000001\\
// Set IBOB 10GbE source port to 4000\\
writeb b x16 x0f\\
writeb b x17 xa0\\
\\
// Set destination MAC address for IP x.x.x.4 to 00:30:48:63:77:c1\\
writeb l x3020 x00000030\\
writeb l x3024 x486377c1\\
\\
writeb b x15 xff\\
write l xd0000000 x0
}

\newpage

\section*{Addendum C: Block Diagram}

\begin{figure}[htp]
	\centering
		\includegraphics[width=2.5in]{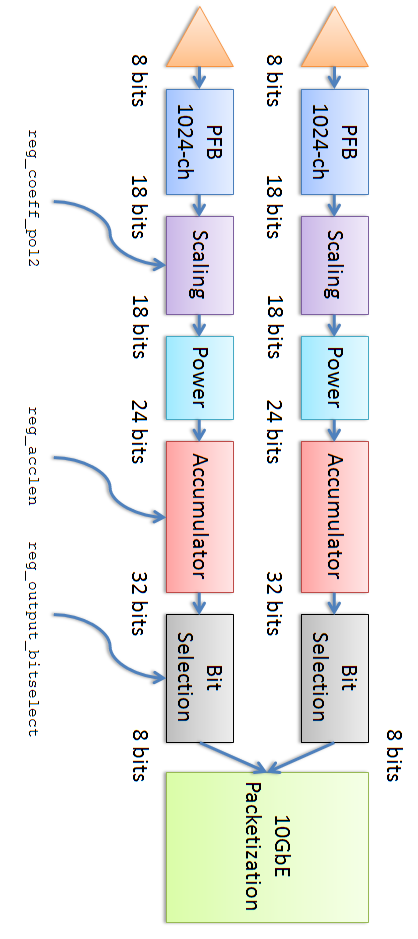}
		\caption{Block diagram of the Parspec FPGA design. The register inputs to the second spectrometer are shown.}
\end{figure}

\bibliographystyle{plain}

\end{document}